\documentclass[11pt,twoside]{article}
\usepackage{fancyhdr}
\usepackage{epsfig}
\usepackage{rotate}
\usepackage{url}
\usepackage{amsmath}

\usepackage[figuresright]{rotating}

\begin{document}
\rightline{DESY 11-259}
\rightline{POL2000-2011-002}
\vspace{2cm}
\begin{center}
{  \Large{\bf{{Polarisation at HERA}}}}

{  \Large{\bf{ -- Reanalysis of the HERA II Polarimeter Data -- }}}


  \vspace{1.5cm}{\large
  B.~Sobloher$^{\,a}$\footnote{{\bf e-mail:} Blanka.Sobloher@desy.de}, 
  R.~Fabbri$^{\,a,b}$\footnote{{\bf e-mail:} r.fabbri@fz-juelich.de},
  T.~Behnke$^{\,a}$\footnote{{\bf e-mail:} Ties.Behnke@desy.de},
  J.~Olsson$^{\,a}$\footnote{{\bf e-mail:} Jan.Olsson@desy.de},
  D.~Pitzl$^{\,a}$\footnote{{\bf e-mail:} Daniel.Pitzl@desy.de},
  S.~Schmitt$^{\,a}$\footnote{{\bf e-mail:} Stefan.Schmitt@desy.de},
  J.~Tomaszewska$^{\,a}$\footnote{{\bf e-mail:} Justyna.Tomaszewska@desy.de}}

  \vspace{1cm}
  $^a${\it Deutsches Elektronensynchrotron DESY\\
  Notkestrasse 85, D-22607 Hamburg, Germany}\\
  \vspace{0.3cm}
  $^b$Now at: {\it Forschungszentrum J\"ulich FZJ\\
  Wilhelm-Johnen-Strasse, D-52425 J\"ulich, Germany}

\end{center}

\vspace{2cm}
\begin{abstract}
\noindent In this technical note we briefly present the analysis of the HERA polarimeters (transversal and longitudinal) as of summer 2011. 
We present the final reanalysis of the TPOL data, and discuss the systematic uncertainties. A procedure to combine and average LPOL and TPOL
data is presented.
\end{abstract}
\newpage

\thispagestyle{plain}
\tableofcontents
\thispagestyle{plain}
\clearpage

\section{Introduction}
After the upgrade of the HERA machine longitudinally polarised lepton beams were available to the HERMES, the H1 and the ZEUS Experiment. 
The degree of polarisation was measured for nearly all available data with two independent polarimeters, the transverse polarimeter TPOL, 
located close to the HERA-West interaction point, and the longitudinal polarimeter LPOL, located close to the HERA-East interaction point. 
Throughout the HERA II running period from fall 2003 to mid 2007 typically $\sim 97.8\%$ of the integrated luminosity used in polarisation dependent 
analyses of the experiments is covered by at least one polarimeter \cite{privcomm:Olsson}.

In addition, for some part of data in 2006 and 2007, a new polarimeter was used in place of the longitudinal polarimeter, the cavity 
polarimeter. The analysis of data from this instrument is covered in \cite{cavity}, and is not included in this note. 

In this report the analyses of the transverse and the longitudinal polarimeters are presented. For the transverse polarimeter, a completely new 
analysis method has been developed and is presented, while for the LPOL an in-depth evaluation of the systematic errors has been performed.
This is followed by a recommendation on how to treat the errors of the polarimeters and how to combine the data from the two devices to obtain 
one HERA II polarisation measurement.


\section{The LPOL Polarimeter}
The main method of the analysis of the LPOL has been unchanged for a number of years. The main focus of the work presented in this note has been a
careful re-evaluation of the systematic errors
as published in \cite{LPOL_paper}. Studies were undertaken to understand the behaviour of key parameters in more detail. Extensive searches have been conducted to look for correlations between variables in the LPOL and the LPOL/ TPOL ratio, to understand potential sources of discrepancy between the two devices. 
To this end the data of the LPOL have been restructured for 
easier access, and additional variables have been included in the 
database~\cite{REPO_DESCRIPTION}.

The values of systematic uncertainties are given in Tab.~\ref{tab:lpol}.
\begin{table}\centering
\begin{tabular}[h]{|l|c|c|}
\hline
Source of Uncertainty & $\delta P / P \,(\%)$ & Class\\
\hline
Analysing Power & 1.2 & IIu \\
\qquad - Response Function & \quad(0.9) & \\
\qquad - Single to multi Photon Extrapolation & \quad(0.8) & \\
Long term Stability & 0.5 & I\\
Gain Mismatch & 0.3 & I\\
Laser Light Polarisation & 0.2 & I\\
Pockels Cell Misalignment & 0.4 & IId\\
Electron Beam / Laser Beam Interaction Region & 0.8 & IIId\\
\hline
Total HERA I uncertainty & 1.6 & \\
\hline
\hline
Extra Uncertainty for new Calorimeter & $\leq 1.2$ & IIu\\
\hline
Total HERA II uncertainty & 2.0 & \\
\hline
\end{tabular}
\caption{\sl \label{tab:lpol} Systematic (relative) uncertainties of the LPOL measurements. The so-called HERA I contributions are described in 
\cite{LPOL_paper}. The extra contribution to the error is estimated from the studies in \cite{lpoloverview}, and should be applied to the LPOL 
values measured from July $2^\text{nd}$ 2004 onwards, after the replacement of the cracked calorimeter crystals. The table is adapted from 
\cite{pol-summer2007}.
The third column indicates the estimated class of systematic error and possible period dependence, see Sect.~\ref{sec:class} for details.}
\end{table}

\subsection{Offline Analysis}
%
The LPOL operates with a pulsed laser, which is triggered externally. 
The trigger is synchronised with the HERA clock. The $3$ns pulse 
has a non uniform time profile, and the laser firing has a sizable 
jitter of $\pm 1.5$ns relative to the HERA clock. These two effects 
generate false asymmetries on the collected Compton photon energy in 
the LPOL calorimeter, and are corrected for. 
This potentially large source of 
systematic uncertainty has been discussed in detail in~\cite{PRC_2009}, 
where no significant dependence of the LPOL/TPOL ratio on the value of 
this correction has been found. The currently released TPOL data 
were used in that analysis and in the study here reported.

\begin{figure*}[t!]
  \begin{minipage}{7cm}
    \includegraphics[ height=5.1cm, width=6.1cm]
                    {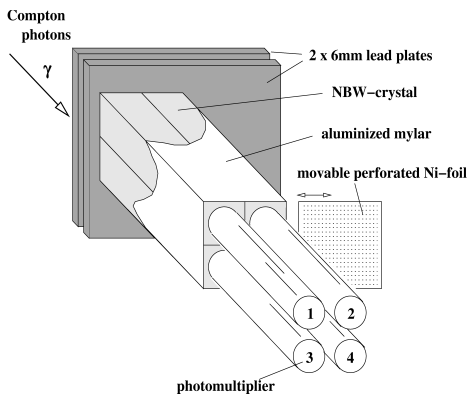}
  \end{minipage}\hfill
  \begin{minipage}{7cm}
    \caption{\sl \label{FIG:CRYSTALS}Schematic drawing of the LPOL calorimeter. 
Visible are the four crystals, used to measure the energy, and the 
four photomultipliers. The number codes of the photomultipliers and 
their location with respect to the HERA beam pipe are denoted.}
  
  \end{minipage}
\end{figure*}

Another effect which can have a potentially significant impact on the 
energy measured in the LPOL calorimeter is the background and pedestal 
subtraction. Each photomultiplier (PMT) channel 
(see Fig.~\ref{FIG:CRYSTALS}) has a pedestal, which can, potentially, 
vary from channel to channel. Each signal from PMTs is split into 
two signal lines, and the second line is installed to an additional 
ADC module channel. Each of these extra channels is delayed by $96$ns, 
so only the actual pedestal instead of the signal plus pedestal 
is gated to the ADC module. 
Since the pedestals are measured with separate ADC channels, a calibration 
between the pedestals in the delayed and the non-delayed lines is needed. 

\begin{figure*}[b!]
  \begin{minipage}{6cm}
    \includegraphics[ height=5.1cm, width=5cm]
                    {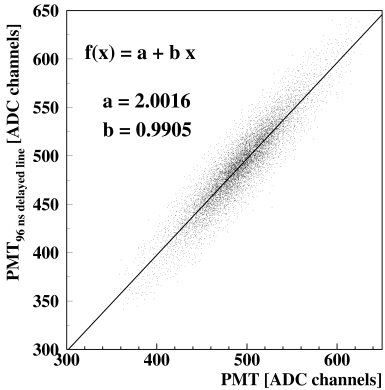}
  \end{minipage}\hfill
  \begin{minipage}{8cm}
    \caption{\sl \label{FIG:Pedestal}Example of the relative calibration of the delayed and the 
             non-delayed channels for the same PMT. 
             A linear unbinned maximum likelihood fit is performed 
             to the ADC values of the delayed versus the undelayed 
             line. The calibration is performed using laser Off events, 
             with filled and empty HERA bunches.}
  
  \end{minipage}
\end{figure*}

The calibration is performed considering laser Off events (including 
both filled and empty HERA bunches). Between the two channels a linear 
dependence is expected, whose offset and slope will give the calibration. 
An unbinned maximum likelihood fit with a linear model is done to the ADC values for the 
delayed versus the undelayed line. 
An example of a fit (taken from~\cite{menden}) is presented 
in Fig.~\ref{FIG:Pedestal}. 
The fit is performed every minute, and the resulting offset and slope 
values are then used to subtract the pedestal event by event, 
separately for each channel: 
\begin{equation}\begin{split}
        S^\text{Compton}_\text{undelayed line}  &=  S^\text{raw}_\text{undelayed line}
              - P_\text{undelayed line}  \\
        P_\text{undelayed line}   &=  
           \big( P_\text{delayed line} - \text{Offset}_\text{fit} \big) / 
           \text{Slope}_\text{fit} .
\end{split}\end{equation}

Several factors can affect the pedestal subtraction. The pedestal
calibration is performed using laser Off events on both 
empty and filled bunches. Events from filled bunches may suffer from 
additional background sources like synchrotron radiation or 
Bremsstrahlung. Events from empty bunches will not be subject to these 
background sources. Comparing the results for the pedestal determination 
for the two classes of events, 
no significant differences are observed, indicating that backgrounds 
from synchrotron radiation and Bremsstrahlung do not play a significant 
role in the pedestal calibration. 

\begin{figure*}[t!]
 \centering
  \includegraphics[height=6.1cm,width=7.0cm]
           {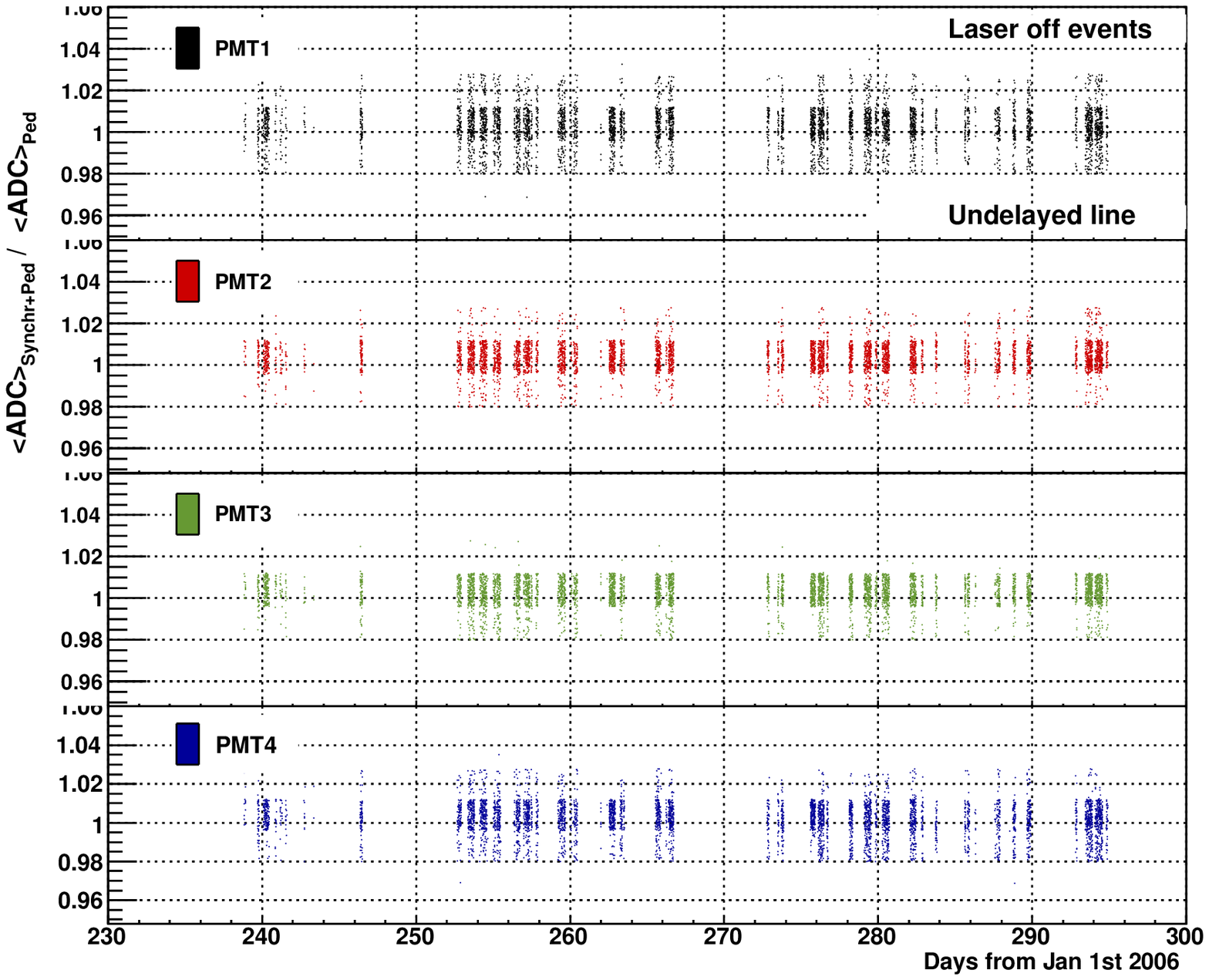}
  \hspace{-1.20cm}
  \includegraphics[height=6.5cm,width=6.1cm]
           {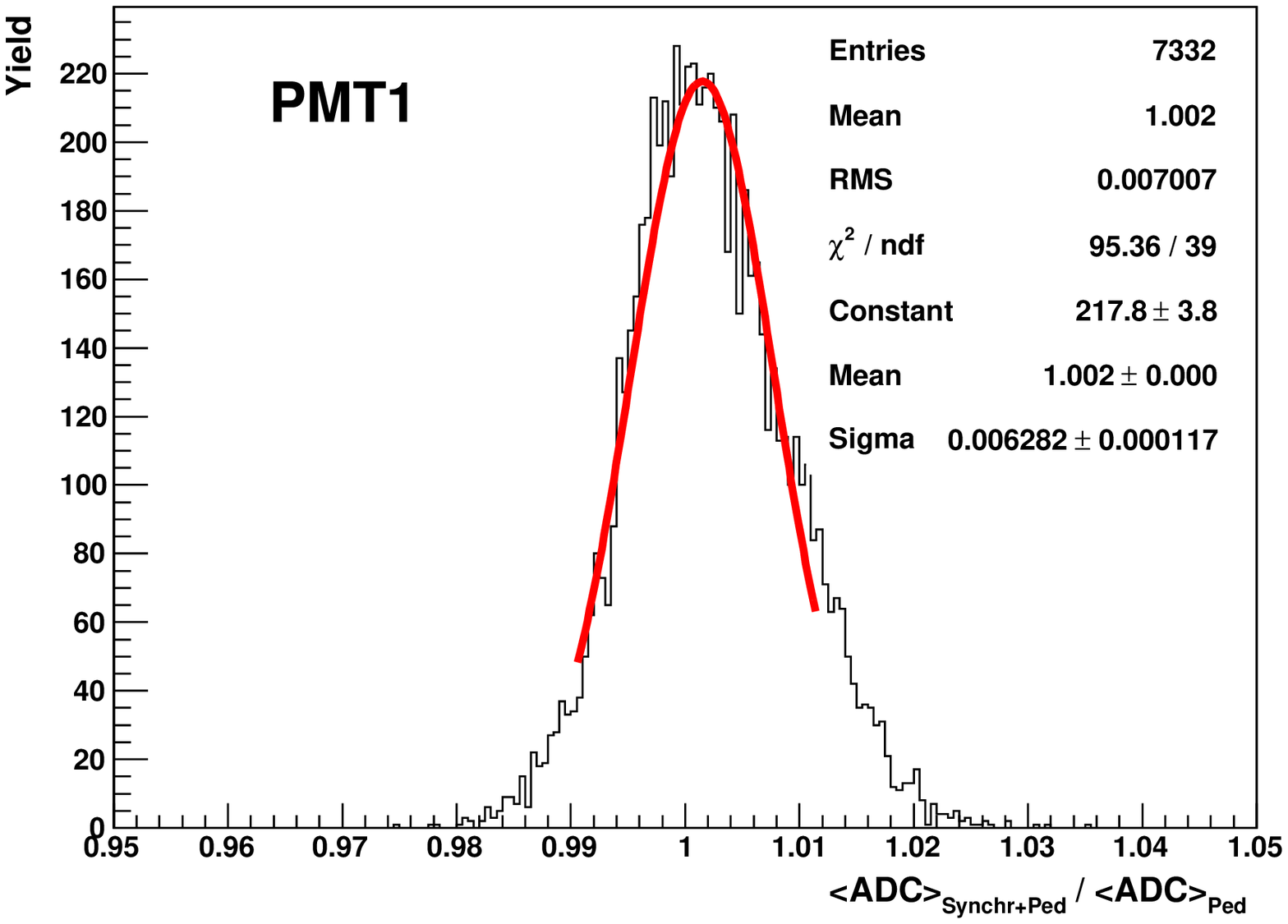}\\
  \includegraphics[height=6.1cm,width=6.8cm]
           {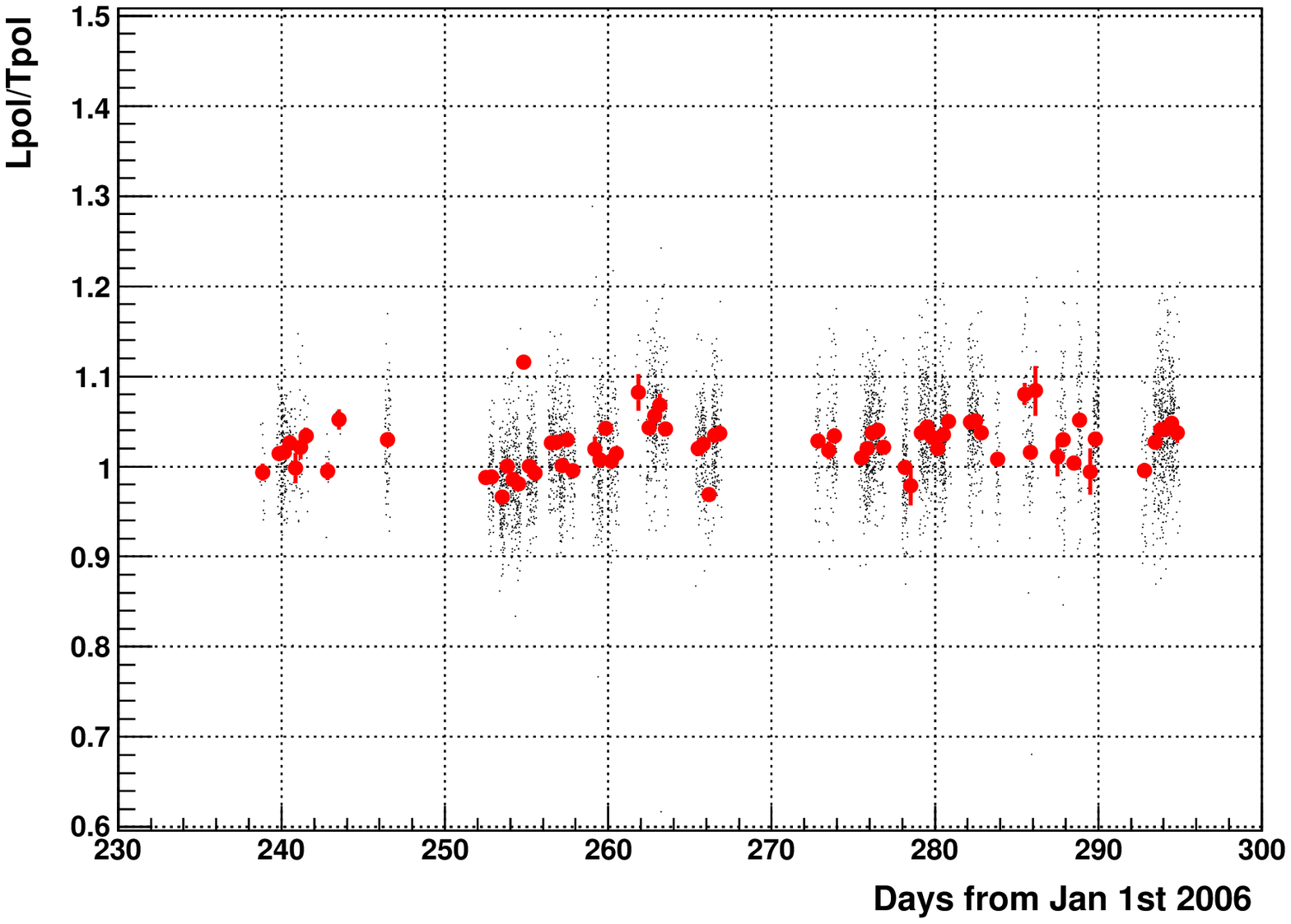}
  \hspace{-0.8cm}
  \includegraphics[height=6.1cm,width=5.8cm]
           {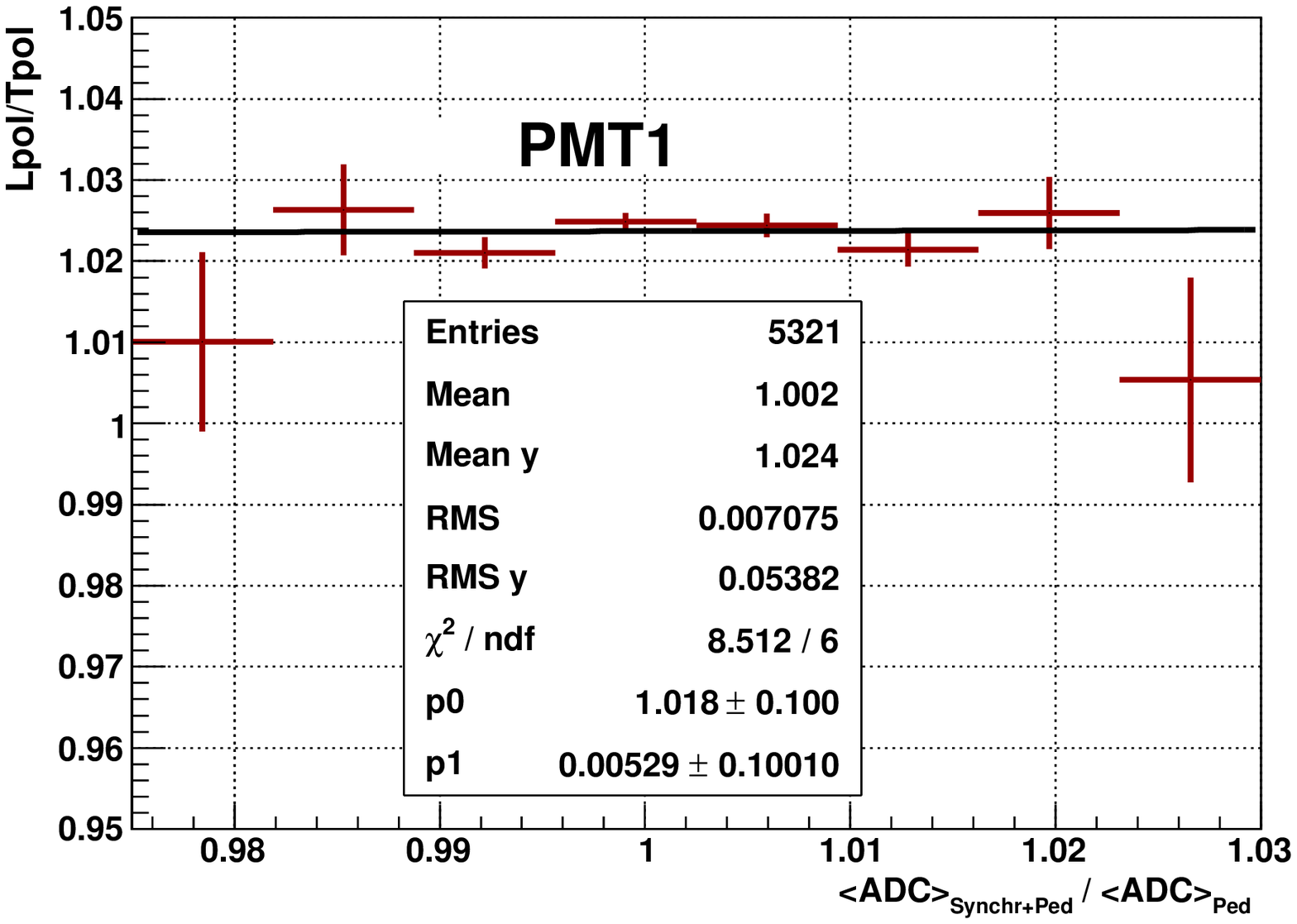}
  \caption{\sl \label{FIG:SYNCH_VS_TIME}Top left panel: Ratio A
           of the mean values for 
           background radiation plus pedestal over pedestal energies
           for a subset of data taken in the second half of 2006, as a function of time. 
           Top right panel: Distribution of the ratio A for the same data sample.
           The spread of the data is below $1\%$.  
           Bottom left panel: The LPOL/TPOL ratio 
           for the same period. 
           Superimposed to the one minute data (black points) 
           are shown the $8$ hour average values (red filled circles). 
           Bottom right panel: Dependence of the LPOL/TPOL ratio on the 
           ratio A. No statistically significant dependence is found. 
           }
  \end{figure*}

In Fig.~\ref{FIG:SYNCH_VS_TIME} the ratio of the mean values for background 
energy plus pedestal over only pedestal is presented for all four PMTs. The 
distributions do not show significant deviations from unity. They have a width 
typically below $1\%$, indicating a negligible amount of synchrotron 
and of other background photons. Also shown in Fig.~\ref{FIG:SYNCH_VS_TIME} 
is the dependence of the LPOL/TPOL ratio on this quantity, with no obvious 
dependence found. 
This is also consistent with results obtained during running, from counting 
the number of high energy photons seen by the calorimeter, but independent 
from the laser trigger. No significantly enhanced rate is observed, 
indicating that the dominant source of high energy photons in the calorimeter 
is from Compton photons. 

\begin{figure*}[t!]
  \centering
  \includegraphics[height=8.1cm,width=14.1cm]
                   {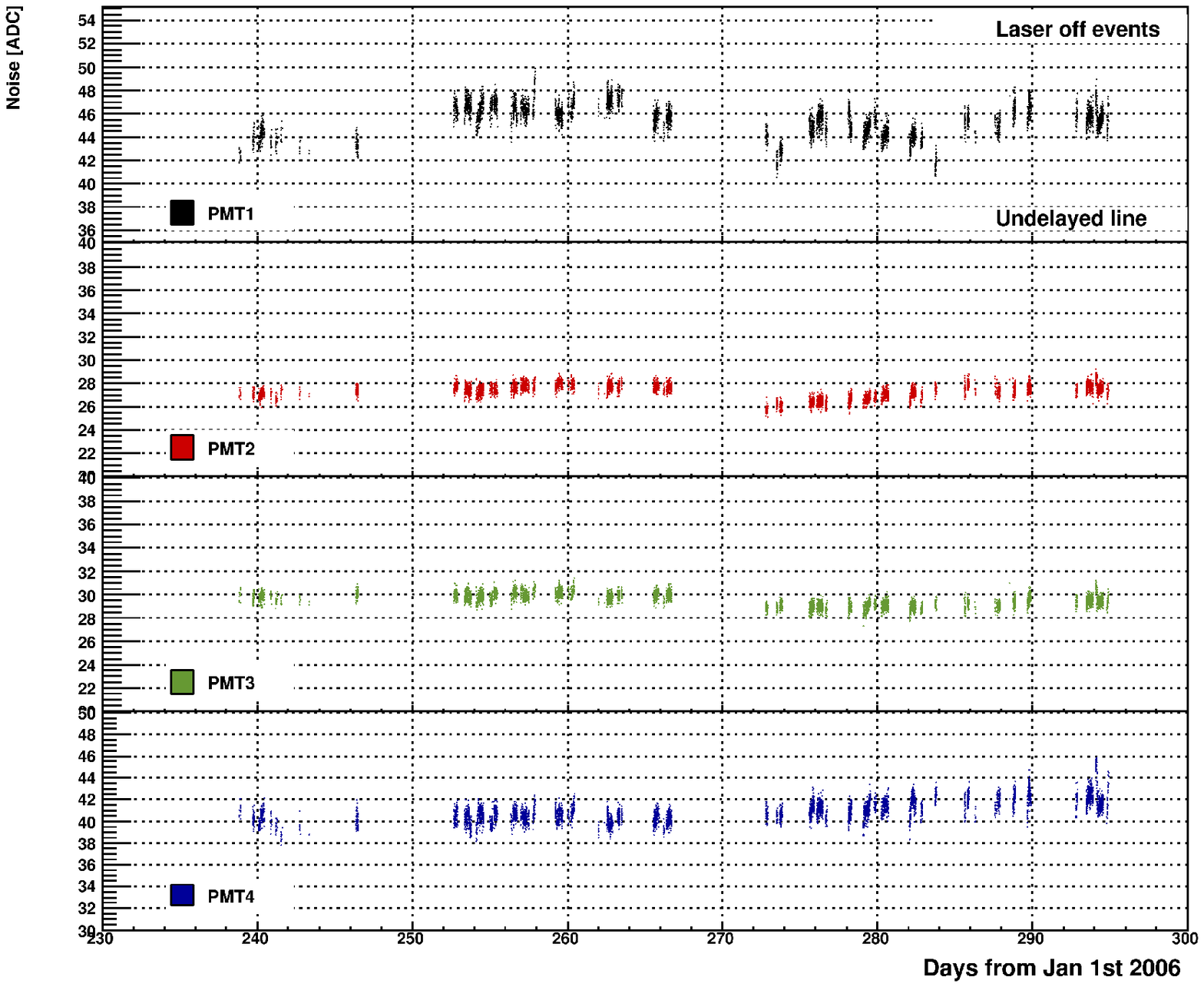}
  \includegraphics[height=8.1cm,width=14.1cm]
                   {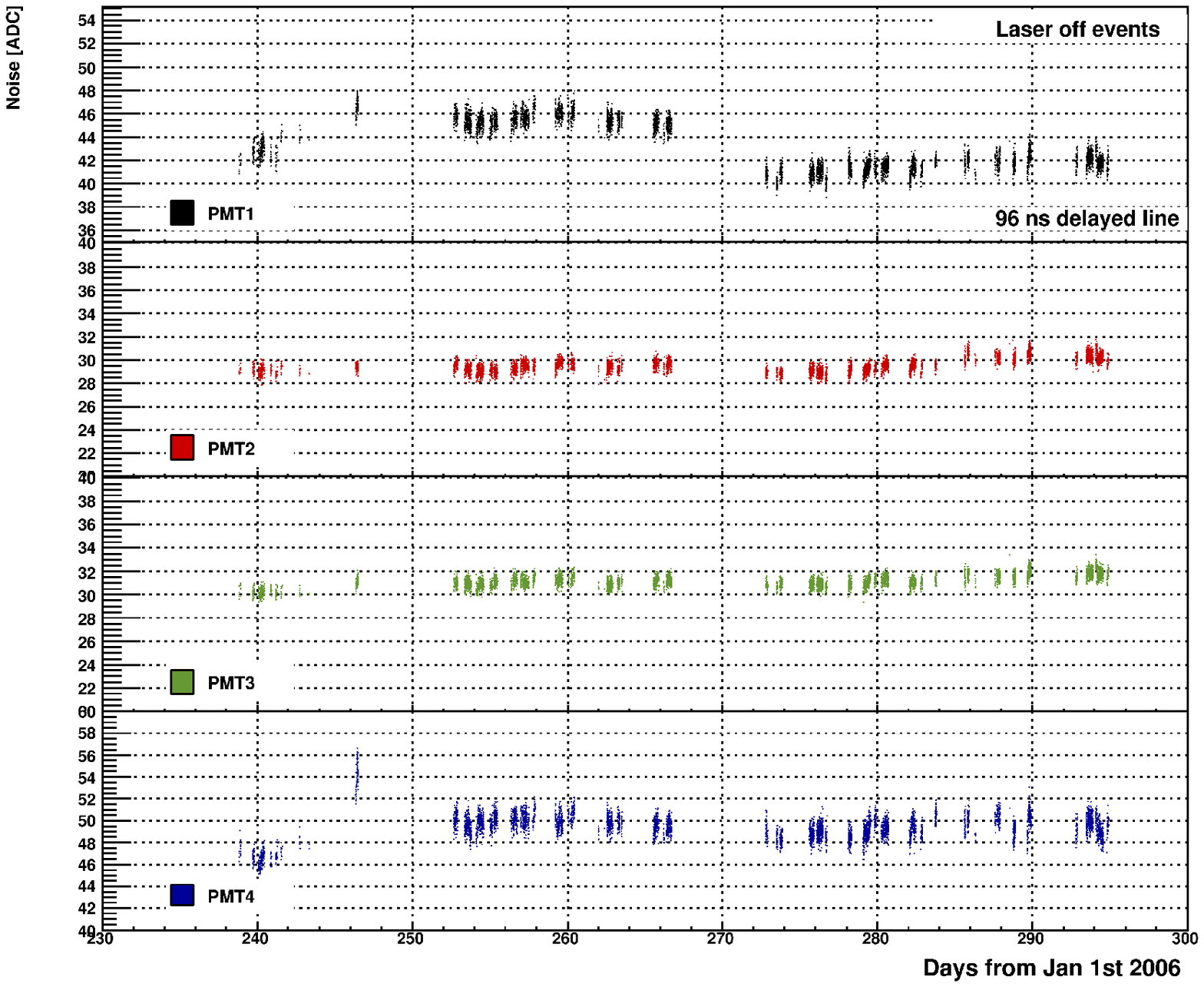}
  \caption{\sl \label{FIG:RMS_VS_TIME}Width of the pedestal distribution for 
laser Off events
           in the undelayed (top panels) and in the $96$ns delayed 
           ADC channels, as a function of the time.}
\end{figure*}

A noisy line would result in broader ADC distributions and could affect 
the calibration parameters extracted from the linear fit, 
thus biasing the pedestal subtraction. 
The spread of the signal for all four PMTs is presented in terms of ADC values in 
Fig.~\ref{FIG:RMS_VS_TIME}, for both undelayed and delayed channels, 
and for laser Off events. 
The PMT channels $2$ and $3$ appear to be reasonably stable, 
while \makebox{channel $1$} (close to the beam pipe) shows significant 
variations. A similar behaviour (although less significant) is observed 
in channel $4$. 
The latter is located far from the beam pipe, thus suggesting the 
source of noise variation to be possibly unrelated to the HERA beam line.

The observed increase of noise might affect the calibration parameters 
extracted from the fit. 
To investigate whether any correlation exist between the increase of the 
noise, and the calibration constants obtained in the fit, the quality of 
the calibration fit is studied as a function of time. 

For unbinned maximum likelihood fits no direct 
goodness of fit quantity is available.  However it is possible to 
calculate a correlation coefficient, which tests the correlation between 
the values used in the fit, and the assumed model used in the fit. 
The correlation coefficient between two measurable quantities $x$ and $y$ 
in a sample of size $N$ is defined as~\cite{STATISTICS}:
\begin{eqnarray}
   r_{xy} = \frac{1}{N - 1} \sum^N_{i=1} 
            \Big( \frac{x_i - \langle x \rangle }{s_x} \Big) 
            \Big( \frac{y_i - \langle y \rangle }{s_y} \Big) \ , 
\end{eqnarray}
with $\langle x \rangle$ and $s_x$ 
($\langle y \rangle$ and $s_y$) are the mean value and the estimated 
variance of the variable $x$ ($y$). 
If one assumes a functional dependence between x and y of $y = f(x)$, the 
deviation between the data point $i$ and the function can be written as
\begin{eqnarray}
   y_i - \langle y \rangle = \Big( f_i - \langle y \rangle \Big) 
                           + \Big( y_i - f_i \Big), 
\end{eqnarray}
which is decomposed into a component explained by the proposed 
linear model $f = a + b\cdot x$, and a deviation not justified 
by the model. After some algebra~\cite{STATISTICS}, one obtains 
that the 
sample correlation coefficient provides a measurement of the 
ratio of the sum of deviations given by the model over the sum 
of the total data deviations, 
\begin{eqnarray}
   r = \sqrt{r_{xy}^2} = \sqrt{\frac
       { \sum^N_{i=1} \Big( f_i - \langle y \rangle \Big)^2 } 
       { \sum^N_{k=1} \Big( y_k - \langle y \rangle \Big)^2 } }\ .
\end{eqnarray}

\begin{figure*}[t!]
 \centering
  \includegraphics[height=10.1cm,width=14.1cm]
                   {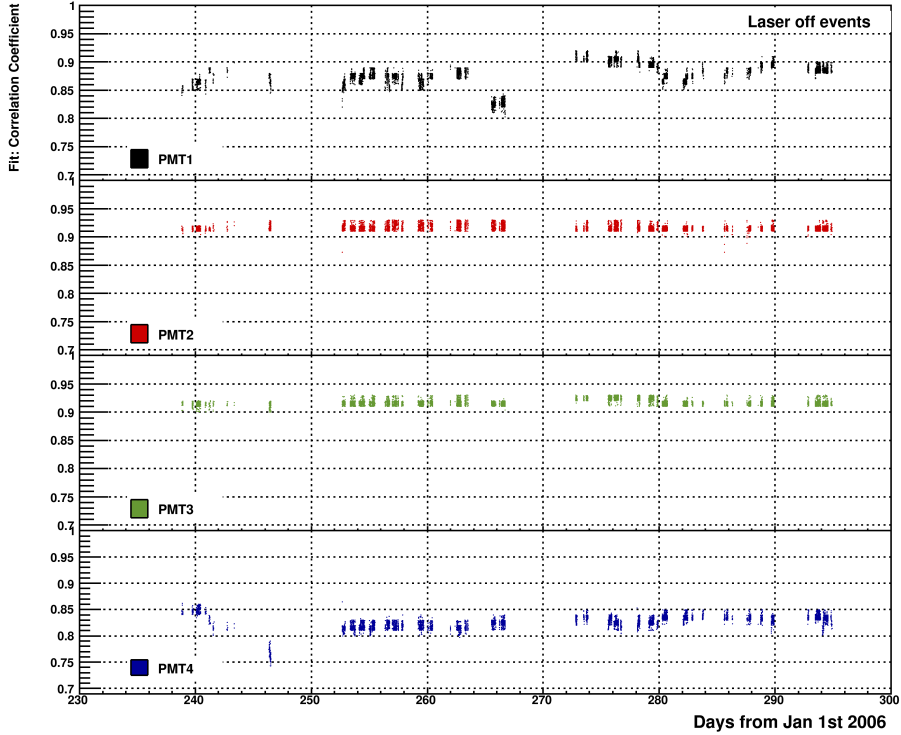}
  \vspace{-0.5cm}
  \caption{\sl \label{FIG:CORRELATION}The behaviour of the correlation coefficient (as defined 
           in the text) for the pedestal calibration is shown 
           for the analysed period in $2006$, separately for the 
           four PMTs.}
\end{figure*}

The calculated values of the linear fit correlation coefficient 
is presented in Fig.~\ref{FIG:CORRELATION} for all four PMT lines. 
As observed for the noise, the coefficients are stable for 
PMT channels $2$ and $3$, while for \makebox{channels $1$}
and $4$ more significant variations are found. Typically the values 
are larger for the less noisy channels. 
Beyond these qualitative observations, no clear quantitative 
correspondence between the noise variation and the correlation 
coefficients can be found.   

\begin{figure*}[t!]
 \centering
  \includegraphics[height=7.cm,width=6.cm]
                   {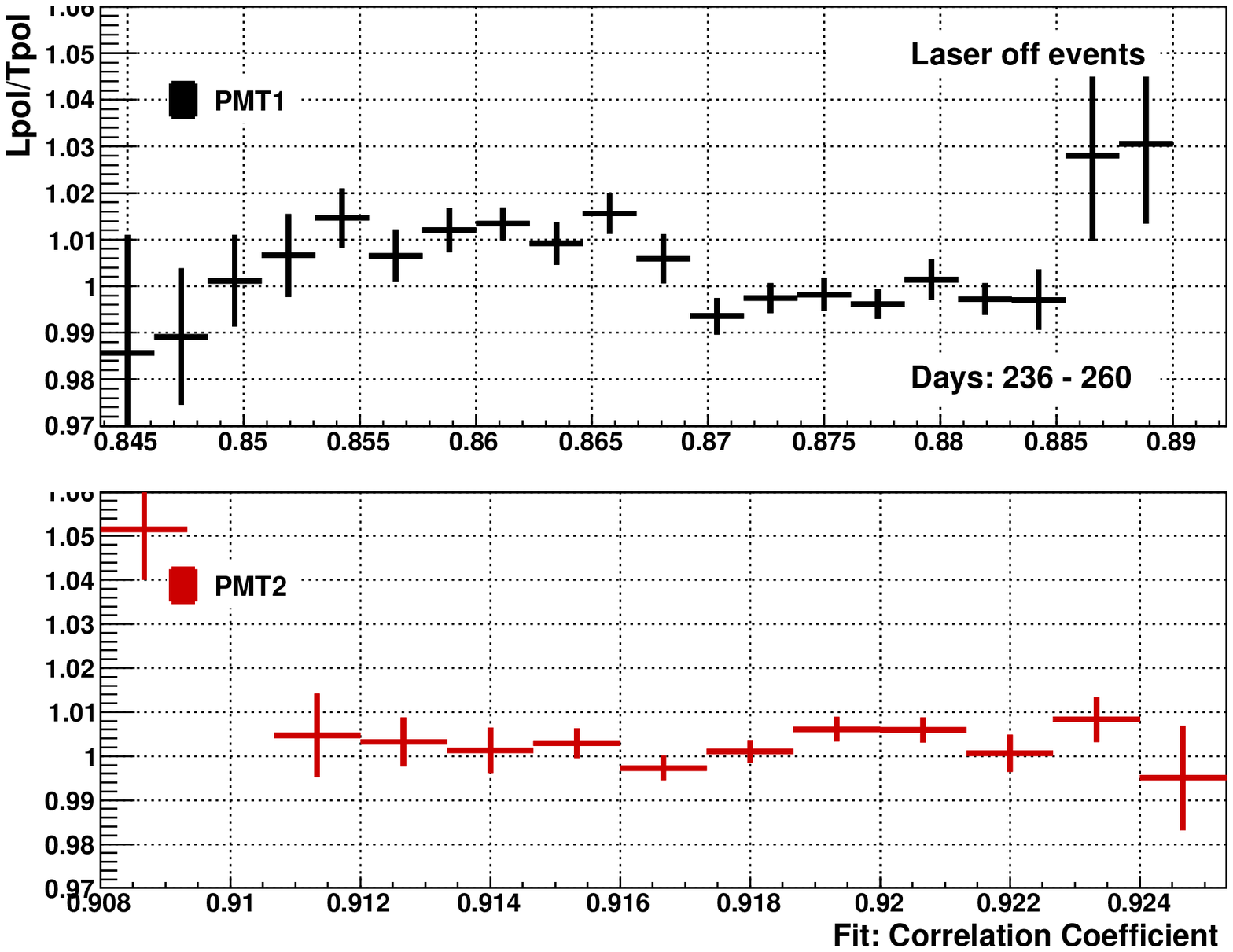}
  \includegraphics[height=7.cm,width=6.cm]
                   {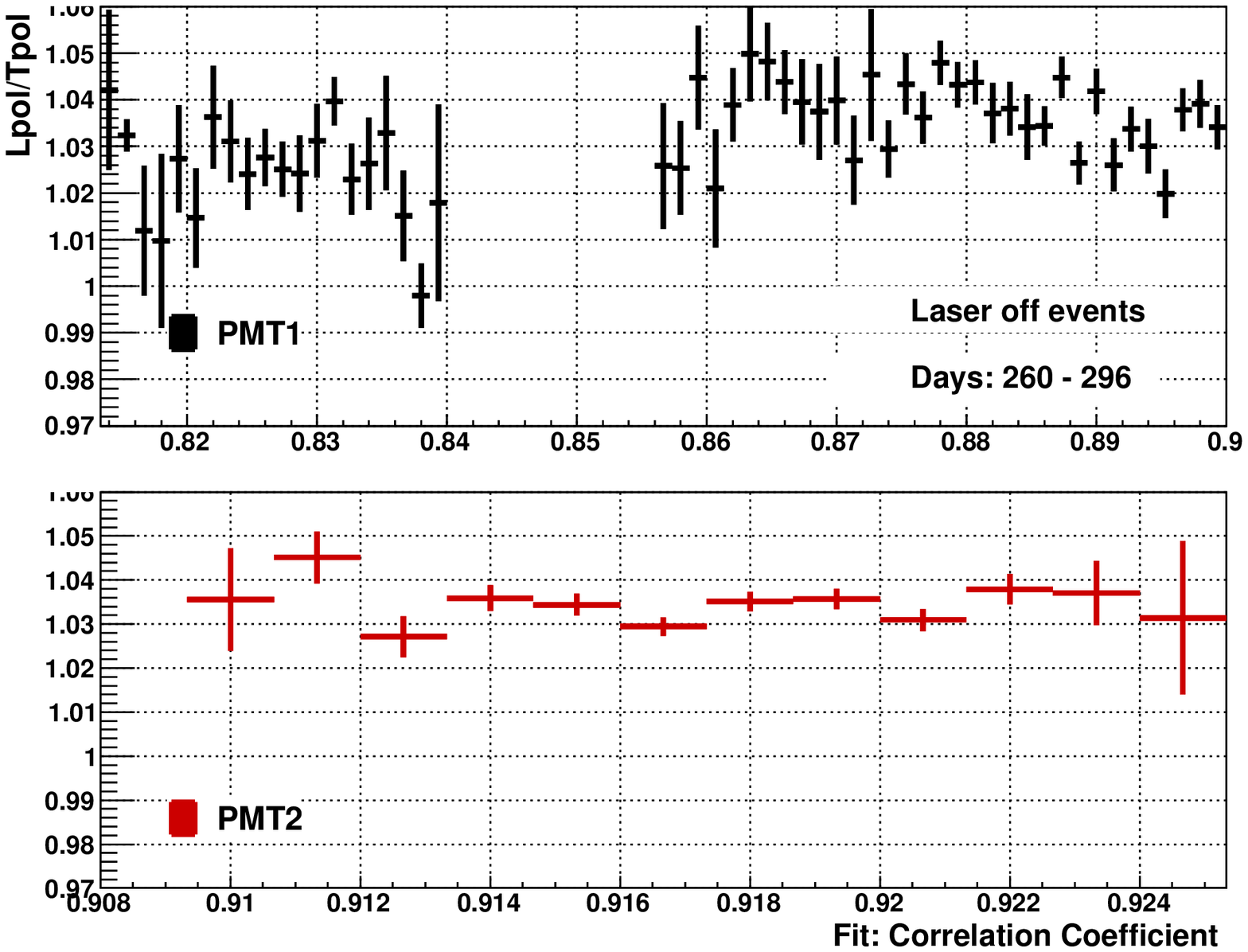}
  \includegraphics[height=7.cm,width=6.cm]
                   {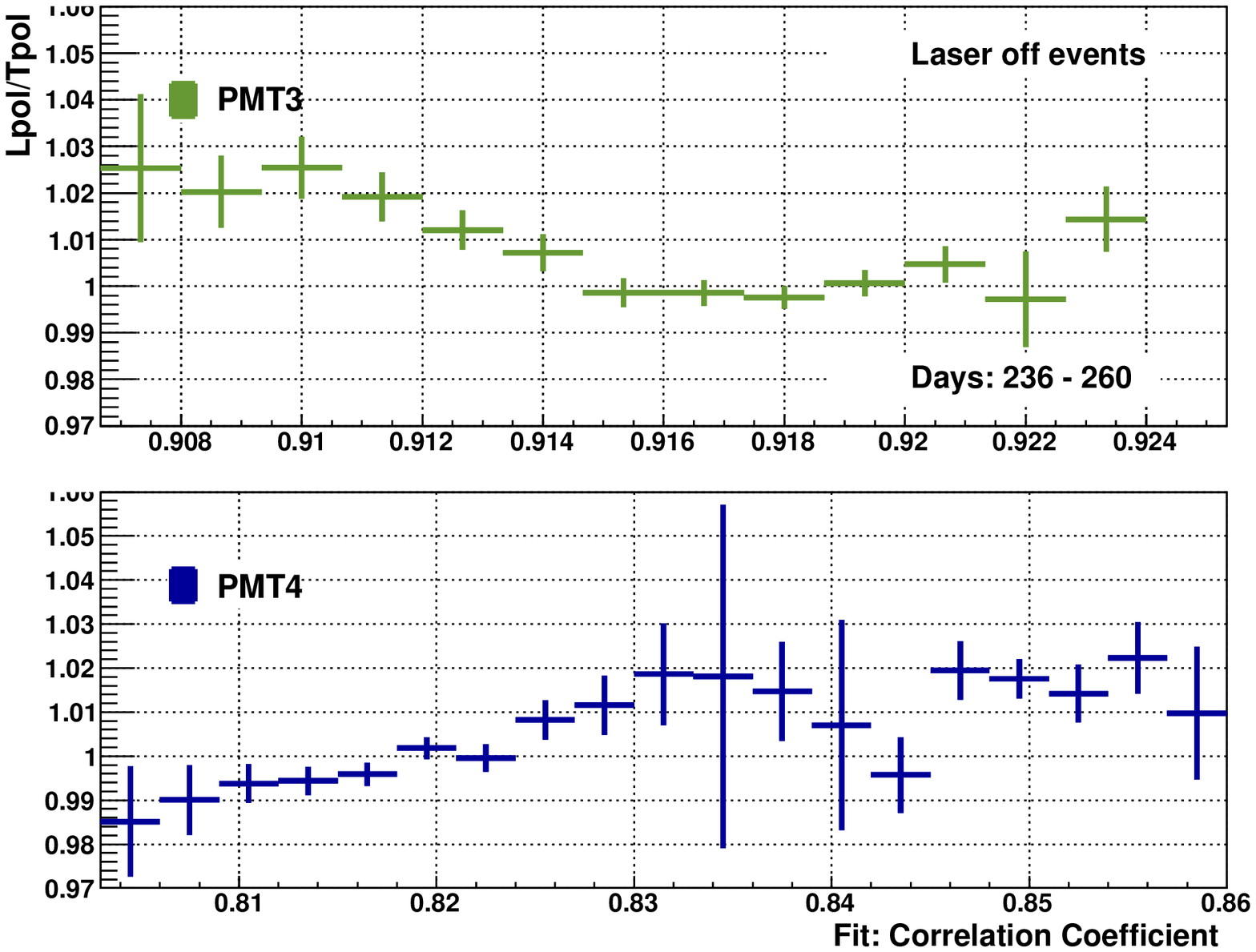}
  \includegraphics[height=7.cm,width=6.cm]
                   {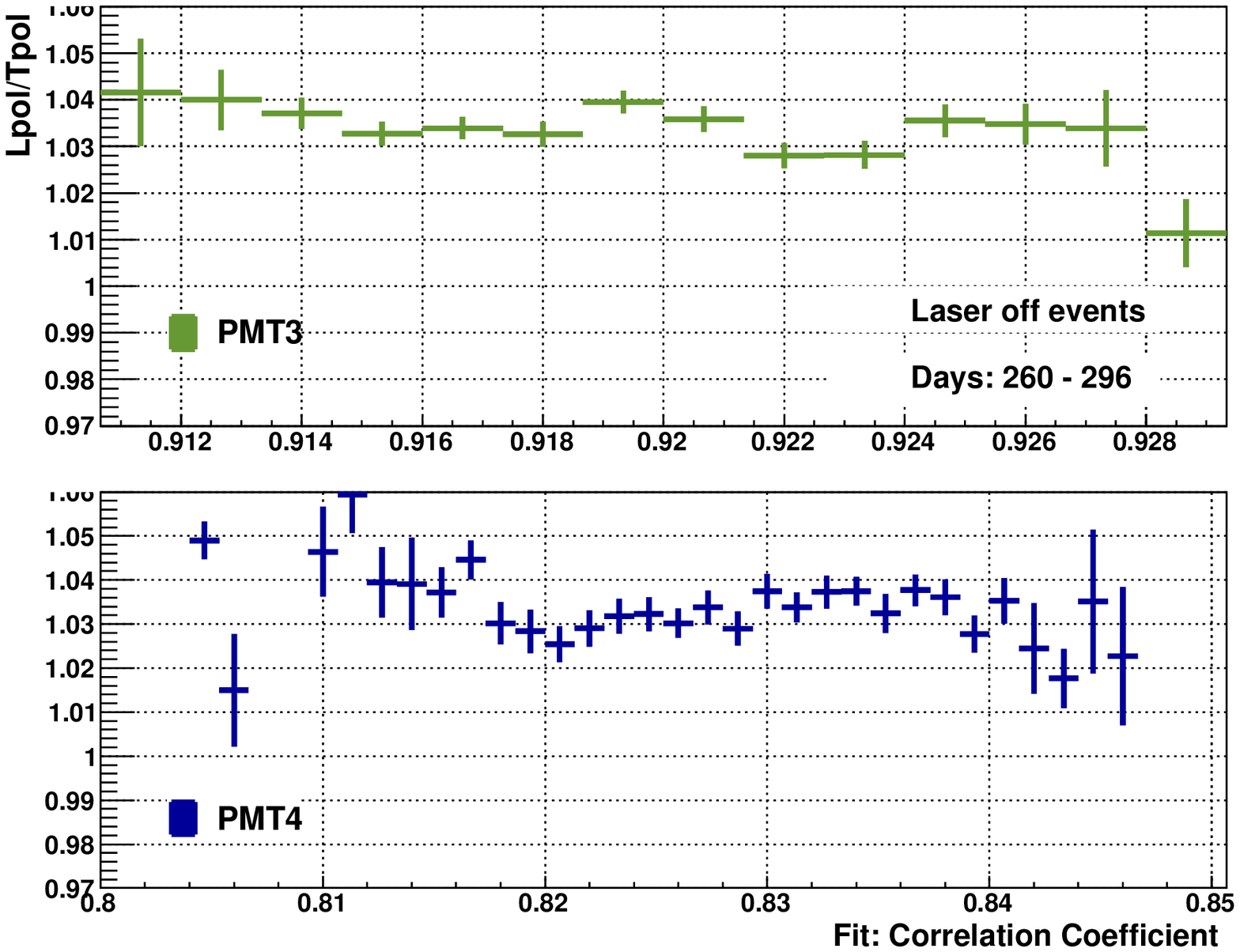}
  \caption{\sl \label{FIG:CORRELATION_VIEW}The LPOL/TPOL ratio is plotted versus the correlation 
           coefficient (as defined in the text) separately for 
           the four PMTs and for two different periods of 
           data taking.}
\end{figure*}

The LPOL/TPOL ratio is investigated versus the correlation 
coefficients in Fig.~\ref{FIG:CORRELATION_VIEW}, for two 
data taking periods of similar size, and independently for all four PMTs.
No sizable correlation is found for channels $2$ and $3$, while 
for channels $1$ and $4$ the results are less stable. 

\begin{figure*}[t!]
 \centering
  \includegraphics[height=8.5cm,width=10.cm]
                   {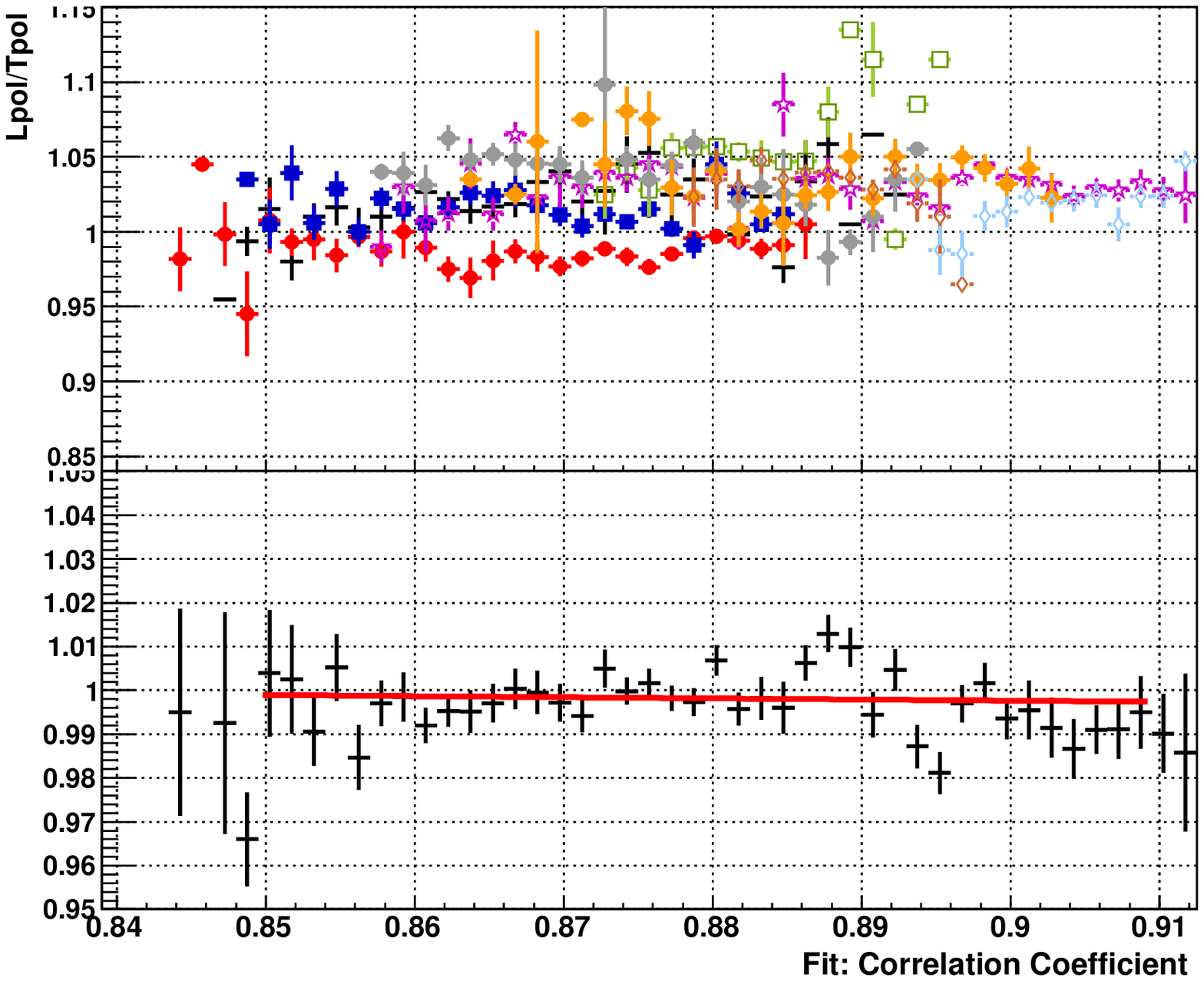}
  \caption{\sl \label{FIG:CORRELATION_VIEW_NORM}Upper panel: The LPOL/TPOL ratio versus the fit correlation 
           coefficient is shown for the PMT channel $1$ and for 
           different contiguous periods (each colour and marker code 
           corresponding to 
           a different period) fulfilling the requirements mentioned in
           the text. Bottom panel: The ratio is presented for all merged 
           data samples after normalisation. A linear binned fit is 
           superimposed to the data, showing no clear trend of the data 
           within the data precision. The extracted fit parameters 
           offset and slope are $1.021 \pm 0.047$ and $-0.026 \pm 0.054$
           respectively, for a $\chi^2/NDF$ value of $71.84/38$.}
\end{figure*}

To investigate whether a net effect is present in the data, events from all 
four PMTs and for different data periods are combined. To avoid biasing 
the data, data are grouped into periods of similar LPOL/TPOL ratio. Each 
group then is re-normalised to a LPOL/TPOL ratio of one at a correlation 
coefficient values in the bin from $0.87$ and $0.88$. The resulting distributions 
are then averaged in bins of the correlation coefficient. The results are 
shown in Fig.~\ref{FIG:CORRELATION_VIEW_NORM}. The upper panel shows
the measured LPOL/TPOL ratio as a function of the correlation coefficient
separately for all four PMT channels, and for the investigated 
data periods. One data sample (with light blue markers) has coefficient values 
outside the common normalisation region, and has been normalised to the 
only data sample overlapping its values (in red markers) in the 
region $0.90 - 0.91$, after prior common normalisation of the latter. 
The dependence of the LPOL/TPOL ratio on the correlation parameter is presented in the bottom panel of the 
picture, after merging together all the normalised data samples.
No significant dependence on the correlations coefficient $r$ is found. 

\subsection{LPOL Conclusions and Outlook}
A comprehensive re-analysis of systematic errors for the LPOL has been conducted.
Care has been taken to minimise the dependence on simulation in this, the emphasis has been on 
understanding the data and comparisons with the other polarimeters at HERA.
Within the precision possible, the background subtraction method has no impact on the polarisation measurement.

Even though the noise in the signal lines was found to vary significantly in the
channels $1$ and $4$, no significant effect on the pedestal 
calibration procedure was found, and no clear evidence for an 
impact of this variation on the LPOL/TPOL ratio was found. 

A number of other effects have been studied, including the effect of the timing of the laser pulse, 
the effect of empty HERA bunches etc., 
but no clear systematic impact on the polarisation determination was found. 

\section{The TPOL Polarimeter}
\subsection{Introduction}
The transverse polarimeter TPOL is located in the straight section West of the HERA tunnel. 
Circularly polarised laser photons are Compton scattered off the lepton beam and are transported $66$m downstream through a beam line to a 
sampling calorimeter. 
The calorimeter has been designed to measure precisely the average position of an electromagnetic shower created by a single photon. To this 
end the calorimeter is split horizontally into two halves, which are read out independently. The energy asymmetry $\eta$ defined as 
\begin{equation}
\eta = \frac{E_U - E_D}{ E_U + E_D} 
\end{equation}
is related to the vertical position of the photon hitting the face of the calorimeter through a non-linear transformation, the 
$\eta(y)$ transformation. The energies used in the definition of $\eta$ are pedestal subtracted.

The interaction rate between laser and lepton beam is such that on average less than $1\%$ of all photons are scattered back into the 
calorimeter, thus ensuring that to a very good approximation only single photons hit the calorimeter. 

The information on the polarisation of the lepton beam is contained in the vertical distribution of the photons, where vertical has been 
defined relative to the plane formed by the circulating lepton beam in the HERA accelerator. The shift in the mean of the distribution 
measured for two different states of light polarisation (positive circular and negative circular) is proportional to the polarisation of the 
lepton beam. A key parameter in this is the so-called Analysing Power, which describes the relation between the measured shift in the mean of 
the distribution and the polarisation. 

A re-analysis of the data taken at HERA during HERA II running period became necessary to optimally use the information from the polarimeters.
The old analysis, described e.g. in \cite{pol-summer2007}, exhibited some unexplained systematic behaviour, showed large fluctuations of the 
ratio between LPOL and TPOL, and did not take some known systematic effects into account. 

The goal of the new analysis is to improve the overall analysis strategy, to make it more stable, to improve the correction for known effects,
and to include corrections for new effects like e.g. the dependence on the distance between the calorimeter and the interaction point.

\subsection{Principle of the Analysis}
The fundamental principle of the analysis of the TPOL data has remained unchanged compared to the older analysis. The polarisation is 
calculated on a minute by minute basis, based on the measured spatial asymmetry between the two halves of the detector, introduced 
by the two helicity states of the laser light.

During the operation of the polarimeters the calorimeter was regularly re-calibrated using an automated procedure. This procedure ensured 
that the calorimeter is centred on the backscattered photon beam, and that the gains of the two halves of the detector are equalised. 

Background comes primarily from Bremsstrahlung photons, from synchrotron radiation and from blackbody radiation events which 
scatter into the calorimeter. Background is subtracted on a statistical basis using spectra recorded where the laser is blocked off and only 
background photons reach the calorimeter. This method removes all background contributions which are present independent of the laser light. 
In addition, the energy spectrum allows for the determination of the Compton edge of the laser backscattering and of the edge from 
Bremsstrahlung photons at the energy of the HERA beam, separately. This can be used to test the energy scale of the calorimeter and the 
pedestal subtraction method. 

An experimental independent determination of the contribution of the other background sources is much more difficult. Since during running no 
data were recorded which were not triggered by a high-energy photon, no unbiased estimator exists for the number of synchrotron radiation 
photons in particular. The level of synchrotron radiation needs to be estimated from the Compton data itself, as described later on in this 
note. 

\begin{figure}
	\centering
		\includegraphics[width=0.48\textwidth]{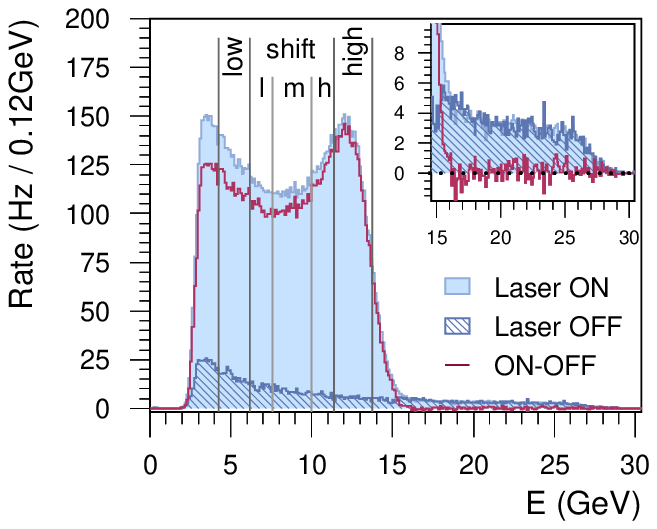}\hfill
		\includegraphics[width=0.48\textwidth]{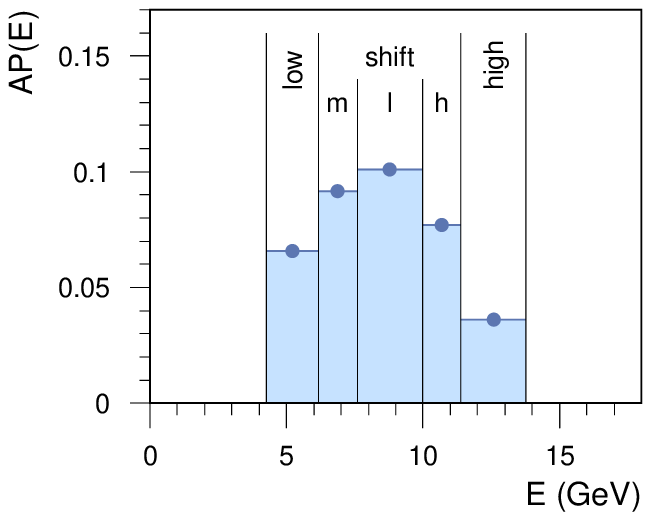}
	\caption{\sl \label{bins} Left: Compton energy spectrum with the energy bins used in the analysis indicated. The insert at the top 
right corner shows the region above the Compton edge in more detail. 
Right: Analysing Power as a function of the energy of the Compton photon, binned into the energy bins indicated on the left.}
	
\end{figure}

The analysis of the polarimeter data is done in several energy bins. In Fig.~\ref{bins} the location of the five bins in the Compton energy 
spectrum is shown. 

The most relevant observable used for the determination of the polarisation is the vertical shift of the mean energy deposition in the 
calorimeter. This shift depends on a number of external factors: 
\begin{itemize}
\item The interaction region between the laser beam and the lepton beam has a finite extension, transversely and longitudinally to the lepton 
beam. The shape of the interaction region has an impact on the Analysing Power of the system. In addition the divergence of the lepton beam 
adds to the photon beam spot size on the calorimeter surface, convoluting with the intrinsic Compton photon beam spread.
The correction was previously known as the focus correction, and was corrected for based on Monte Carlo studies \cite{focusnote}.
\item The distance between the calorimeter and the interaction point has a direct impact on the Analysing Power. For larger distances, the 
distribution of the photons on the calorimeter at fixed values of the polarisation become broader, for smaller distances more narrow. The position 
of the interaction point moves around from fill to fill and within fills. The effect of this is correlated to the dependence on the beam spot 
size. Photons in the low energy bin are sensitive to this effect as well, in addition to the beam spot size effect, photons at high energy 
are primarily dependent on the beam spot size, and not on the interaction point (IP) distance. This effect can be used to disentangle the 
two contributions. 
\item The data acquisition system of the transverse polarimeter performed an online pedestal subtraction to the photomultiplier signals of the
four calorimeter channels using a late off-time sample of the signals. There are indications that this subtraction does not completely remove 
all contributions from the pedestals, especially when being generated on-time with the lepton beam pass. 
In case a small amount of pedestal shift is present in the data this will induce a systematic reduction of the energy asymmetry, and thus the 
Analysing Power.
\item Although the laser light is measured to be $100\%$ circularly polarised at the location of the laser, imperfections in the transport 
optics result in a small residual linear light polarisation. A non-vanishing linear light polarisation biases the measured value of the 
polarisation. In the older analysis no correction based on this effect was applied, but a systematic error was assigned instead. 
During the low energy running in May and June of 2007, values of the linear polarisation larger than usual have been observed, due to some 
damaged optics element. The light polarisation was measured in between the fills. These values are used in the new analysis to correct the
polarisation measurement to the measured value of linear light polarisation.
\end{itemize}

\begin{figure}
	\centering
		\includegraphics[width=0.48\textwidth]{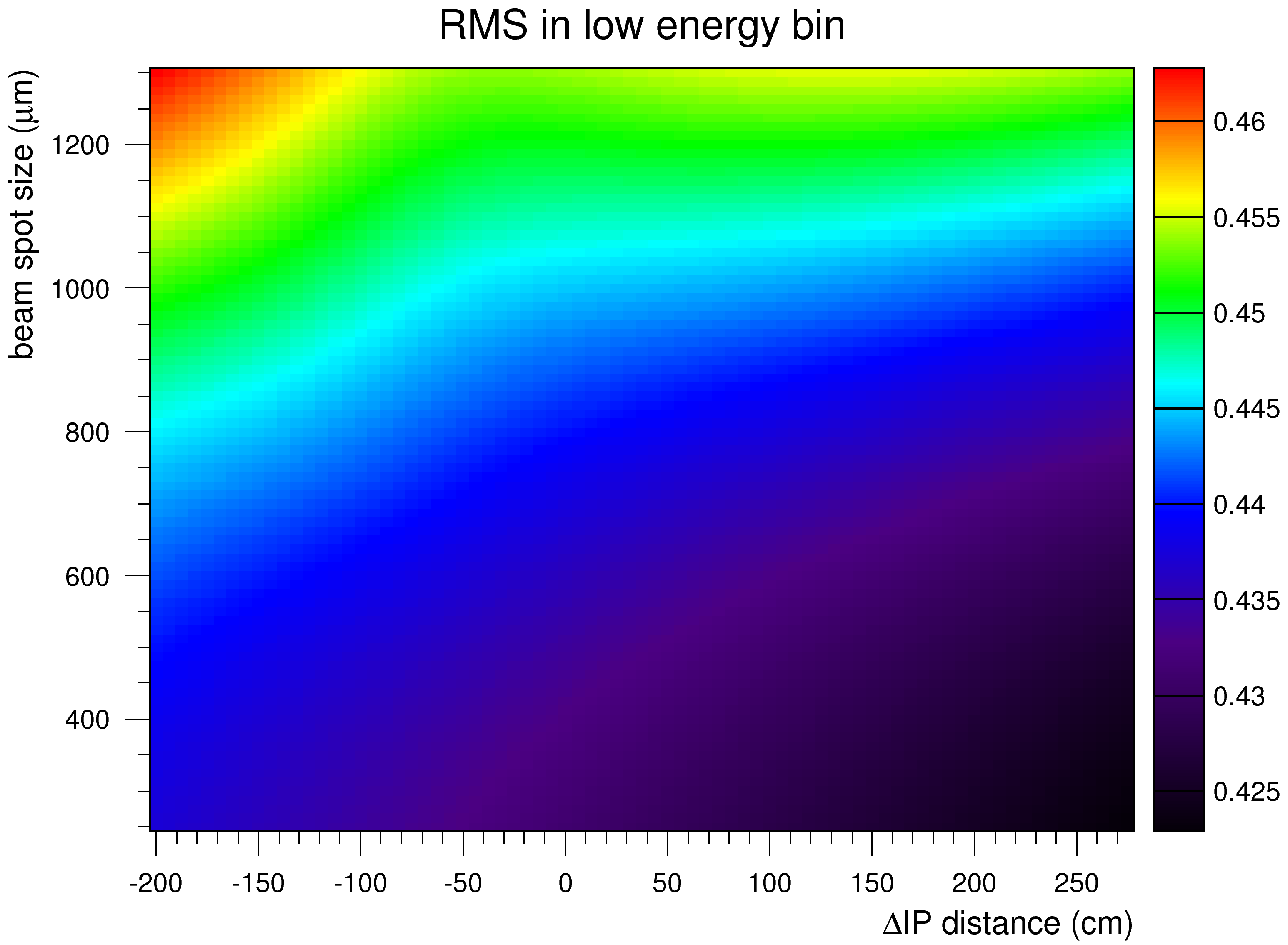}\hfill
                \includegraphics[width=0.48\textwidth]{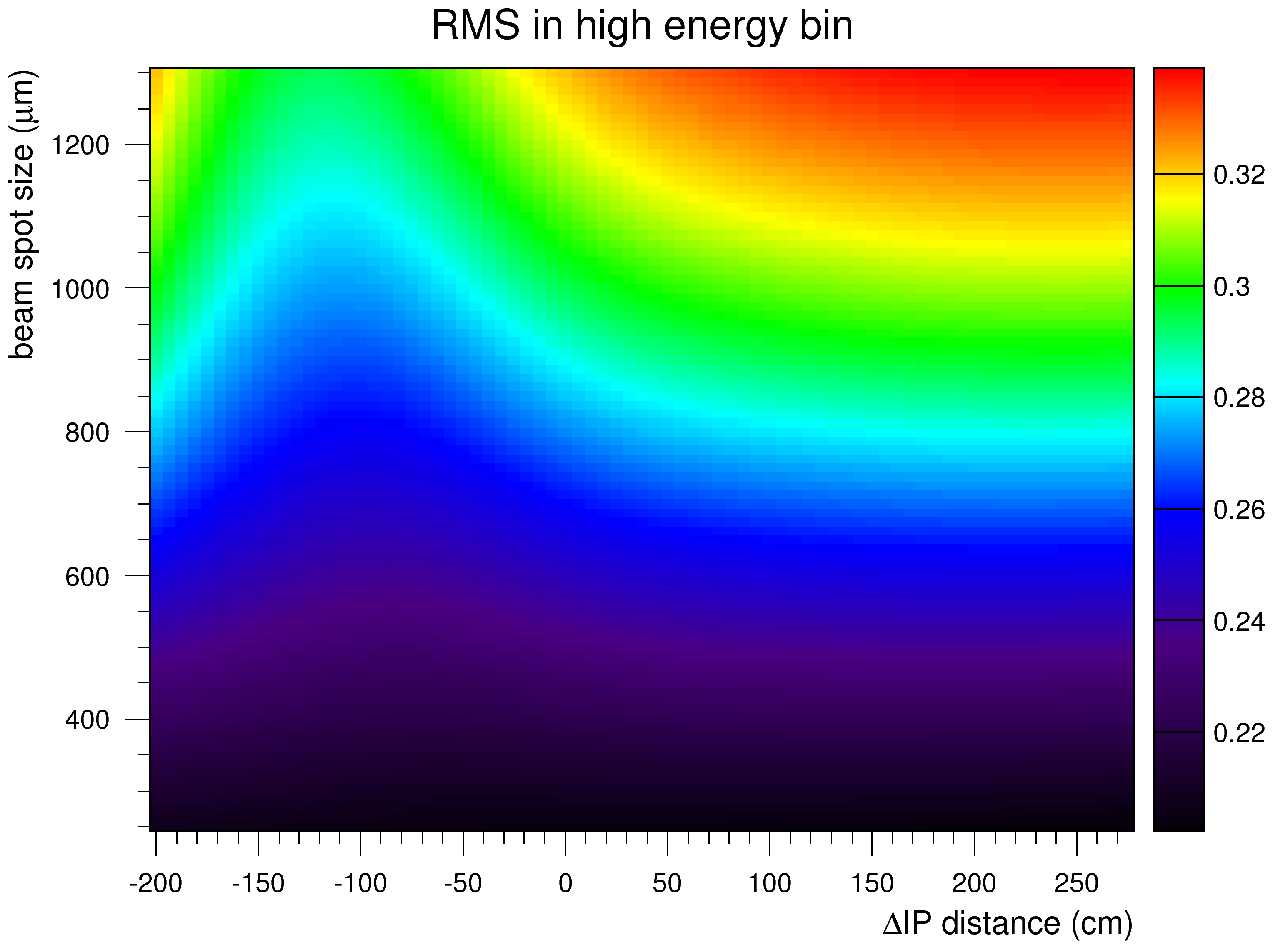}	
	
  \caption{\sl \label{map}Two-dimensional map (from Monte Carlo) of the RMS values of the $\eta$ distributions as function of the IP distance 
versus the beam spot size on the calorimeter, for the low energy bin (left) and the high energy bin (right) and $0$MeV pedestal shift.}
\end{figure}

\begin{figure}
	\centering
\begin{minipage}[t!]{0.48\textwidth}{
		\includegraphics[width=\textwidth]{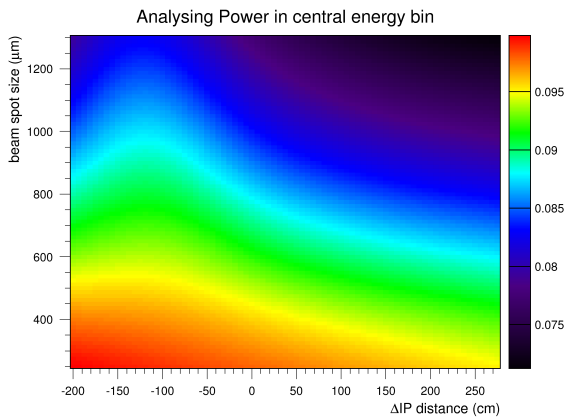}
}\end{minipage}\hfill
\begin{minipage}[t!]{0.48\textwidth}{
  \caption{\sl \label{map2}Two-dimensional map (from Monte Carlo) of the Analysing Power as function of the IP distance versus the beam spot 
size on the calorimeter, for the central energy bin and $0$MeV pedestal shift.}
}\end{minipage}
\end{figure}

In Fig.~\ref{map} the root mean square (RMS) values of the $\eta$ distributions of the upper and the lower energy bin as a function of the 
IP distance and the beam spot 
size for a pedestal shift of $0$MeV are plotted. The different level of correlation between the two variables is clearly visible. 
Fig.~\ref{map2} shows the corresponding Analysing Power. 

The shift and the width of the distribution are measured in several energy bins. The central energy bin has been optimised to give maximal 
sensitivity to the determination of the polarisation. The low and the high energy bin are much less sensitive to the polarisation, but carry 
sensitivity to other parameters of the setup. In total six energy bins are used in the current analysis (see Fig.~\ref{bins}).

\subsection{Analysis Steps}
The new analysis is done in the following steps: 
\begin{itemize}
\item Based on data taken at HERA with the Silicon detector in front of the calorimeter the response of the calorimeter is calibrated over a 
wide range of nominal impact points of the Compton photon. This is used to derive an $\eta(y)$ transformation function. As part of the 
determination of the $\eta(y)$ function a detailed parametrised model has been developed which can describe the shower in the calorimeter, 
for a range of vertical offsets, and for a range of photon energies. This model describes the average energy depositions in the upper and the 
lower half of the calorimeter and thus the $\eta(y)$ function as well as the total energy response $E_U+E_D$.

In the Silicon detector only photons which converted in the lead converter in front of the Silicon detector can be measured. Photons which do 
not convert do not leave a signal. The electromagnetic shower of converted photons however is slightly different from the one of unconverted 
photons, resulting in small differences for both the $\eta(y)$ transformation as well as the total energy response for both classes. 
In the polarisation measurement all data are accumulated, being a mixture of converted and non-converted photons. 

The $\eta(y)$ function determined from data combining both Silicon detector and the calorimeter for converted photons is shown in Fig.~\ref{etay}, the total energy 
response as determined from the same data is shown in Fig.~\ref{Etune}. A combined fit to both data sets is used to determine all relevant 
parameters of the analytical model.
\begin{figure}
	\centering
		\includegraphics[width=0.7\textwidth]{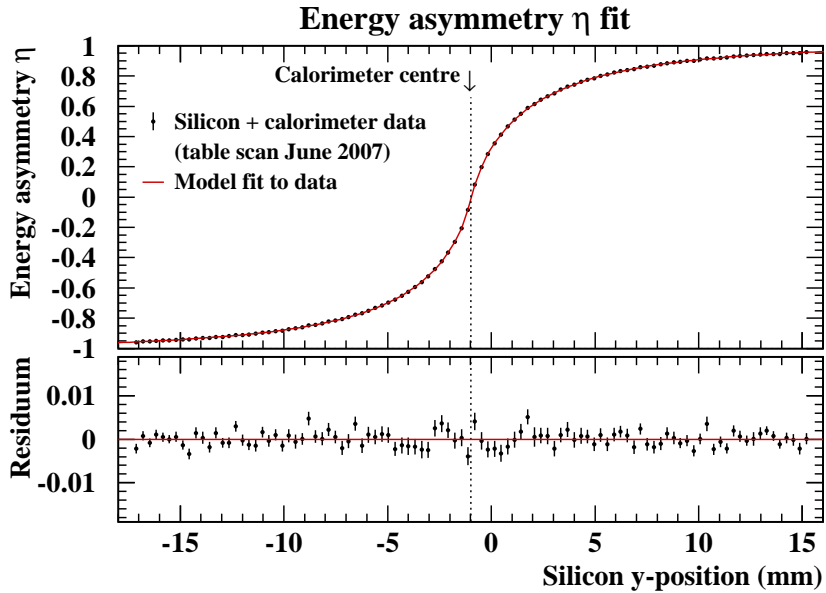}
	
	\caption{\sl \label{etay}$\eta(y)$ transformation function as determined from Silicon calorimeter combined data. Points are 
measurements, the line represents the description for converted photons used in the parametrised Monte Carlo. The bottom plot shows the 
deviations between the points and the fit.}
\end{figure}

\begin{figure}
	\centering
		\includegraphics[width=0.7\textwidth]{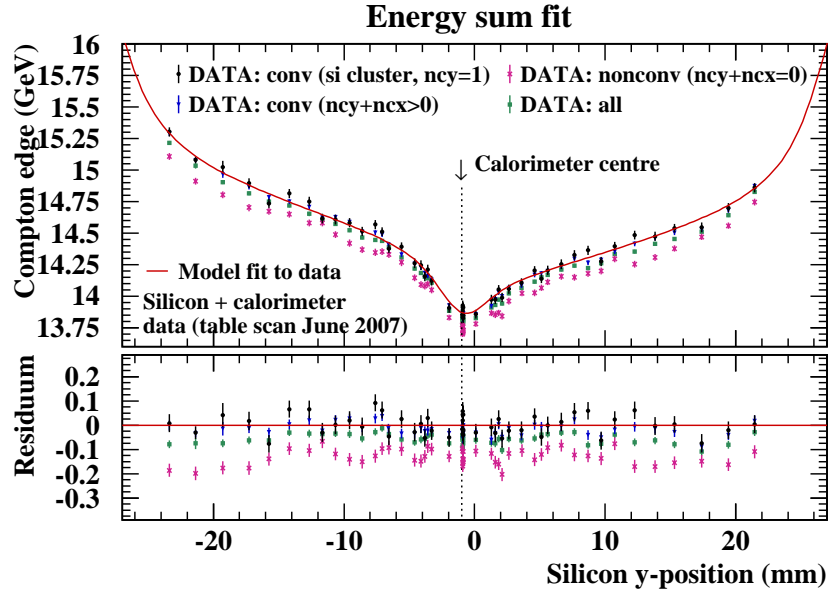}
	
	\caption{\sl \label{Etune}The measured Compton edge in the calorimeter as a function of y, as measured with Silicon calorimeter 
combined data, for converted and non-converted photons. The bottom plot shows again the difference between the measured data and the 
parametrisation for converted photons used in the analysis.
The number of clusters in the horizontal and the vertical Silicon detector are denoted with ncx and ncy.}
\end{figure}
\item The analytical physical model of the electromagnetic shower used to measure the $\eta(y)$ transformation for converted photons from the 
Silicon calorimeter combined data allows for the extrapolation to the one of non-converted photons as described in more detail below. 
The difference between the two curves is confirmed by detailed GEANT3 simulations \cite{geant3}, as is indicated in Fig.~\ref{DiffGeant}.
\begin{figure}
	\centering
		\includegraphics[width=0.7\textwidth]{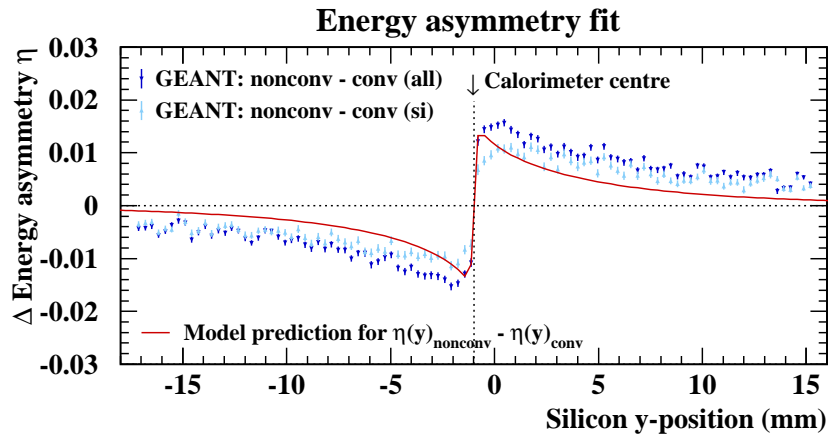}
	
	\caption{\sl \label{DiffGeant}The difference between the $\eta(y)$ function of non-converted photons to the one of converted photons as 
is predicted by the analytical shower model (line). The markers represent the results of detailed GEANT3 simulations.}
\end{figure}
\item  The energy resolution of the calorimeter has been tuned between measurements from Silicon calorimeter combined data and detailed GEANT3 
simulations. Resolution correlations between the two calorimeter halves need to be taken into account as the two halves share the same shower. 
The resolution correlations do not influence the resolution of the total response $E_U+E_D$ but have an impact on the $\eta$ resolution. The 
correlation coefficients as function of $\eta$ and energy are derived from detailed GEANT3 simulations. Their qualitative and quantitative 
behaviour is confirmed by test beam data of the calorimeter taken at CERN.
\item A detailed description of the beam line around the interaction point has been developed. It contains all relevant optical elements of 
the machine, and all aperture limiting elements. A detailed three-dimensional simulation of the interaction point between the laser beam and 
the lepton beam is available. 
\item Based on the previous steps a parametrised Monte Carlo (PMC) has been developed, using the parametrised shower and resolution models, 
which are applied to calculate an $\eta(y)$ distribution, taking properly into account the fraction of converted and non-converted photons. 
The model also includes the tuned resolutions and takes into account the beam line and apertures. It is used to calculate the Analysing Power 
for a given location of the interaction point between the laser and the lepton beam. The Monte Carlo then is used to describe the Analysing 
Power as a function of the location of the actual interaction point, the emittance of the lepton beam, and of the linear light polarisation 
component of the laser light. 
\item Since the location of the interaction point is a priori not known, the parametrised Monte Carlo is used to generate template 
distributions on a regular grid covering basically the complete phase space in IP distance, beam spot size and pedestal shift. From this the 
observables (mean and RMS values) of the energy asymmetry distributions in all energy bins are derived and mixed by reweighting methods to 
represent the desired status of linear polarisation of the laser light. Fluctuations are reduced using Savitzky-Golay filters along IP distance 
and beam spot size and cubic splines smoothing along the IP distance. The template values are then interpolated using basic splines algorithms 
and linear regression methods to generate 3-dimensional continuous, smooth and differentiable mapping functions for the observables mean and 
RMS as functions of the physics parameters IP distance and beam spot size (a function of the lepton beam emittance), a possible pedestal shift, 
valid for the linear light polarisation as measured in-between the fills.
Examples of such maps are shown in Fig.~\ref{map}.
\item Using the measured RMS values in the different energy bins, and the linear light polarisation measured for each fill, the Monte Carlo 
maps are used to find for each set of data values the best set of parameters IP distance, beam spot size and pedestal shift, which describe 
the data in each energy bin. The effective Analysing Power is then taken from the Monte Carlo maps for the set of three parameters and applied 
to the shift of mean in the large central energy bin to calculate the polarisation for this minute of data.
\end{itemize}
Even though the new analysis heavily relies on parametrised Monte Carlo, the input to the simulation has been derived to a large extent from 
data. A central role is played by the Silicon calorimeter combined data, which are used to calibrate the spatial response of the calorimeter. 
Only at points where no data are available, the parametrised Monte Carlo has been tuned to the results of detailed GEANT3 simulations using 
setups which are tuned as much as possible to describe the available data.

In the following the ingredients of the new analysis are described with additional detail. 

\subsubsection{Laser - Lepton Beam Interaction Region}
\begin{figure}
	\centering
 		\includegraphics[width=0.45\textwidth]{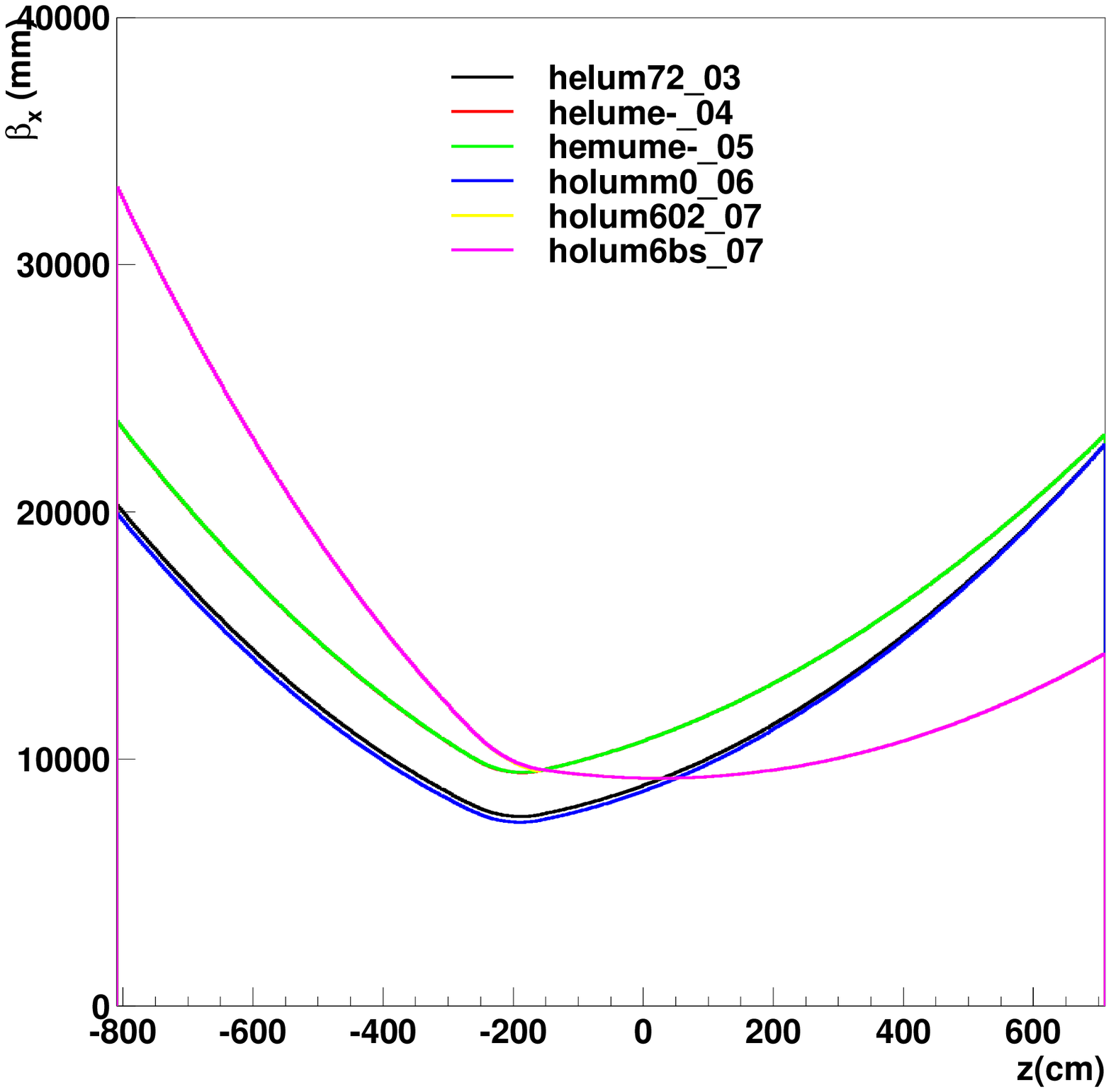}\hfill
		\includegraphics[width=0.45\textwidth]{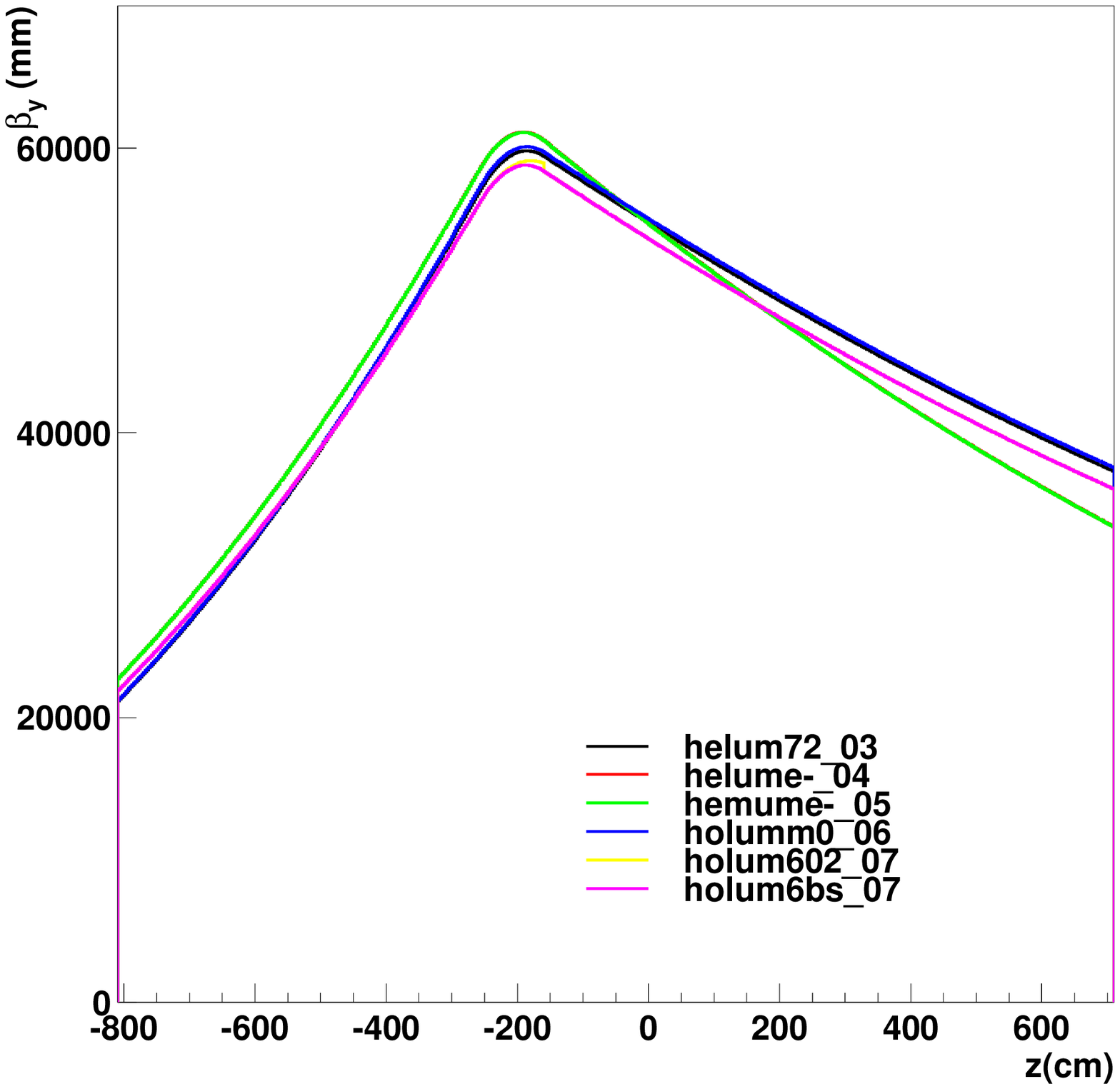}
	\caption{\sl \label{fig-beam}Predicted horizontal (left) and vertical (right) beta function for the different HERA beam optics (colours 
and names correspond to the different optics in use during HERA II running). }
\end{figure}

In the previous analysis the interaction region has been modelled in a very simplistic way assuming a Gaussian interaction region, and no real 
model of the beam line. This has been replaced by a detailed model of the beam line over several meters before and after the interaction 
region. The simulation of the behaviour of the polarimeter takes this beam line into account, and is done for one representative set of optics 
parameters for HERA. This has an impact on the expected distribution of the interaction region between the laser and the lepton beam as a 
function of time. In Fig.~\ref{fig-beam} the different beam optics are compared in terms of the vertical and horizontal beta function. The 
model used for the measurement assumes that the beam is not displaced from the nominal beam orbit. Studies with beams displaced in both the 
horizontal and the vertical direction have been made, and have shown no significant effect. 
In total seven sets of optics parameters have been in use over the HERA II running from 2003 to mid 2007. The difference to the other sets of
parameters has been studied and found to be small.

\subsubsection{Linear Laser Light Polarisation}
\begin{figure}
	\centering
		\includegraphics[width=0.6\textwidth]{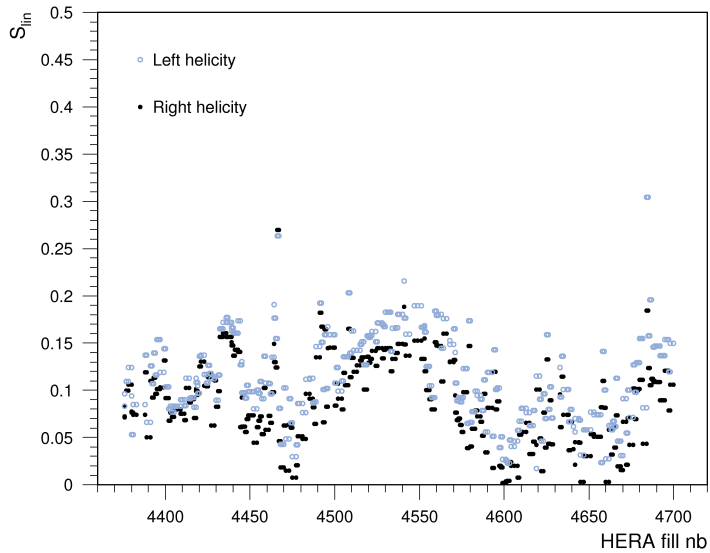}
	
	\caption{\sl \label{linlight}Values of the measured linear light polarisation for the second half of 2006, for left and right circular 
light polarisation of the laser.}
\end{figure}

The linear polarisation of the laser light is measured in between fills by optical means. In Fig.~\ref{linlight} the measurements taken 
in the fall of 2006 are shown. There are clear changes as a function of time. 
In the new analysis the maps derived from the parametrised Monte Carlo have been calculated for both full circular polarisations $S_3 = \pm 1$
and both full linear polarisations $S_1 = \pm 1$. From this the unpolarised state as well as any mixture of circular and linear polarisation 
for each helicity can be calculated by reweighting techniques to derive maps applicable to a given helicity with a certain component of linear 
light polarisation as measured before a fill. For each fill with a new linear light polarisation measurement the new analysis calculates in 
this way the weighted maps from the basic ones, thus taking the effect of linear light polarisation from the measurement into account. As the 
changes in linear light polarisation from fill to fill are typically small the light polarisation is assumed to be constant over the fill.

\subsection{Pedestal Shift}
\begin{figure}
	\centering
		\includegraphics[width=0.8\textwidth]{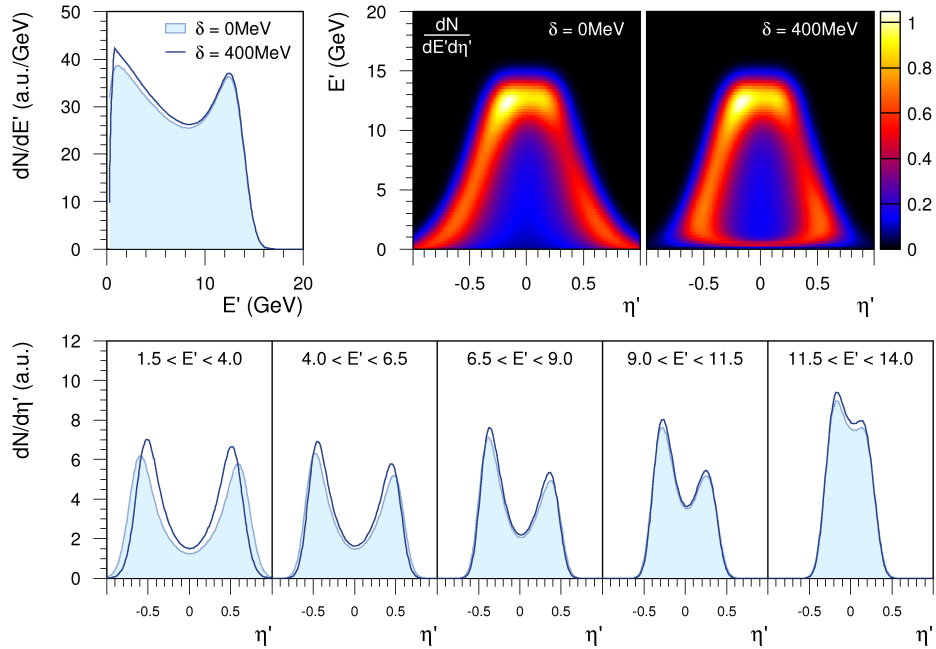}
	
  \caption{\sl \label{pedestal}Energy and energy asymmetry $\eta$ distributions without (left surface plot) and with (right surface plot) 
an assumed pedestal shift present. The effect is most pronounced in the low energy bin, but also visible in the higher energy bins. The 
histograms show how the $\eta$ distribution for different energies changes.}
\end{figure}

The photomultiplier signals from the four calorimeter channels are online pedestal subtracted. The latest sample of the signals contains
approximately only about $3\%$ of the signal and is thus dominated by the time-independent electronical pedestal. However, contributions
arising together with the actual signals in-time with the lepton beam pass cannot be subtracted with this method.
A possible source for such contributions could be additional low energy photons like synchrotron radiation or scattered black-body radiation, 
but could also arise from technical artefacts in a non-ideal pedestal subtraction routine.
The result of this type of in-time pedestals would shift the zero point of the energy scale of the calorimeter. As no independent measurements
of the zero scale exist, as no untriggered events have been recorded, there is no independent measure of the actual zero point of the energy 
scale.
If the measured energies $E_U$ and $E_D$ are different from the true energies by a small contribution $E_p$, the measured energy asymmetry 
$\eta'$ is always smaller than the true energy asymmetry $\eta$:
\begin{equation}
\eta' = \frac{(E_U + E_p) - (E_D+E_p)}{(E_U + E_p) - (E_D+E_p)} = \frac{E_U-E_D}{E_U+E_D+2 E_p} < \eta 
\end{equation}
The result would be a reduction in the Compton distribution widths, the effect being higher the lower the Compton energy. In consequence the 
RMS values in the energy bins would be systematically smaller, the low energy bin being affected most as is shown in Fig.~\ref{pedestal}, 
resulting in systematically smaller reconstructed interaction point distances and Analysing Powers.
Introducing an additional energy component symmetric in $\eta$ significantly improves the agreement between data and Monte Carlo. 
The effect that an added pedestal shift will have on the relevant energy distributions is shown in Fig.~\ref{pedestal}.
The centre plots show the correlation between energy asymmetry $\eta$ and energy, without (left) and with (right) an added energy shift. 
The histograms show how the $\eta$ spectra in several energy bins differ, without (filled histogram) and with (solid line) an added energy 
shift. To take this effect into account the Monte Carlo maps in IP distance and beam spot size are extended by adding a possible pedestal 
shift as a third free variable.

\subsubsection{Analysing Power}
\begin{figure}
	\centering
		\includegraphics[width=0.92\textwidth]{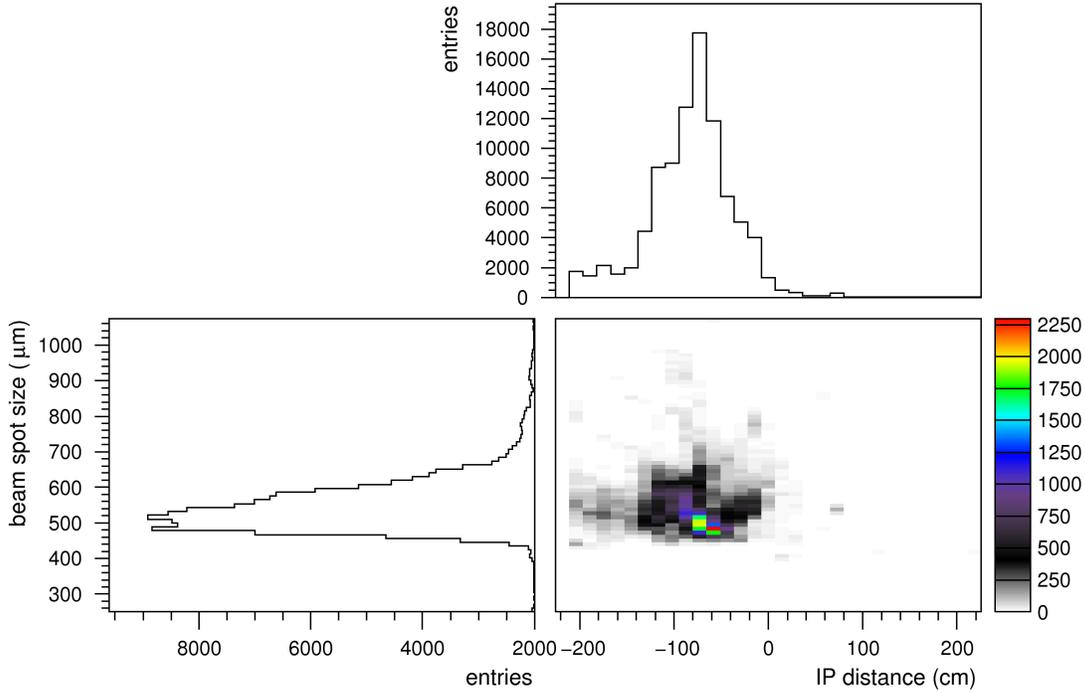}
	
	\caption{\sl \label{results-IPBS}Scatter plot of the reconstructed beam spot size on the face of the calorimeter, for different values 
of IP distance. Data are for the second half of 2006, where the HERA ring was operated with positrons and with one stable optics setup. The 
projections of the beam spot and the IP distance are shown as well. Data shown are for colliding bunches only.}
\end{figure}

\begin{figure}
	\centering
		\includegraphics[width=0.47\textwidth]{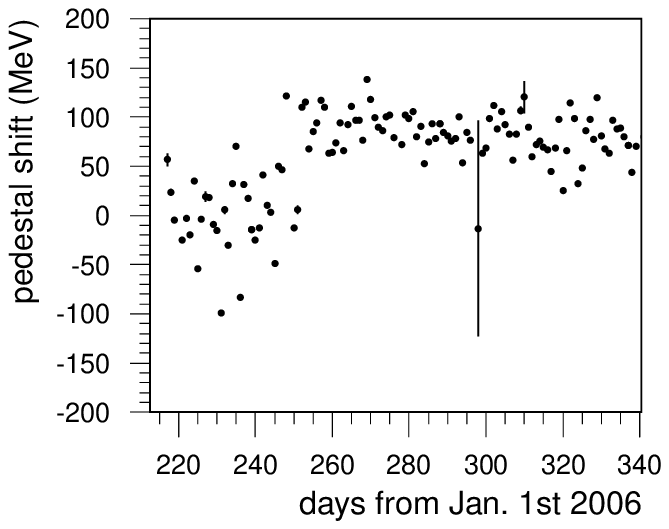}\hfill
		\includegraphics[width=0.47\textwidth]{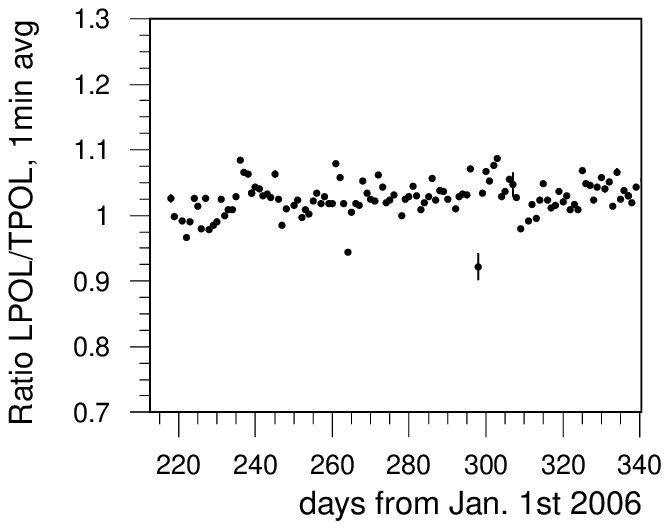}
	        
	\caption{\sl \label{ped-shift} Measured amount of pedestal shift as determined in the combined fit (left). The data represent colliding 
          bunches and was collected during the second half of 2006. If the pedestal shift is interpreted in terms of synchrotron radiation, a 
signal of $100$MeV visible in the calorimeter corresponds to less than $3$MeV in actual synchrotron radiation, as the sampling fraction of the 
calorimeter for very low energies is significantly different than for high energy photons. The corresponding ratio of LPOL over TPOL is shown on 
the right. It can be seen that the jump in pedestal shift is very well absorbed and does not introduce systematic changes to the Analysing Power 
of TPOL which are not accounted for by the analysis.}
\end{figure}

The most critical part of the analysis is the estimation of the IP distance, the beam spot size and the auxiliary pedestal shift from the 
width of the $\eta$ distributions over the five smaller energy bins. Extensive Monte Carlo has been produced to describe the behaviour of the 
system over the full phase space of the three variables, taking the full effect of linear light polarisation into account. For each set of RMS 
values in the different energy bins from one-minute data of TPOL, the best corresponding values of IP distance, beam spot size and pedestal 
shift are evaluated in a multi-dimensional fit and the corresponding value of the Analysing Power is determined from the Monte Carlo maps.
In Fig.~\ref{results-IPBS} a scatter plot of the reconstructed IP distance versus the reconstructed beam spot size is shown, together with 
the projections of the two variables on their respective axes. 
The degree of pedestal shift as determined in the fits at the same time for the same data set is shown in Fig.~\ref{ped-shift}. 
The analysis described so far is based on a detailed model of the transverse polarimeter including calorimeter, the beam line and the overall 
laser-lepton beam interaction. Nevertheless a number of parameters are not precisely known and need to be calibrated. 

\subsection{Results}
\begin{figure}\centering
  \includegraphics[width=0.97\textwidth]{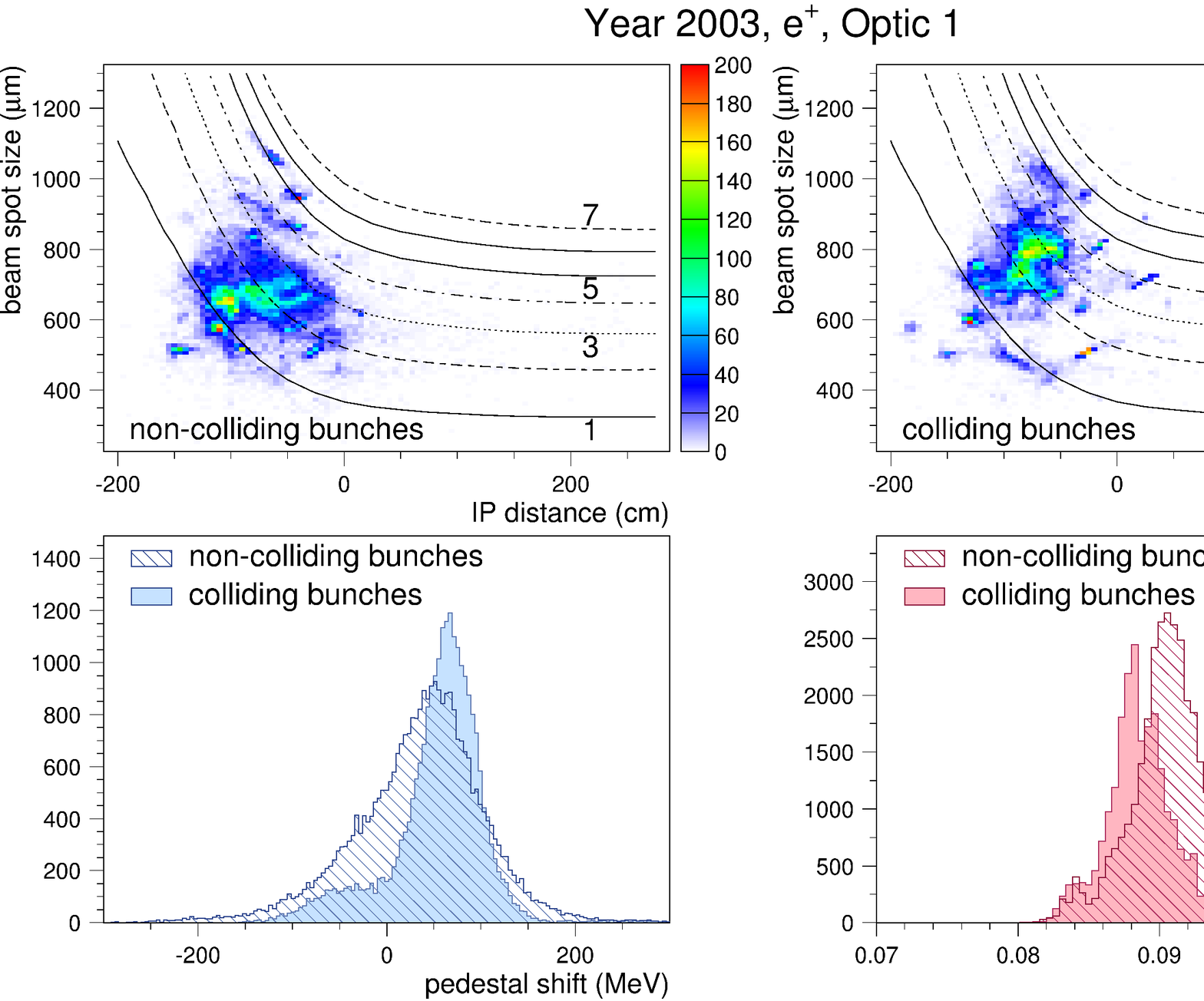}\vfill     
  \includegraphics[width=0.97\textwidth]{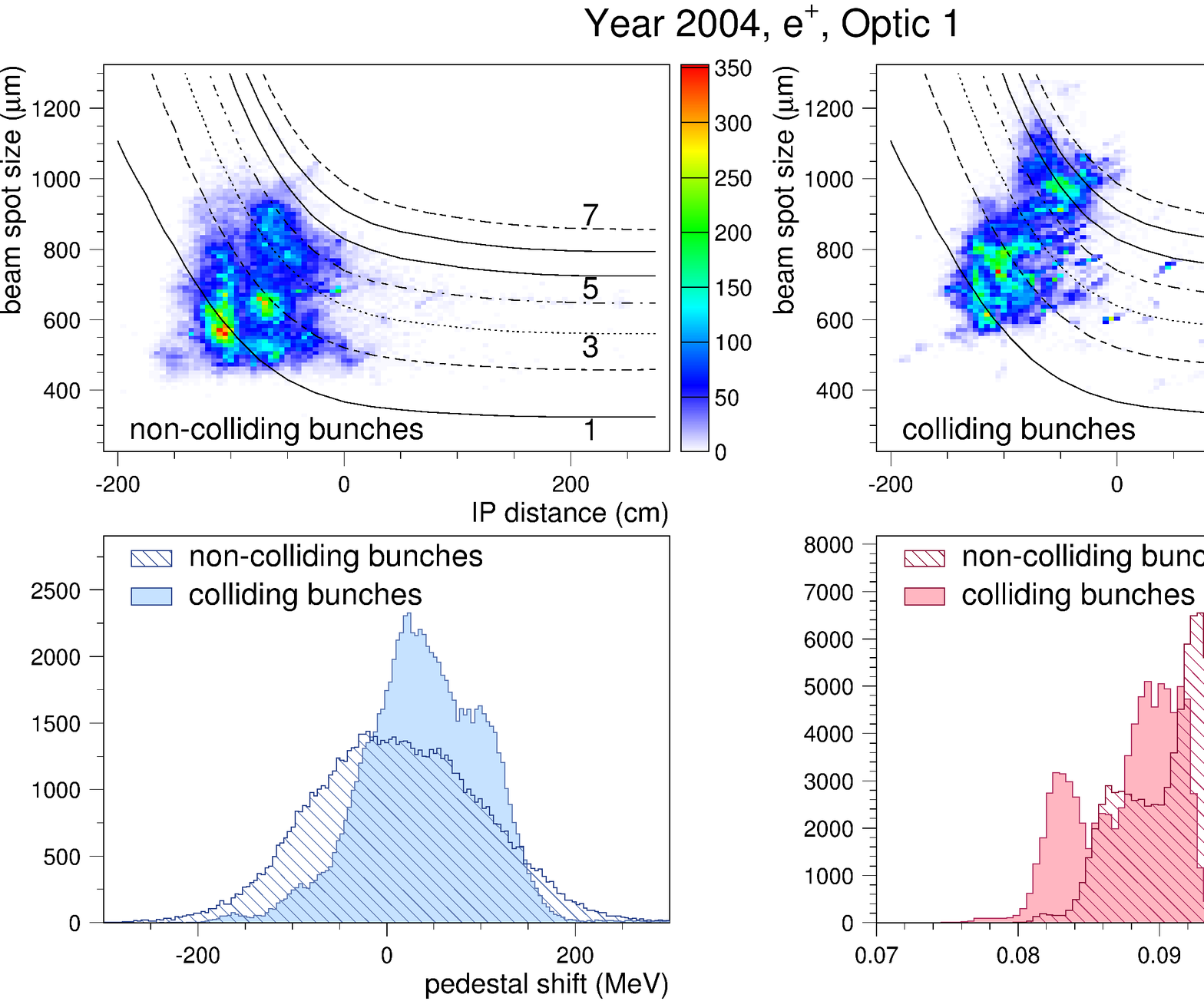}        
  \caption{\sl \label{recoy1} Results of reconstruction for the years 2003 and 2004. In each set of four plots the upper plots show the 
cumulation of reconstructed IP distances versus the beam spot size for non-colliding and colliding bunches. The black lines denote contour 
lines of constant emittances $1-7$nm.
The lower plots in each set show the corresponding pedestal shifts from reconstruction and the derived Analysing Powers, both for 
non-colliding and colliding bunches.}
\end{figure}

\begin{figure}\centering
  \includegraphics[width=0.97\textwidth]{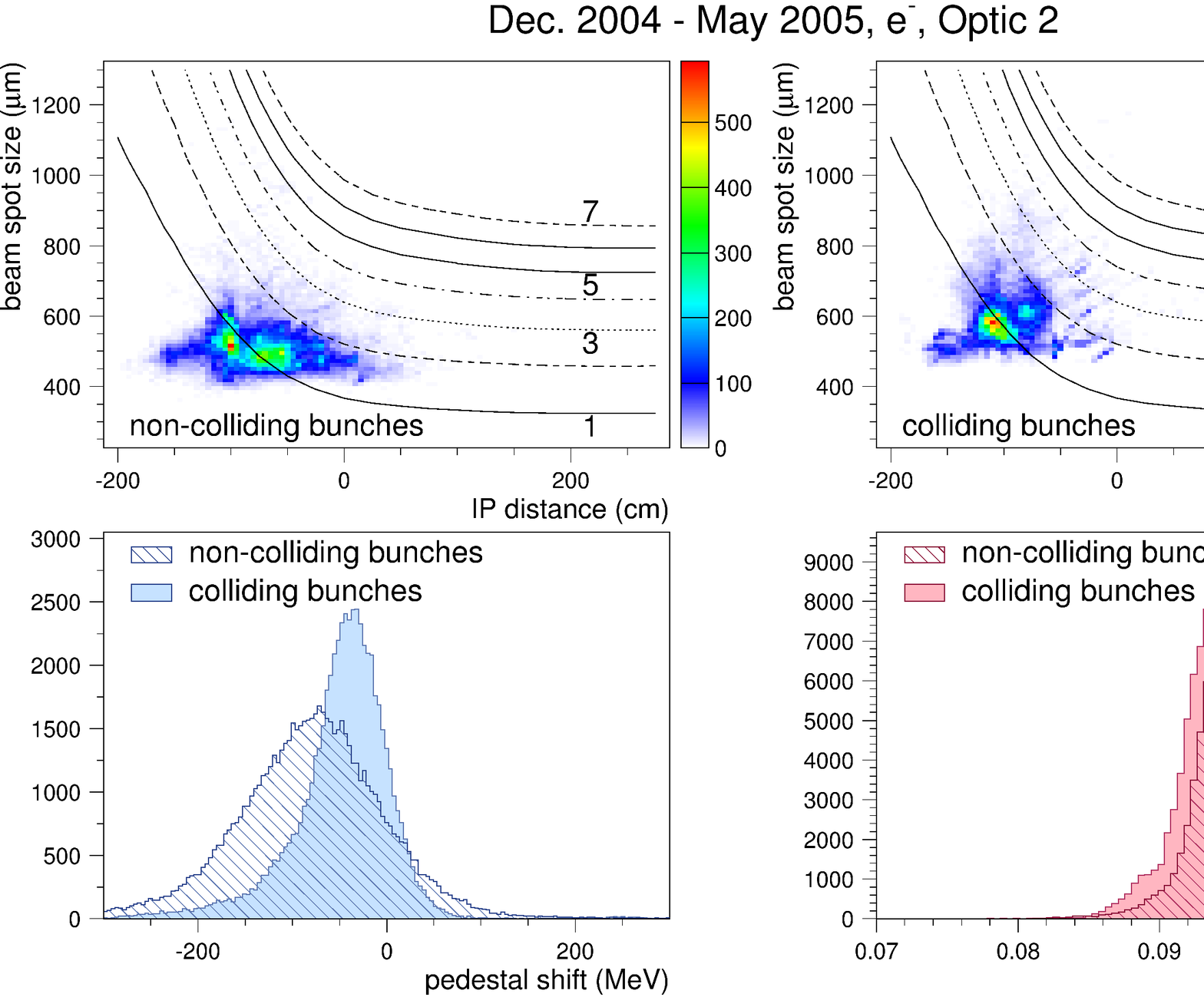}\vfill 
  \includegraphics[width=0.97\textwidth]{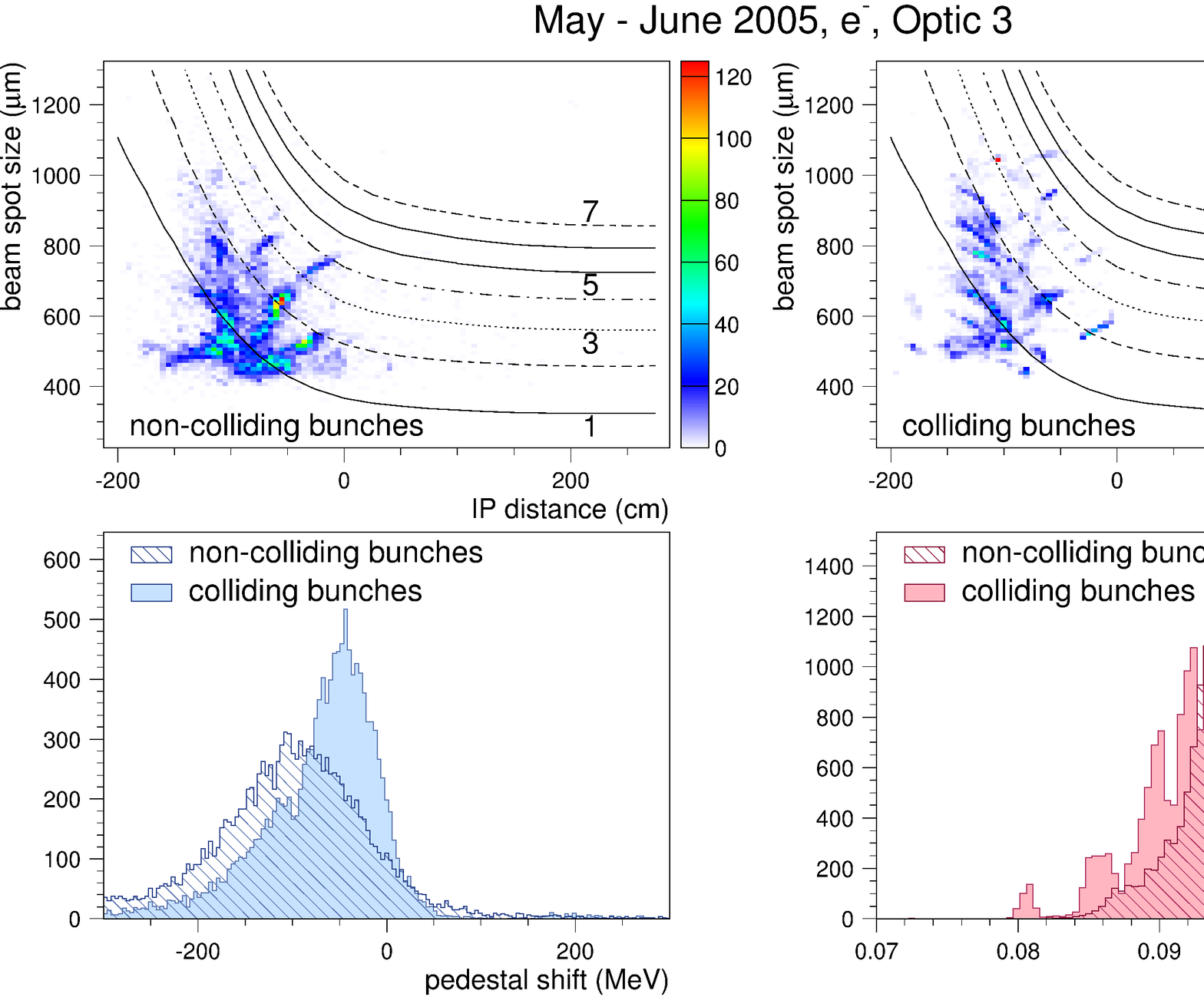}        
  \caption{\sl \label{recoy2} Results of data reconstruction for Dec.~2004 -- May 2005 and May -- June 2005. See Fig.~\ref{recoy1} for details.}
\end{figure}

\begin{figure}\centering
  \includegraphics[width=0.97\textwidth]{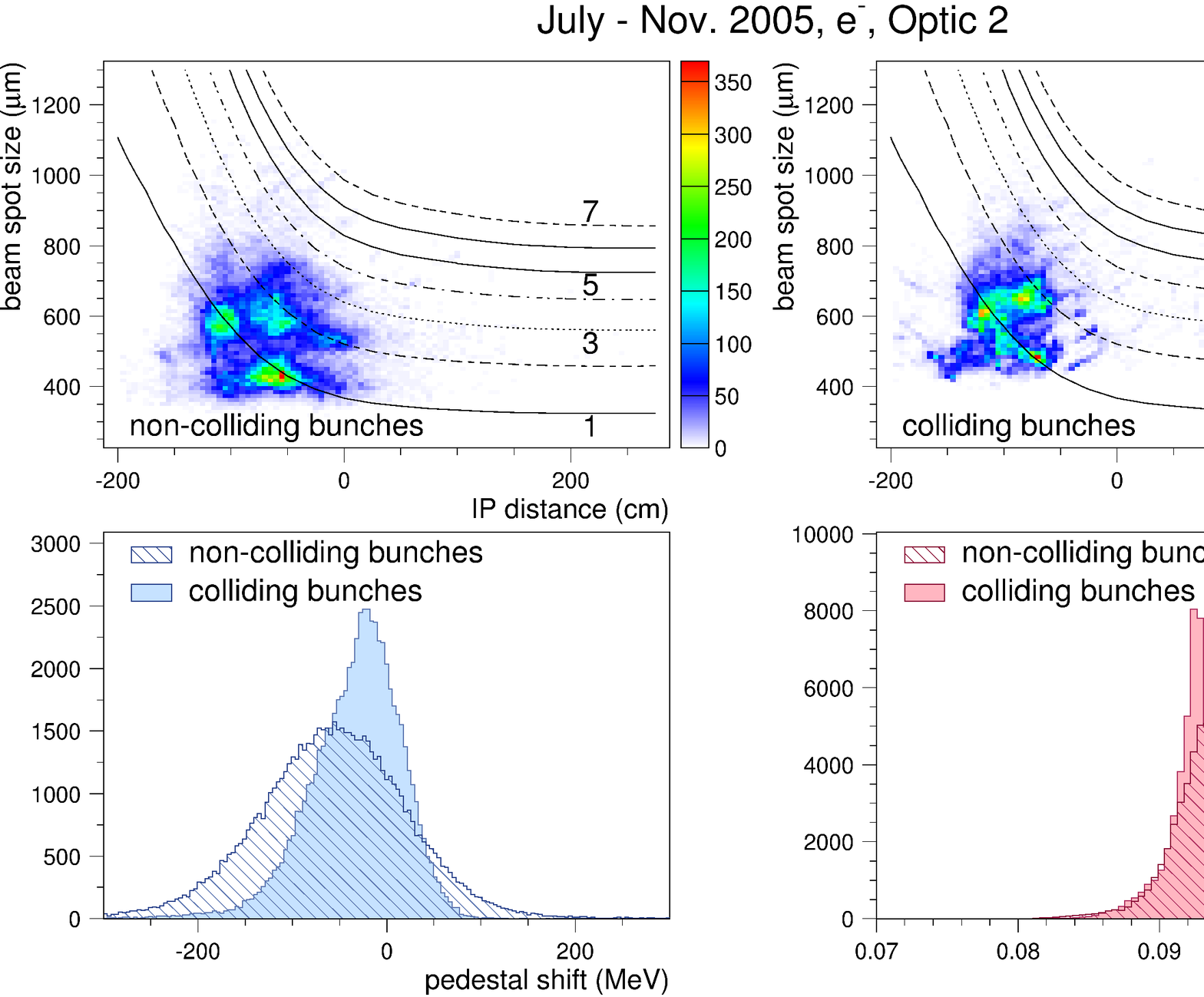}\vfill  
  \includegraphics[width=0.97\textwidth]{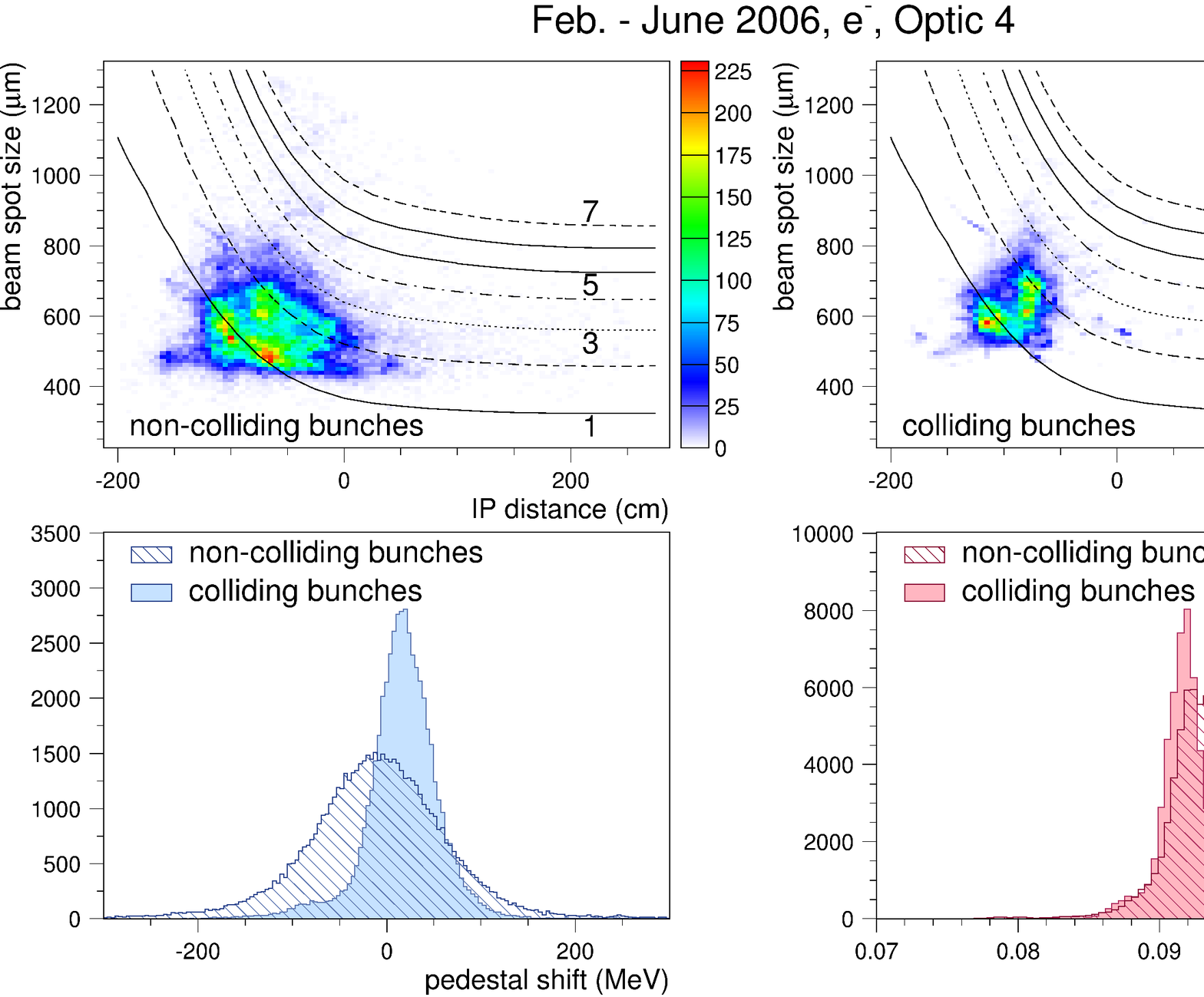}        
  \caption{\sl \label{recoy3} Results of data reconstruction for July -- Nov.~2005 and Feb.~-- June 2006. See Fig.~\ref{recoy1} for details.}
\end{figure}

\begin{figure}\centering
  \includegraphics[width=0.97\textwidth]{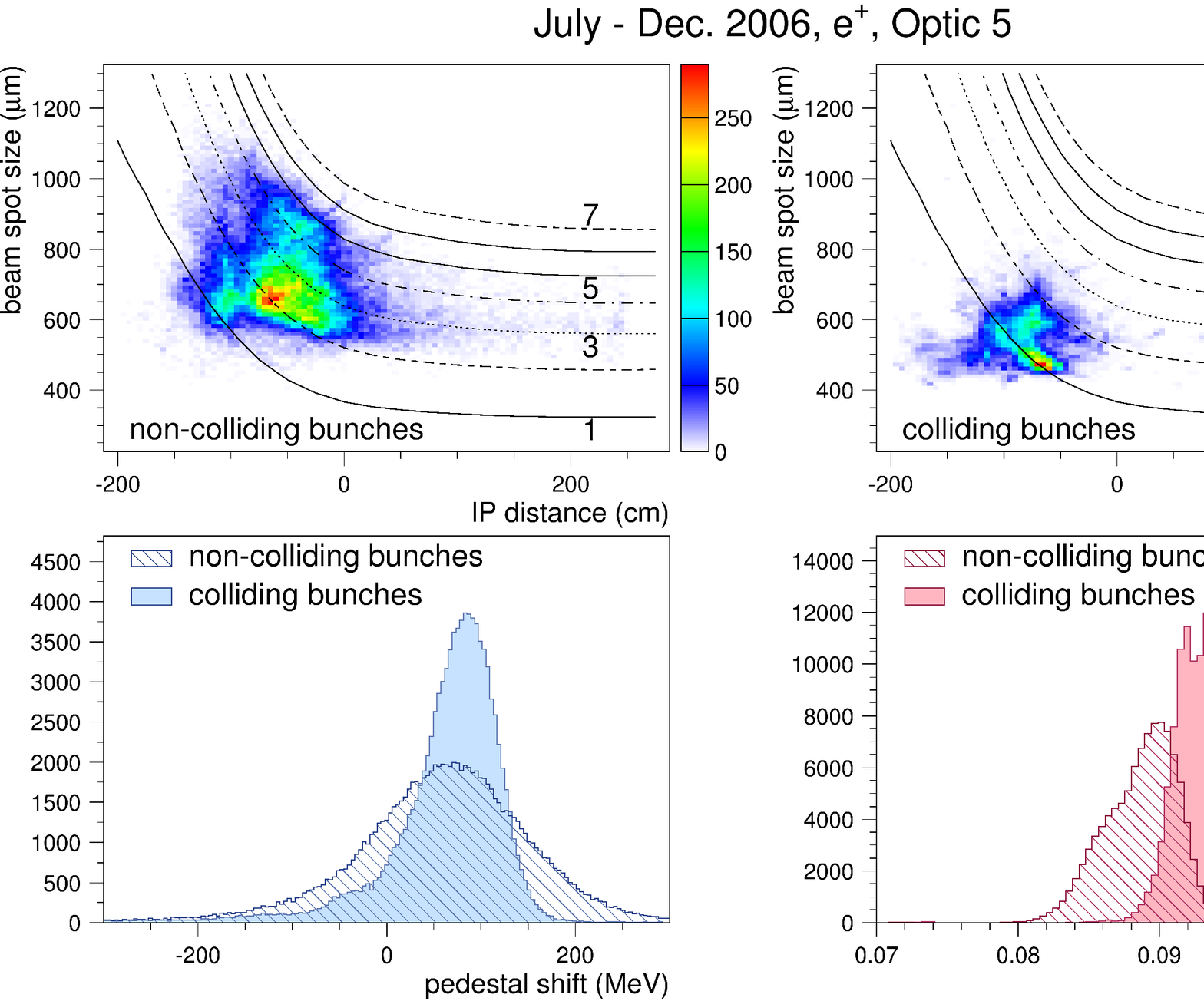}\vfill   
  \includegraphics[width=0.97\textwidth]{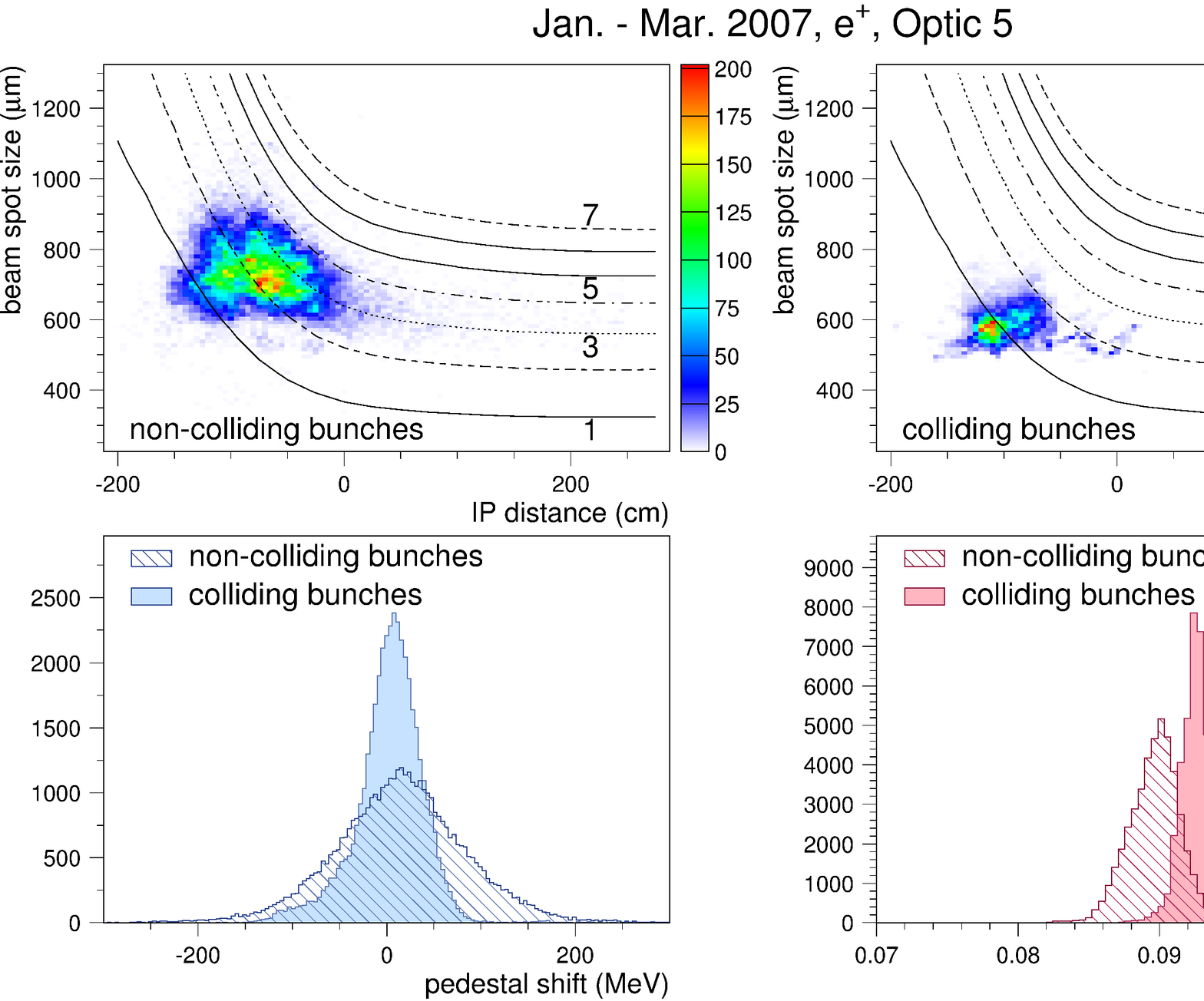}        
  \caption{\sl \label{recoy4} Results of data reconstruction for July -- Dec.~2006 and Jan.~-- Mar.~2007. See Fig.~\ref{recoy1} for details.}
\end{figure}

\begin{figure}\centering
  \includegraphics[width=0.97\textwidth]{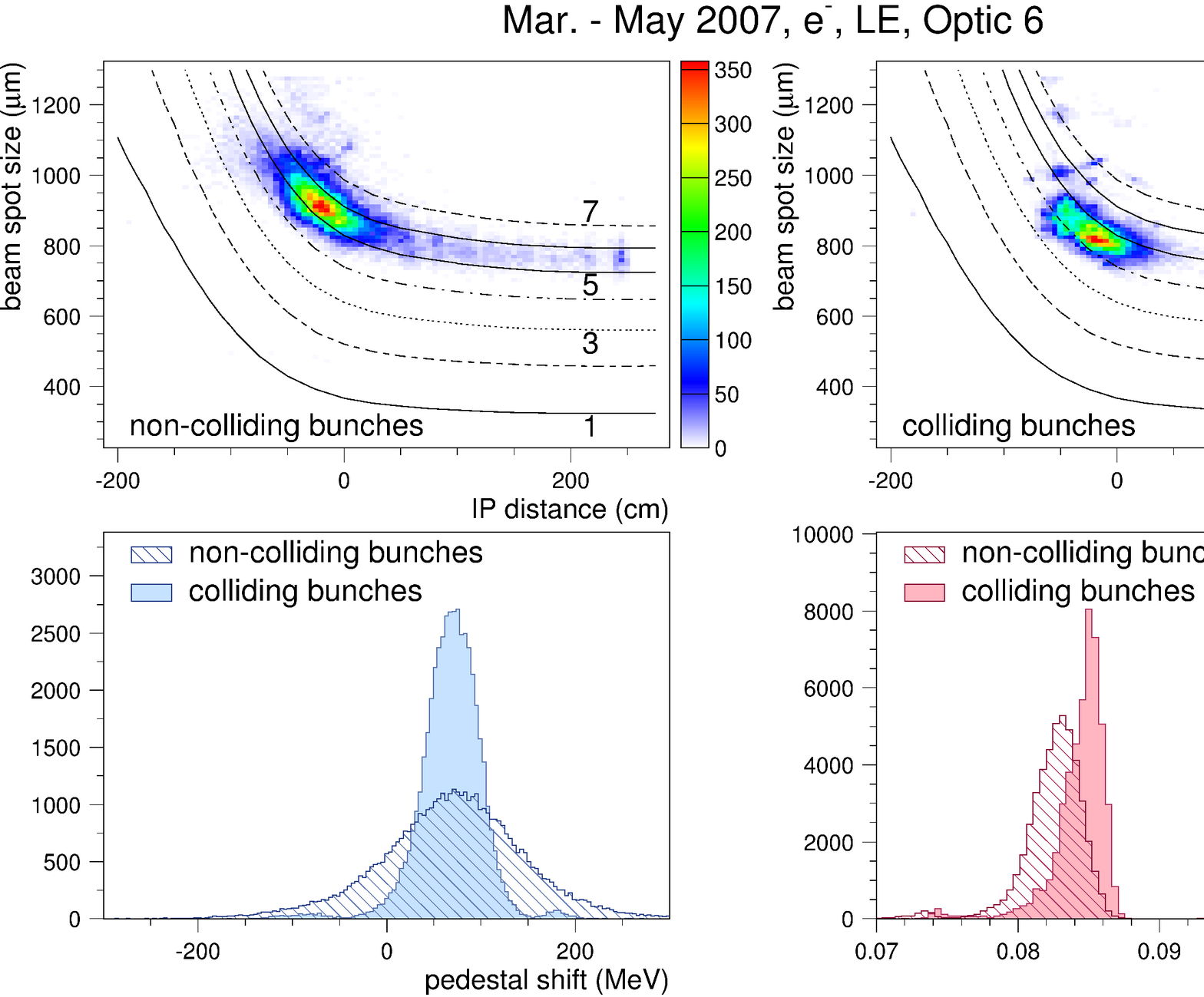}\vfill  
  \includegraphics[width=0.97\textwidth]{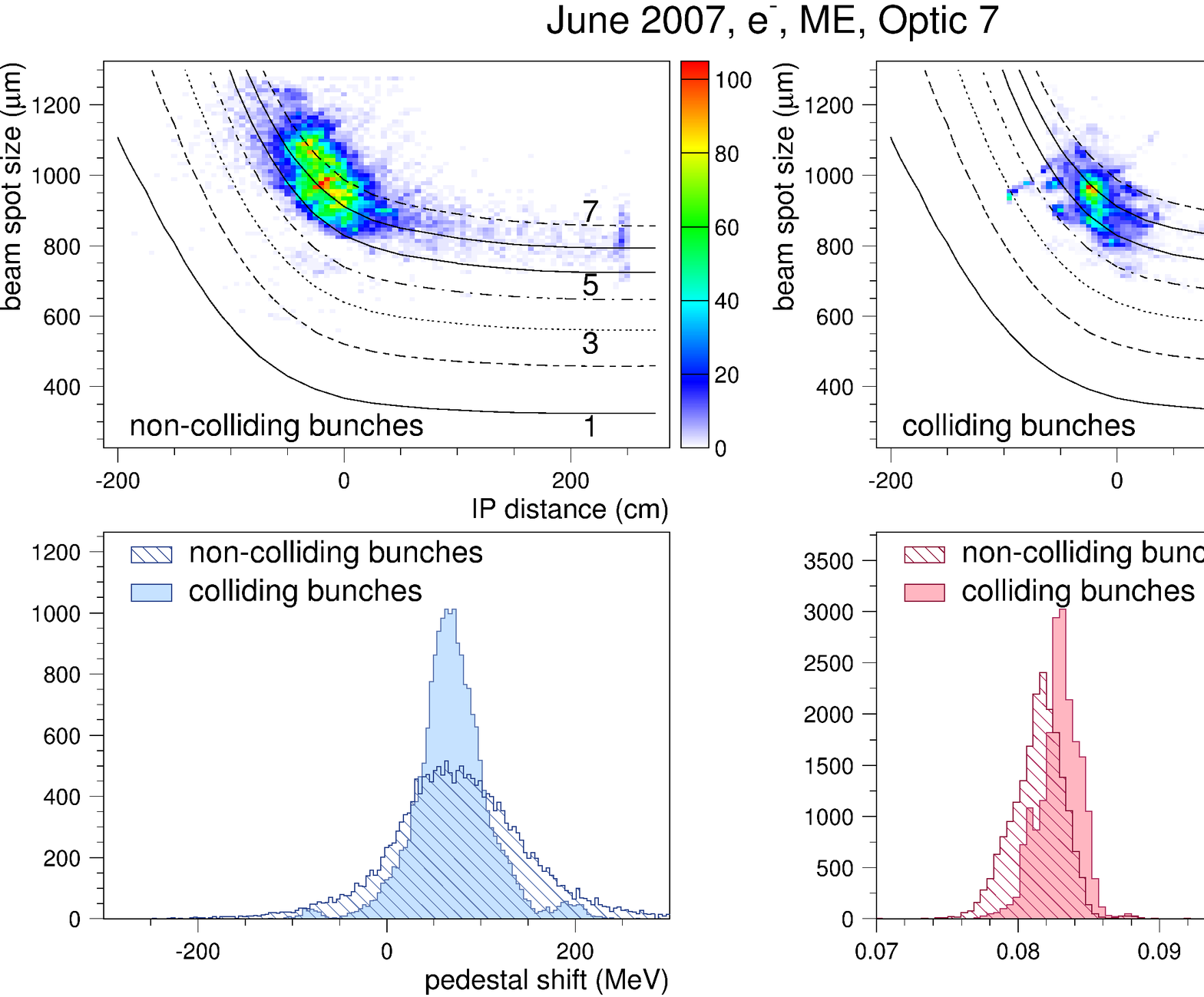}        
  \caption{\sl \label{recoy5} Results of data reconstruction for Mar.~-- May 2007 and June 2007. See Fig.~\ref{recoy1} for details.}
\end{figure}

The results from the fit are used to determine the polarisation values for the HERA II running. 
Results of the reconstruction are shown in Fig.~\ref{recoy1} to Fig.~\ref{recoy5}. For this the whole HERA II running period has been divided 
into ten periods according to year, type of beam ($e^+$ or $e^-$) and HERA optics set as is shown in Tab.~\ref{tab:HERA}. 
For each period a set of four plots is shown. The upper panels show cumulation plots of the reconstructed IP distances versus the beam spot size, 
for non-colliding and colliding bunches separately. The lower panels show one-dimensional projections of the reconstructed pedestal shifts and
the Analysing Power values derived from the fitted parameters, again for non-colliding and colliding bunches.

\begin{table} \centering
\begin{tabular}[h]{lcl|rc}
\hline
Time Period & $e^+ / e^-$ & HERA optic & Nb Meas. & Rec.~Eff. $(\%)$\\
\hline
Feb.~+ Oct.~-- Dec.~2003 & $e^+$ & 1 (helum72\_03)     &   43498  & 98.9\\
Jan.~-- Aug.~2004        & $e^+$ & 1                   &  102306  & 99.9\\
Dec.~2004 -- May 2005    & $e^-$ & 2 (helume-\_04)     &   83335  & 99.5\\
May -- June 2005         & $e^-$ & 3 (helume-\_05)     &   23502  & 94.7\\
July -- Nov.~2005        & $e^-$ & 2                   &   84345  & 99.3\\
Feb.~-- June 2006        & $e^-$ & 4 (helumsx\_06)     &   69948  & 99.8\\
July -- Dec.~2006        & $e^+$ & 5 (holumm0\_06)     &  117318  & 99.0\\
Jan.~-- Mar.~2007        & $e^+$ & 5                   &   45727  & 99.7\\
Mar.~-- May 2007         & $e^+$, LE & 6 (holum602\_07)&   56667  & 99.7\\
June 2007                & $e^+$, ME & 7 (holum6bs\_07)&   21231  & 99.3\\
\hline
\end{tabular}
	\caption{\sl Different periods of HERA II running, as used in the analysis. Data are divided by year, particle type ($e^+$ or $e^-$) 
and HERA II optics set. In total seven different optics sets have been employed. At the end of HERA II the proton beam energy was 
lowered. These two running phases with low and middle proton energies are denoted with LE and ME. The last two columns give the total number of
measurements analysable and the reconstruction efficiency of the new analysis. The latter is defined as the fraction of measurements for 
which the fit of the new analysis converged within the allowed phase space ranges and a reliable polarisation measurement could be provided.
	\label{tab:HERA}}
\end{table}

In the cumulation plots also the vertical emittance of the lepton beam, used in the PMC to generate the templates, is denoted by contour 
lines. The bending of the emittance isolines is induced by the focusing quadrupole located at $[-200,-100]$cm from the nominal interaction point.

For most of the HERA II running phases an emittance of $2-3$nm has been expected, except for the low and middle energy proton runs, 
where an emittance of $6-7$nm has been expected \cite{privcomm:Voigt}.
As the emittance is expected to be roughly constant over a fill as well as over some time with similar machine conditions, most data arrange 
nicely along the emittance contours. Periods characterised with extensive tuning of the HERA machine (e.g.~in May -- June 2005, see 
Fig.~\ref{recoy2}) show a more varying emittance from fill to fill. 

The mirrors of the laser beam path have been adjusted from time to time to find the point of highest luminosity. It had been estimated from 
mirror scans, that this point lies about half a metre behind the point where the laser beam hits the analyser box and which is defined here as 
the nominal interaction point. As a consequence, most of the data lie behind the nominal IP distance shortly before the 
quadrupole.
For the low and middle energy proton runs the emittance has been significantly higher and the laser beam has been moved forward on purpose to 
recover the beam spot size at TPOL. This is also nicely shown by Fig.~\ref{recoy5}.

However, compared to the expectation for the lower emittance range the reconstruction seems to suggest too small emittance values. No 
explanation could be found for this and a systematic uncertainty is assigned instead.
At high emittances in 2007 the data is nicely reconstructed around the expected emittance values. 

The pedestal shifts are reconstructed at most times near zero. The width is typically broad. As no independent measure of the amount of 
synchrotron radiation or electronic on-time pedestals exist, the pedestal shift can be regarded as an auxiliary to the analysis with limited 
physical meaning. Therefore a systematic uncertainty has been assigned to describe the global influence as can be derived from comparisons of
results derived from analysing with and without this auxiliary.

The derived Analysing Power values can vary up to $20\%$ relative, the largest contributions are given by a 
varying beam spot size. The IP distance effect contributes only on the level of $1-3\%$. The effect of the beam spot size has formerly been 
known as \emph{focus}. The focus correction of the old analysis attempted to correct for this very strong effect, which is by comparison with the 
results of the new analysis, known to have been quite successful. In total it is the most dominant effect influencing the Analysing Power.

Tab.~\ref{tab:HERA} shows the amount of available TPOL measurements which can be reanalysed and the reconstruction efficiency for the new 
analysis. For a small fraction of data the fit has not converged within the allowed 3-dimensional phase space and no new polarisation values 
can be derived. As can be seen, reconstruction efficiencies are typically higher than $99\%$. 

It has to be noted that a small fraction of the raw data (estimated to be around 3\%) was not stored due to technical problems, and is lost. 
For these data only results from the old online analysis exist.

\subsection{Systematic Errors}
In this section we present and discuss the systematic uncertainties assigned to the recalculated TPOL measurement. The results are summarised in
Tab.~\ref{tab:TPOLsys1} and Tab.~\ref{tab:TPOLsys2}.
\begin{table} \centering
\begin{tabular}[h]{|l|c|c|l|}
\hline
Source of Uncertainty & $\delta P/ P \,(\%)$ & Class & Comment \\
\hline
\multicolumn{4}{|l|}{\bf Description of Photon Generation, IP and Photon Beam Line}\\
HERA Beam Optics & 0.5 & IIId & 7 different optics \\
Lepton Beam Line & 0.5 & IId & \begin{minipage}[t]{4cm}{Mainly beam position in quadrupole}\end{minipage} \\
Lepton Beam Horizontal Emittance & 0.1 & IIId & \\
Laser Beam Line & 0.2 & IId & \\
Lepton Laser Beam Crossing & 0.1 & IIId & \\
Tilt of Photon Beam Ellipse & 0.1 & IIId & Mostly $\approx 2^\circ-4^\circ$ \\
Photon Pileup: Multi Photon Interaction & 0.1 & I & \\
\hline
\multicolumn{4}{|l|}{\bf Calorimeter Response}\\
Average Response & 0.6 & IIu & \\
\qquad - $\eta(y)$ and $E(y)$ & \quad(0.2) & & Up and Down channels \\
\qquad - \begin{minipage}[t]{5cm}{Difference converted to non-converted Photons}\end{minipage} & \quad(0.2) & & \\
\qquad - Linearity of Calorimeter Response & \quad(0.2) & & \\
\qquad - Effective $\eta(y)$ Calibration & \quad(0.5) & & Eff. Silicon strip pitch \\
\qquad - \begin{minipage}[t]{5cm}{Horizontal and LR-channels Response}\end{minipage}& \quad(0.1) & & \\
Energy Resolution & 0.7 & IIu & \\
\qquad - Total Energy Resolution & \quad(0.4) & & Fits to Compton edges\\
\qquad - Central spatial Description & \quad(0.2) & & \\
\qquad - \begin{minipage}[t]{4cm}{Difference converted to non-converted Photons}\end{minipage} & \quad(0.1) & & \\
\qquad - Resolution Correlations & \quad(0.5) & & \begin{minipage}[t]{4cm}{Channels sharing the same shower}\end{minipage}\\
Signal Modelling & 0.3 & IIu & \\
\qquad - Digitisation & \quad(0.1) & & \\
\qquad - Cross Talk and Non-linearity & \quad(0.3) & & \\
Horizontal Beam Position & 0.2 & IId & \\
\hline
\end{tabular}
	\caption{\sl Table of systematic uncertainties of the TPOL (first part). 
The third column indicates the estimated class of systematic uncertainty and possible period dependence, see Sect.~\ref{sec:class} for details.
	\label{tab:TPOLsys1}}
\end{table}

\begin{table} \centering
\begin{tabular}[t]{|l|c|c|l|}
\hline
Source of Uncertainty & $\delta P / P \,(\%)$ & Class & Comment \\
\hline
\multicolumn{4}{|l|}{\bf Data Calibration}\\
Absolute Gain & 0.3 & I & Beam energy changing with time\\
Gain Difference & 0.3 & I & Channels Up vs Down\\
Vertical Table Centring & 0.1 & I & \\
Background Subtraction & 0.1 & I & \\
\hline
\multicolumn{4}{|l|}{\bf Fitting Procedure}\\
Method Uncertainty & 0.5 & I & Covering complete phase space\\
Quality of Maps & 0.2 & I &  \begin{minipage}[t][5ex]{5cm}{MC Statistics, smoothing and interpolation}\end{minipage}\\
Impact of Starting Values & 0.2 & I & \\
IP Distance Reconstruction & 0.5 & I & Random jumps in data\\
Pedestal Shift Impact & 0.5 & IId & Global impact estimated from data\\
\hline
\multicolumn{4}{|l|}{\bf Laser Light Properties}\\
Linear Laser Light Polarisation & 0.2 & IId & \\
\hline
\multicolumn{4}{|l|}{\bf Trigger Threshold}\\
Bias at low Energies & 0.2 & IId & \\
\hline
\multicolumn{4}{|l|}{\bf Machine Performance}\\
Emittance Reconstruction & 0.9 & IId & \begin{minipage}[t][5.5ex]{5cm}{Comparison with expected emittances}\end{minipage}\\
\hline
\end{tabular}
	\caption{\sl Table of systematic uncertainties of the TPOL (second part). 
The third column indicates the estimated class of systematic uncertainty and possible period dependence, see Sect.~\ref{sec:class} for details.
	\label{tab:TPOLsys2}}
\end{table}

In the simulation of the TPOL setup, the GEANT3 based modelling of the calorimeter has been finally replaced by a detailed parametrised model 
of the calorimeter response and resolution. This has become necessary for several reasons.
First of all, extensive studies with the GEANT3 based model have been unsuccessful to tune the GEANT calorimeter response to the ones 
measured from Silicon calorimeter combined data. 
Although apparent modelling parameters as well as inherent GEANT parameters have been studied, the $\eta(y)$ transformation as found from 
GEANT turned out to be always much steeper, no matter what has been done for tuning. Resolutions, measured differences between converted and 
non-converted photons, as well as non-homogeneous response in horizontal directions have been difficult to tune as well. 
But the main reason for this step have been simply time considerations. Compared to GEANT based simulation of the calorimeter the parametrised 
one is faster by factors of the order of $10^3$. Only with the parametrised simulation has it been possible to generate the necessary Monte 
Carlo templates on such a fine grid and with such high statistics to produce high quality mapping functions.

The parametrised response of the calorimeter, as has been used in the simulation of the TPOL polarimeter setup (PMC) is derived and tuned to 
data in most parts. Only in those parts of the simulation where no or scarce data exist to derive parametrisations from, 
the according parametrisations and/or parameter values have been derived from GEANT simulation using the best setup achieved.

The influence of most systematic uncertainties assigned has been studied using the PMC, which includes the full photon generation and beam line
simulation, the parametrised calorimeter description and a digitisation module. 
Within the same simulation it is possible to switch between Compton photon generation and the generation of Bremsstrahlung photons to simulate
the main background at TPOL.
An interface allows for the collection of the generated events into histograms identical to those in data collection and to write out those 
histograms in the same format as is used in the data analysis. Here, both Compton and Bremsstrahlung events need to be generated to simulate 
the cycle structure of laser On and Off sequences.
Simulations performed in this way can be fed through the original software of data analysis and be analysed in the same way as real data 
(referred to as \emph{full chain}) in order to find the influences on reconstruction and Analysing Power as induced by the systematic effect 
under study. Only in some cases analysis is done in a simpler way using histograms of Compton events only, collected directly during event 
generation. And only in one case the analysis is based on theoretical considerations only, respectively on calculations performed using a very 
simple simulation, the results being reliable enough not to necessitate an iteration with the full chain.\\

\subsubsection{Interaction Point Description}
{\noindent\bf HERA Beam Optics}\\
Throughout the HERA II running phase from 2003 to mid 2007 seven different nominal beam optics have been employed. The mapping functions for 
the analysis have been produced employing an optic setup from the positron high energy run 2006/2007 (after switching from electrons, before changing
the proton beam energy towards the end of HERA). The influence of different nominal optics is tested at several different points in the 
3-dimensional phase space of IP distance, beam spot size and pedestal shift. This also represents a test for the robustness against varying 
beam conditions. The observed changes in Analysing Power are typically small and $\delta AP / AP < \pm 0.5\%$, the limiting value is therefore 
assigned as a conservative estimate of the uncertainty from this source. \\

{\noindent\bf Lepton Beam Line}\\
The lepton beam line is modelled using the nominal HERA optics given at several points in the TPOL straight section at HERA-West. This straight
section consists of two weak bending dipoles, separating the straight section in-between from the rest of the straight section West. The nominal
interaction point, where the lepton and the laser beam are crossing, lies in the middle between these two dipoles. From this point the 
backscattered Compton photons travel $66$m downstream of the lepton beam to the TPOL calorimeter. 
Inside the short straight section a focusing quadrupole is located at 1m in front of the nominal interaction point.
The HERA optics include the bending dipoles, the quadrupole and the drift regions in-between. Transformations are applied to model the general 
bending of the beam in the dipoles. To this point the beam line is modelled for a lepton beam passing the machine elements on the nominal axis,
no off-centre beam position in the beam elements is assumed.
If, however, the beam passes the quadrupole with an offset to the nominal axis it receives a kick and the beam direction will be altered. 

To study this effect, the beam position has been implemented assuming a dipolar feed-down of the beam when passing the quadrupole with an offset to 
the nominal axis. The size of the total kick induced should be proportional to the quadrupole strength and the size of the offset. The 
resulting dipole radius is therefore a function of the position itself and changing throughout the quadrupole, leading to complicated and 
iterative trajectory solutions, albeit a dipole curvature would occur for both the vertical and the horizontal direction.
The following assumptions have thus been made: 
\begin{itemize}
\item The horizontal curvature for horizontal offsets is neglected. Those would lead to horizontal offsets and increase of dispersion on the 
calorimeter surface. The trajectory with two-dimensional curvature would imply path integrals on an ellipsoid. Neglecting the horizontal 
curvature this reduces to a simple dipolar description in the vertical direction. The influence of the horizontal description of the beam is 
anyway of minor interest, as apertures are quite open, the calorimeter integrates over the horizontal direction and the energy dependence along
the horizontal is quite small. 
\item A constant dipole radius is assumed by calculating an average offset over the quadrupole length without the curvature effect and calibrating 
the radius such, that the total dipolar feed-down corresponds to the total kick reported in literature. The curvature of the beam trajectory 
will thus be not entirely correct as well as the total induced offset between the beginning and the end of the quadrupole (being anyway small), 
while the change in beam direction has the correct size. 
\end{itemize}
The effect of the beam position has been studied as a function of vertical offsets $y \in [0.00,0.11]$cm and different IP distances 
$IP \in [-120,20]$cm with PMC, full chain, covering at least $200$ simulated data samples each.
Upon passing the quadrupole at an offset, the lepton beam will be bent, changing effectively its direction. In consequence the generated 
photon spot will move depending on whether the interaction takes place after the quadrupole, inside or before. A movement of the centre of gravity 
of the photon beam spot is absorbed by centring the calorimeter table automatically onto the photon beam, and movements of the beam inside the 
quadrupole and movements of the interaction point should be sufficiently slow. In addition to the general movement additional dispersion is added 
when interaction takes place after the quadrupole, increasing the photon beam spot on the calorimeter surface, compared to beam dispersion before 
the quadrupole. Inside the quadrupole a mixed state depending on the size of the interaction region occurs leading to distorted beam spots.

The more the IP enters the quadrupole the larger are the observed effects. Typical expected beam offsets in a HERA quadrupole are not larger than $
30\mu$m. At a vertical offset of $30\mu$m, the total effect on the fitted RMS values, namely the ratio between assumed RMS values from fit and 
the RMS values calculated from varied simulation, is just about the maximal size observed in data. The corresponding change in Analysing Power 
$AP_{fit}/AP_{found}$ shows an increasing discrepancy in the energy dependence (over the energy bins), which is roughly opposite to the 
quality observed in data, indicating that $30\mu$m can be regarded as a sensible upper limit of possible offsets. 
At this offset the total influence on the Analysing Power derived from the large central bin is $\delta AP / AP < 0.5\%$.

It has to be noted, that the beam position as measured by the beam position monitors in data does not correspond to the simulated offsets. The 
beam position according to the beam position monitors (BPMs) is influenced by the offset and calibration setting of the BPM which change over time. 
The zero position of the BPM is not necessarily identical to the nominal axis of the quadrupole. The response of the BPMs is known to be nonlinear. 
In addition the BPMs show long term drifts.

The effect of an off-centre beam pass in the quadrupole is estimated to be the dominant uncertainty in the complete lepton beam line 
description. The limit found in this study is therefore assigned as the uncertainty from the lepton beam line description.\\

{\noindent\bf Lepton Beam Horizontal Emittance}\\
The coupling factor of the vertical emittance to the horizontal one at HERA II is estimated to be $\approx 0.15$ \cite{privcomm:Voigt}. 
The mapping functions for the analysis are produced using this factor. To study a possible influence, this coupling factor has been changed 
by factors $0.5$ and $2$ to generate a narrower and broader beam along the horizontal direction while the vertical size stays constant. These 
variations are then tested at several different points in the phase space of IP distance and beam spot size. The observed changes in Analysing 
Power are small $\delta AP / AP < 0.1\%$, confirming that the horizontal distribution of the beam is of minor importance as long as the beam is 
well contained in the beam line apertures (and thus in the calorimeter surface).\\

{\noindent\bf Laser Beam Line}\\
The laser beam is modelled as a Gaussian beam, characterised by three parameters: the waist size, the waist position and the size at the last 
mirror located at a distance of $18.34$m before the nominal interaction point. The basic parameters are estimated from data provided 
\cite{privcomm:Schueler}, measuring the approximate laser size in Nov.~1999 and by information provided by \cite{privcomm:Vahag}. Additional 
parameters of the laser are the laser photon energy of $2.41$eV ($514.5$nm, dark green), a vertical crossing angle between laser and lepton beam 
of $3.1$mrad (laser crossing from above) and a possible horizontal crossing angle $\phi$ between the two beams.

Using PMC, with own histogramming, different configurations of waist size, position and size at mirror are tested, varying sizes by factors of 
$0.5$ and $2$ and the waist position by $\pm 100$cm. These configurations are tested at several different points in the phase space of IP 
distance and beam spot size, for a constant pedestal shift value. The observed changes in Analysing Power are $\delta AP / AP < 0.2\%$.  

The laser photon energy is changed to the second main line of Argon-Ion: $2.54$eV ($488$nm, light blue). Consequently, the gain factors 
change, the measured  RMS values decrease, reconstruction from the fit and Analysing Power change accordingly and significantly, leaving no 
doubt that a possible running on the second main line of the laser can be excluded.\\

{\noindent\bf Lepton Laser Beam Crossing}\\
The interaction region is modelled using the information of both the lepton and the laser beam modelling, 
calculating a full 3-dimensional interaction probability for the crossing region of both beams. The nominal vertical crossing angle is 
$3.1$mrad, which is given by the height of the last mirror above the lepton beam line. This mirror is located at $18.34$m distance from the 
nominal interaction point. Depending on the status of the preceding mirrors the position of the laser beam on the last mirror can change, 
resulting the laser beam to point to different interaction point distances than the nominal one. Together with the vertical movement of the 
lepton beam this is the main reason for a possible movement of the interaction point, changing the IP distance to the calorimeter. However, 
when changing the IP distance in the simulation it is generally not assumed if this is due to a movement of the lepton or the laser beam, so 
no change of crossing angle is usually applied.

The principal influence of the crossing angle is studied by deliberately changing it within $\pm 0.5$mrad, which corresponds roughly to the 
change occurring if the interaction region moves between the two weak dipoles of the TPOL short straight section by a pure change of laser 
angle. The resulting change in Analysing Power induced by a changing form of the interaction region is small and estimated to maximally 
$\delta AP / AP = 0.1\%$. 
The horizontal crossing angle $\phi$ changes luminosity, but does not have any correlation with the vertical asymmetry as a homogeneous 
distribution of spins in the beam can be assumed.
As there is no data from which the actual crossing angles can be measured, the limit of $0.1\%$ serves as an estimate of the systematic 
uncertainty arising from the interaction region modelling.\\

{\noindent\bf Tilt of Photon Beam Ellipse}\\
Silicon data suggest that the photon beam ellipse is rotated with respect to horizontal and vertical direction on the calorimeter surface 
with rotation angles changing over time. 

The rotation is not part of linear beam dynamics and thus no feature of the beam line. Rotation is studied by rotating the 
generating lepton beam before convoluting it with the laser beam in PMC, full chain. The beam rotation is therefore strictly a beam rotation 
without rotating the spin direction (and thus the size of polarisation) with respect to the calorimeter.
For each rotation the ratio of vertical and horizontal emittances is adapted to reproduce two different ratios of vertical to horizontal beam 
spot sizes similar to those observed in Silicon data, and then varied as a function of rotation angle $\alpha \in [0,16]^\circ$. 

Observed angles in data are mostly about $2^\circ-4^\circ$, sometimes larger, e.g.~Feb.~2004 ($8^\circ-10^\circ$), Mar.~2004 -- Oct.~2005 
($5^\circ-6^\circ$) with ratios of effective Silicon beam spot sizes $\sigma_y :\sigma_x \approx 1.3:4.5$. 
From April 2007 the beam ellipse has been rotated into the other direction with angles between $-2^\circ$ and $-4^\circ$
and with ratios $\sigma_y : \sigma_x \approx 1.4:5.4$.

No effect is observed in simulation except for very high angles, where the constant beam size require a very small emittance, leading to 
Compton photons at very high energies near the Compton edge to shift downwards into the high energy bin, the average response of the 
calorimeter having a dip at the very centre due to the gap of the optical decoupling. This migration distorts the reconstruction and the 
derived Analysing Power. This means that it is an artefact of the scan particulars, not of the rotation itself.

If the spin is rotated together with the beam, the Compton scattering process is rotated against the calorimeter and the Analysing Power 
degrades with $\cos(\alpha)$. For various collider related reasons the spin cannot be rotated more than a few degrees from transverse in the 
arcs of the collider. Additional rotations by the TPOL straight section beam elements could be possible though, e.g. by off-axis paths in the 
quadrupole that rotate in the $(y,z)$ plane. This would generate longitudinal polarisation at the location of the TPOL, which is not affecting the TPOL 
measurement besides degrading the transverse polarisation again by $\cos(\alpha)$.
For a rotation of $<3^\circ$ from the transverse direction in the $(x,y)$ or $(y,z)$ plane the degradation is found to be $< 0.1\%$.

It has to be noted that this tilt of the beam ellipse discussed here does not imply a rotation of the calorimeter with respect to the HERA 
plane. Upon installation in the HERA tunnel, the calorimeter has been carefully aligned in tilt and roll angles. Additionally, it can be shown 
that the table moves to a good degree horizontally and vertically with respect to the optical slid of the calorimeter. Also, a tilt of the 
beam image does not imply a rotation of the horizontal and/or vertical Silicon planes. Their alignment has been studied separately and no 
significant tilt of the two planes to each other, to the calorimeter or to the table movement has been found, implying that all components are 
sufficiently aligned. It can be concluded that the rotation of the beam ellipse on the surface of the calorimeter must be a pure effect of the 
shape and characteristics of the interaction region.\\

{\noindent\bf Photon Pileup: Multi Photon Interaction}\\
The fraction of multi photon interaction events increases with photon rates following Poisson statistics.
The effect is studied analysis independent without cuts on the energy asymmetry and energy continuous, i.e.~unbinned.
The response of more than one photon is deduced from the single photon response by adding up energies for each channel.
Energy and asymmetry spectra are studied for up to three Compton photons (1C, 2C, 3C), two Bremsstrahlung photons (1B, 2B) and the mixture of 
two such photons of both types (1C1B). 

Combining more than one photon shifts the centroid of the energy distribution to higher energies and thus beyond the interesting energy range 
roughly between $5$GeV and $12$GeV, only a small fraction of such pile-up can contribute. The effect is studied for total photon rates per 
bunch crossing of $\langle n \rangle=0.01$ (TPOL at $100$kHz, single photon mode) and $\langle n \rangle =1$ (few photon mode). The first case 
shows a maximal degradation of $\delta AP / AP < 0.1\%$ in asymmetry robust with all types of pile-up as studied. As photon rates in typical 
TPOL running conditions are much lower than $100$kHz, that value representing a sort of upper limit for stable running, it can be concluded 
that pile-up from multi photon interactions can be estimated to degrade the Analysing Power at most by the mentioned limit of $0.1\%$.

\subsubsection{Calorimeter Description}
{\noindent\bf Spatial Response: $\eta(y)$ Transformation}\\
The vertical energy asymmetry curve $\eta(y)$ is derived from Silicon calorimeter combined data. For this an overlay of data from a table scan 
with different vertical table positions is used to illuminate the $\eta(y)$ over a large range of $y$. Under normal conditions only the centre 
part within roughly $\pm 1$mm is illuminated. The step sizes for the table scan are optimised to guarantee a homogeneous illumination over the 
chosen range. From the Silicon data single cluster events are chosen with some quality cuts on charge and noise to give a measurement of the 
$y$ position of a given high energy event with an associated energy asymmetry $\eta$ from the calorimeter.

Single cluster events represent the cleanest subsample of converted photons and provide the highest resolution in the allocation between the 
Silicon cluster position and the impact point of the converting photon. The data for laser On and laser Off events is histogrammed in $(\eta,y)$ with 
cuts on the total energy measured in the calorimeter. Special care is taken to avoid cross effects between binning and regular strip pitch 
distance. 

Bremsstrahlung and other photon background is subtracted on a statistical basis by subtracting the histogram for laser Off events from that with 
collected laser On events using normalisation constants derived from the energy distributions for the two laser states, keeping only events 
with a Compton photon in the calorimeter.
Further background arising from uncorrelated clusters are subtracted on a statistical basis from the histogram too. Such background may be due to 
pileup with synchrotron radiation or other low energy background in the Silicon detector with a high energy photon in the calorimeter, as well as 
two high energy photons, where only one of which is converting.
This type of background is roughly independent of the $y$ coordinate, generating a linear contribution along $y$ with a small slope. This linearity 
is being used for subtraction. The result is a clean picture of $\eta$ as a function of $y$, the resolution
of the scattering in the 2-dimensional histogram being non-Gaussian, as it is given by the energy resolution of the calorimeter cut into two 
halves and sharing the same shower.

The measurement of the $\eta(y)$ curve is the performed using the following steps.
\begin{itemize} 
     \item Primarily the same energy range as in the polarisation measurement is chosen, adding the high energy range of the \emph{focus} 
           determination (the former beam spot size), i.e.~$[5.2,13.8]$GeV.
           The 2-dimensional histogram is then sliced along $y$ to generate $\eta$ distributions for constant $y$ values. Simulation studies 
           using a primitive parametrised Monte Carlo with arbitrary $\eta(y)$ functions and simple modelling of the energy resolution showed 
           that the peak value of the non-Gaussian asymmetry distributions corresponds to the input $\eta$ position
           for the given $y$ value of the histogram slice. Discrepancies between the distribution maximum and the nominal $\eta$ position can 
           be generated by a non-homogeneous distribution of data over the $y$ range over which a certain slice is integrating. The applied 
           energy resolution model can have an influence on a shift of the distribution maximum from the nominal $\eta$ position too.
     \item Nominally, the centre of gravity of the $y$ distribution in a slice is used instead of the bin centre. Results are cross-checked 
           with fits using the bin centre instead and resulting changes to the fitted $\eta(y)$ parameters are found to be negligible.
     \item Different resolution models have been applied to study possible shifts of the distribution maximum from nominal $\eta$ values, 
           typically showing only small shifts compared to the resolution width of the distribution itself.
           The maximum position of the $\eta$ distributions in the $y$ slices are then found by fitting a Gaussian to the peak within 
           $\pm 2 \sigma$. Special care is taken for off-centre $y$ values, where the $\eta$ distribution reaches soon its natural borders 
           at $\pm 1$. 
           The result are measurements of $\eta_i$ in slices $i$ at positions $y_i$ with errors $\delta \eta_i$.
     \item The fitting ranges and the number of iterations are varied, and the results are found to be most stable in the applied fit range of 
           $\pm 2 \sigma$ and some neighbourhood. For fit ranges below $\pm 1.5 \sigma$ a Gaussian fit becomes unstable due to the finite 
           binning of the $\eta$ distribution and above $\pm 2.5 \sigma$ the results get distorted due to the non-Gaussian shape of 
	   the $\eta$ distribution.
\end{itemize}

In a second step the Compton edges of the energy distributions $E_{UD}$ of all Silicon calorimeter combined data samples of the table scan are 
determined. 
\begin{itemize}
  \item The energy calibration factors for the central runs are chosen such that the Compton edges of those runs lie at the 
        expected value given by the HERA beam energy. The same calibration factors are then applied to each data sample at any other table position.
        The position of the Compton edge is derived from a fit using a function of the Compton cross section convoluted with an energy resolution 
        function. Together with the average measured Silicon $y$ position, derived by fitting a Gaussian within $\pm 2 \sigma$ to the peak of the 
        Silicon cluster distribution of each table scan sample, this gives a measurement of the spatial response $E(y)$.
  \item Energy distributions are derived for the clean single cluster events as used for the $\eta(y)$ measurement, as well as for multi cluster 
        events, which have any cluster number larger than zero, for so-called no cluster events and for any number of clusters including zero, to 
        represent the various subsamples of converted and non-converted photons. 
  \item The energy distributions show a very similar behaviour for all types of photons whereas the absolute level varies between converted and 
        non-converted ones. 
        The level of the distributions for any number of clusters is consistent with the expectation when mixing $46\%$ of non-converting with 
        $54\%$ converting photons. These fractions can be expected from classical electromagnetic shower theory taking a thickness of $1 X_0$ 
        radiation length of the lead preradiator in front of the calorimeter into account \cite{Wigmans}. The difference in level between converted 
        and non-converted photon distribution types can be explained with energy leaking from the calorimeter at its back 
        plane, which is different for the two types, as the showers of non-converted photons start later than those of converted ones.
\end{itemize}

The measurements of $\eta$ in the single slices are then fitted using a parametrisation of the $\eta(y)$ curve based on a detailed analytical 
physical model within large ranges of $y$. The fit is performed simultaneously to a fit to a parametrisation of $E(y)$ given by the same 
physical model, as $\eta = (E_U - E_D) / (E_U + E_D)$ and $E = E_U + E_D$.

The physical model assumes a radial exponential energy deposition around the impact point of the photon and properly integrates this ansatz 
for $d^2E/{drd\phi}$ over $x$ and $y$ to calculate the energy depositions in the two calorimeter halves. Calorimeter effects are taken into 
account from the beginning, increasing the integration effort. The main details of the model are:
\begin{itemize}
   \item A two-component electromagnetic shower induced by a single high energy photon with a point-like impact point on the calorimeter. The first 
         component, the so-called {\emph core}, has a relatively short shower length and dominates at the early stage of the shower development.
         As the energies of shower particles are progressed down to lower energies in the evolving shower, the second component, the 
         so-called {\emph halo}, emerges with a much longer shower length, dominating the shower at the later stages of the shower 
         development. 
   \item Light attenuation in the scintillators, here an exponential decay of light intensity as a function of the vertical distance of the 
         integration variable to the readout position at the outer ends of the scintillators is assumed.
   \item A change of shower radius of both halo and core, together with a change in sampling fraction, when crossing the tungsten lead border at 
         $y=\pm 27.5$mm.
   \item Gain difference between the calorimeter halves.
   \item A gap in the centre of the calorimeter due to the optical decoupling of the two halves, leading to some energy loss of energy deposited 
         in the very centre.
   \item Energy leakage at the back side of the calorimeter affecting mostly the late halo component of the shower and thus changing the relative 
         fraction of energy content between core and halo.
   \item An initial spread induced by conversion of the photon in the preradiator, convoluting with the two-component non-converted single photon 
         shower.
   \item Convolution effects of $\eta$ due to the integration of $y$ over a certain range in a slice are taken into account. This effect mainly 
         dilutes the steep effect of the gap in the centre of the calorimeter, where $\eta(y)$ and $E(y)$ change fast.
   \item Convolution effects of the photon beam spread on the measured Compton edge.
   \item A finite end of the shower energy integration at the end of scintillator and lead frames at $y=\pm 55$mm, leading possibly to leakage
         through the sides.
   \item Free offsets $(\eta_0, y_0)$ for the symmetry point of the $\eta(y)$ curve.
   \item A possible shift of the lead tungsten border in $y$, as the DENSIMET15 plates might be set asymmetrically into the lead frames.
   \item Possible pedestal shifts in $E_{UD}$.
   \item Convolution effects in the $E(y)$ curve due to the large width of the beam spot.
\end{itemize}

The fitted curve of $\eta(y)$ corresponds to the response of converted photons.
The $\eta(y)$ curve of non-converted photons is extrapolated from this by adapting the initial spread length to be zero to get back a shower 
starting at a single point and by a different total energy sum, the difference of converted and non-converted photons in total energy mainly 
being given by the differences in hind leakage due to the different depth of shower start.

The vertical fitting ranges are varied, the important parameters of the fit like the shower lengths and fractions are found to be pretty 
stable, if the $y$ range is at least large enough to get sensitivity to both shower lengths, which means that it has to be larger than the 
core length, only afterwards its influence is small enough to disentangle the two lengths.
    
At very high $y$ ranges the parameters start to become unstable again, which can be explained by an increasing systematic distortion of the 
fitted $\eta$ position near the natural borders of $\pm 1$. Simpler $\eta(y)$ models like e.g.~constructions of two or three exponentials only 
show large dependencies of the fitted parameters with the chosen $y$ range, indicating that such models are not able to describe the curvature 
of $\eta(y)$ properly. The stability of the main parameters over a large $y$ range of the chosen physical model indicates that the curvature 
of the data is reproduced accurately by the model, and gives therefore confidence in its validity.

Results of the measurements for $\eta(y)$ and $E(y)$ together with the best adaption of the physical model as determined by a simultaneous fit
to both data sets is shown in Fig.~\ref{etay} and Fig.~\ref{Etune}.
The analysis of the different steps leads to the following conclusions: 
\begin{itemize}
\item No difference in $\eta(y)$ is found between the two different laser helicity states.
\item No significant difference in $\eta(y)$ is found using Bremsstrahlung events only from laser Off.
\item The Silicon $y$ position is not biased compared to the corresponding position on the calorimeter face as the spread angles of the 
starting shower behind the preradiator is small and the Silicon detector is mounted as close to the calorimeter surface as possible. Possible 
biases are estimated to be in the order of nanometre.
\end{itemize}

Possible biases and systematic errors introduced by the $\eta(y)$ analysis are checked by simulating the table scan in PMC and feeding it 
through the $\eta(y)$ analysis full chain. 
The underlying input $\eta(y)$ is reproduced bias-free and stably. The energy response curves $E(y)$ are reproduced bias-free and stably for 
different event classes which are distinguished by cutting on Silicon clusters.
The analysis is repeated for various table scans with different types of $y$ spacing. It could be shown that any differences found are due to
the scan particulars, i.e.~the $y$ spacing. This has been done by simulating also the other table scans in PMC, feeding it through the same
analysis and comparing the occurring differences with those found in the data.
It can be concluded that no biases to the $\eta(y)$ curve are introduced by the special $y$ spacing of the table scan used for the fit.
In addition, as the range of the available table scans covers a time period of over three years with the $\eta(y)$ being stable over the years, it 
can be concluded that neither the response of the calorimeter nor that of the Silicon detector has changed significantly over time.

The basis for the modelling of the $\eta(y)$ curve is given by a detailed analytical physical model of the electromagnetic shower shape and 
various detector effects. The modelling and the general behaviour of this model is tested using Silicon calorimeter combined data in different 
configurations as well as GEANT simulation. The fitted parameters connected to some physical meaning are checked to be meaningful and 
compatible to expectations from theory and the knowledge concerning detector related facts. The 1-dimensional analytical integration approach, 
neglecting the lead tungsten change along $x$ as well as the finite size of the calorimeter in $x$, is checked by detailed 2-dimensional 
discrete integrations where those details are taken into account too.

The $\eta(y)$ curve is fitted from Silicon calorimeter combined data for converted photons. The underlying detailed physical model of the 
electromagnetic shower predicts a difference in the behaviour of non-converted photons with respect to converted ones when adapting the 
parameters with related physical meaning. This extrapolation is checked using GEANT simulation, quantitative and qualitative behaviour of the 
extrapolation is fully confirmed, although the $\eta(y)$ curve of the best GEANT setup found still shows differences to the measured curve.

The $\eta(y)$ analysis is repeated on Silicon calorimeter combined data for different energy ranges of the detected photons. The observed 
absolute differences of the derived $\eta(y)$ curves is on the level of $0.001$ and is thus negligible. This energy independence is confirmed by applying 
the same analysis to GEANT simulation. The energy asymmetry as a function of the vertical impact point $y$ can therefore be regarded as energy 
independent. 

Taken together the effects from the modelling of the $\eta(y)$ curve are estimated to be $0.2\%$.\\

{\noindent\bf Linearity of the Energy Response and Difference of converted to non-converted Photons}\\
The linearity of the average energy response $E_{meas}(E_\gamma)$ of the calorimeter is derived from GEANT simulation. Different calorimeter 
models are employed to compare Compton and Bremsstrahlung edge positions to measurements from Silicon calorimeter combined data and to derive a 
parametrisation for the non-linearity of the response. It is well understood that this non-linearity is driven by leakage at the back plane of 
the calorimeter. The difference between converting and non-converting photons is driven mainly by leakage too, but also by a small offset for 
vanishing energies, which is the difference in energy deposited in the preradiator upon conversion. Non-converting photons also have a small 
offset, which corresponds mainly to the average energy deposited in the gap between the two calorimeter halves, but also in the aluminium 
front plate and the first absorber when the photon is non-converting, i.e.~is not converting in the preradiator.

According to GEANT simulation the response is totally linear for non-converted photons in the absence of leakage as expected, but shows a 
small slope as the calibration of the mixture at the Compton energy does not take the initial offsets into account. Converted photons show a 
slight non-linearity, which can be understood by the relative difference of the structure and material of the preradiator compared to the 
calorimeter: for small energies, the shower start is considerably disturbed by the preradiator, while for higher energies the shower maximum 
moves further into the calorimeter and the energies deposited in the very first layers including the preradiator become more and more 
unimportant. Consequently the slope approaches that one of the non-converted photons for high energies. The more leakage is included the less 
energy is reconstructed on average in the calorimeter. As the electromagnetic showers of non-converted photons start later in the calorimeter they loose more energy through 
the back plane than converted photons, explaining why the two non-linearities cross each other. 

The non-linearity is parametrised to be $1$ at the Compton edge for the proper mixture of converted and non-converted photons. The second hook 
point of calibration is that zero energy should stay zero. This provides relative non-linearities for the two photon classes, where the 
relative offsets at zero are taken from GEANT and the difference between converted and non-converted photons is calibrated with values from 
the measured Compton edges. The same parametrisation form can be applied to both types of photon classes and inherits a linear term for the 
general response and $log(E)$ and $log^2(E)$ terms for the leakage. It has also been shown, that the form and curvature is independent of $y$.
Due to the calibration being applied between the two points zero and the Compton edge, the Bremsstrahlung edge at the HERA beam energy (roughly 
twice the Compton edge) is shifted downwards with respect to its nominal value. Its exact position turns out to be highly sensitive to the 
calibration state, i.e.~the Compton edge position, due to the long lever arm. 

The main uncertainties in this non-linearity determination from GEANT simulations are therefore induced by
\begin{itemize}
\item the applied offsets (which are understood physically and checked theoretically to be meaningful),
\item the difference of converted and non-converted photons and their mixture at the Compton edge as measured from Silicon calorimeter combined 
data (driven by purity of the applied selection cuts) and
\item the GEANT model determining the curvature by changing the leakage of the calorimeter model.
\end{itemize}
The influence of the applied non-linearity model is studied in PMC, full chain, by varying the applied constants for offsets and curvature in 
factor ranges of $[0.5,1.5]$ and by varying the difference between the two photon classes at the Compton edge in form of the relative energy 
loss between non-converted and converted photons in the absolute range of $[0.98,1.0]$, the nominal value determined from Silicon calorimeter 
combined data to be $0.9915\pm0.0005$.
The changes in Analysing Power for the applied offsets and the curvature as well as for the energy loss factor are both $\delta AP/AP < 0.2\%$,
the limit of which is applied as uncertainties for the two sources.\\

{\noindent\bf Effective Silicon Detector Calibration}\\
The vertical energy asymmetry curve $\eta(y)$ is measured with a Silicon detector with a nominal strip pitch of $80\mu$m and the fitted $\eta(y)$ 
parameters are valid for this pitch. However, data analysed with maps generated from PMC using this $80\mu$m $\eta(y)$ indicate a strong 
energy dependence of the Analysing Power as the measured polarisation varies systematically with the average energy of the chosen energy bin 
(polarisation falls with rising energy). It turns out that this energy dependence can be effectively flattened when adapting the $\eta(y)$ to 
represent a different Silicon strip pitch, which is equivalent to an effective re-calibration of the Silicon length scale. PMC with full chain 
is used to generate simulated data with changed pitch values and is analysed with generated maps for $80\mu$m and different configuration 
states like e.g.~cross talk values, pedestal shifts and RMS scale factors. The best Silicon pitch is chosen by comparing the energy dependencies 
of simulation under these analysis conditions of data. The most consistent and similar behaviour to data is found for an effective Silicon 
pitch of $86\mu$m $\pm1\mu$m. New maps for this central value have been generated and applied to data. 
By this effective calibration the observed strong energy dependence of the Analysing Power has been flattened considerately, leaving no effect 
larger than expected from the combined systematics as described in this note.

The remaining systematic uncertainty of this calibration is estimated by generating simulated data using a range of $84-88\mu$m effective 
pitch and analysing it with the $86\mu$m maps. The changes to the Analysing Power induced within $\pm 1\mu$m are $\delta AP/ AP = 0.5\%$.

Extensive studies have been conducted to find a possible source for the necessity of such an effective calibration.
Various other types of possible sources to generate such an energy dependence have been studied, none of which proving to be able to describe 
data like the Silicon pitch calibration. Studies have included various types of crosstalk between cables, pedestal shifts, non-linearities in 
the energy measurement of the calorimeter and in digitisation, changes in $\eta(y)$ shower lengths, free changes of the $\eta(y)$ curve, as 
well as any of the other studied systematic error sources.
The complete setup of the polarimeter has been revised, ranging from the hardware in the tunnel to the implementations in simulation, none of 
which giving a hint to the source of this systematic discrepancy. Although the origin of this effective calibration is not understood, its 
size could be determined relatively precisely directly from data itself, without any assumption on the polarisation scale, simply by requiring 
the polarisation to be a constant over all energies. No additional scale calibration has been employed, the calibration relying completely on 
the observed energy dependencies. By attributing the effective calibration to the Silicon pitch, no inherent internal argumentation chains are 
broken and all analyses of detector response remain valid. 
This effective Silicon pitch calibration has been the only calibration found which leaves a completely self consistent picture of all analyses 
compared to other possible calibration choices and is therefore preferred. Due to the relatively precise knowledge of the size of calibration 
needed, only the remaining uncertainty on its size is assigned as a systematic uncertainty arising from this calibration.\\

{\noindent\bf Possible Biases in Silicon Position Measurement}\\
A tilt of the Silicon $y$-plane with respect to the vertical direction of the calorimeter could have affected the $\eta(y)$ transformation 
measurement by biasing the measured $y$ positions. If the Silicon $y$-plane is mounted with a rotation angle towards the vertical direction of 
the calorimeter, the effective Silicon pitch would appear to be stretched and the distance between two points on the vertical axis would 
appear to be smaller. The effect would be that the $\eta(y)$ transformation appears to be compressed along $y$, giving a steeper rise in
the centre region, leading effectively to larger Analysing Powers. 
Possible rotations of the Silicon $y$-plane have been studied using table scans in the horizontal and vertical direction, comparing table 
movement, the calorimeter response and the measured $x$ and $y$ positions from Silicon. It is found, that the table moves vertically and 
horizontally with respect to the calorimeter optical plane as well as to the Silicon $x$ and $y$ planes to a very good degree. By comparing 
the response changes of both calorimeter and Silicon planes at different horizontal and vertical positions, tilts of the Silicon $y$ plane 
larger than $1^\circ$ can be excluded. Mechanical considerations of the mounting of the Silicon detectors in front of the calorimeter confirm 
that the planes can not be rotated by more than the above limit, which has no significant impact on the $\eta(y)$ transformation measurement.

Radiation damage in the Silicon detector could have affected the $\eta(y)$ transformation measurement, especially in the centre where the beam 
spot hits the detector and where the Analysing Power is most sensitive to changes in the $\eta(y)$ curve. Radiation damage in the Silicon 
detector induces the measured charge clusters to be biased, leading to errors in the measurement of the associated vertical position. The 
scintillating fibre in front of the Silicon detector has been used to monitor the performance of the device. Over one year of running with the 
Silicon detector permanently in front of the TPOL calorimeter, no hint for possible biases of the vertical position measurement from the 
Silicon detector has been found and Silicon data from later years do not suggest such radiation damages to have appeared.\\

{\noindent\bf Horizontal Response and Response of LR-channels}\\
Besides the average response of the Up and Down channels as a function of the vertical position, the so-called energy asymmetry 
$\eta(y)=(E_U(y)-E_D(y))/E(y)$ and the total energy measured $E(y)=E_U(y)+E_D(y)$ also the horizontal response and the vertical and horizontal 
response of the Left and Right channels are needed to fully describe the average response of the calorimeter. 
Under the assumption that the average response is fairly homogeneous and good-natured, the responses along the horizontal and the vertical
directions decouple, leaving eight average response functions: $\eta_{UD}(y)$, $E_{UD}(y)$, $\eta_{UD}(x)$, $E_{UD}(x)$, $\eta_{LR}(y)$, 
$E_{LR}(y)$, $\eta_{LR}(x)$, $E_{LR}(x)$, where the vertical functions of Up and Down are of most importance. The influence of the other 
response functions is studied in PMC, full chain.
All eight response functions are derived from Silicon calorimeter combined data, though the vertical functions of Up and Down have 
unprecedented accuracy compared to the others. The assumption of horizontal and vertical decoupling has also been checked to be valid.

The influence of the horizontal Up-Down responses are studied by applying a constant instead of the parametrisations adapted to Silicon 
calorimeter combined data. This variation is identical to an absent simulation of this response, giving thus a maximal estimation of its 
influence.
The horizontal Up-Down influence becomes only important when moving the beam horizontally. With a stable beam and after calibration the 
polarisation measurement is found to be stable over large parts of the horizontal range for the applied dependencies, changing the Analysing 
Power $\delta AP / AP <0.1\%$ with the above mentioned variation.

The influence of the vertical as well as the horizontal Left-Right responses is studied by varying them similarly as in the case of 
horizontal Up-Down responses.
The Left and Right channels influence the polarisation measurement mostly via the calibration path, as the Left and Right channels are used to 
derive the total and the relative calibration of the Up and Down channels after centring the beam on the calorimeter. Though the final 
calibration constants might vary when varying the Left-Right response the polarisation measurement is affected only little, when a stable, 
centred beam and valid calibration is assumed. When applying a constant instead of the measured dependencies the derived Analysing Power 
is found to change $\delta AP / AP <0.1\%$.

The influence of the response functions other than the vertical Up-Down responses can therefore be estimated to be maximally 
$\delta AP /AP =0.1\%$.\\

{\noindent\bf Total Energy Resolution}\\
The energy resolution of the calorimeter is modelled in the PMC using the classical approach including a statistical term $a$, a constant 
term $b$ and and energy linear dependent term $c$ following the known formula 
\begin{equation}
 \left( \frac{\sigma_E}{E} \right)^2 = \left( \frac{a}{\sqrt E} \right)^2 + \left( \frac{b}{E} \right)^2 + c^2 \label{eqn:resolution}
\end{equation}
The statistical term inherits the statistical fluctuations of the shower development including photon statistics, the constant term arises 
mainly from leakage, namely the leakage at the back plane of the calorimeter. Values $a$ and $b$ for converted as well as non-converted 
photons are input values to the simulation. The linear term $c$ is not added explicitly, but arises indirectly from the digitisation process. 
Its size has been been determined to be $c \approx 0.08$ and the linear energy dependence of this contribution verified. Both size and the 
linear energy dependence are typical for an effect arising from digitisation.

The total resolution of the simulated response has been tuned via adapting the statistical terms $a$ to reproduce the measured values at the 
Compton edge when simulating Compton spectra including the full PMC. For non-converted and converted photons only slightly different values 
are applied, the ones for the Left and Right channels being a bit higher than those of the Up and Down channels. This can be understood from 
the geometry of the readout of the scintillator plates.

The size of the constant terms $b$ for converted and non-converted photons has been determined from GEANT simulations, the converted photon 
value has been checked to be in agreement with measurements from test beam data using electrons and positrons and a preradiator. There also the
statistical terms and the shape of the energy dependence have been measured, and the measured total resolution at the Compton edge is in agreement 
with the test beam data as well as the general classical shape forming the basis of the simulation.
The influence of the energy resolution of the Up-Down and Left-Right channels has been studied in PMC, full chain, varying the $a$ and the 
$b$ term, as well as both together within factors of the nominal values. Within the range $[0.96,1.04]$, comprising variations well beyond the 
measurement errors of the Compton edge resolution in Silicon calorimeter combined data, the Analysing Power changes by no more than 
$\delta AP/AP = 0.4\%$, which is assigned as a conservative estimate for the systematic uncertainty arising from the modelling of the total energy 
resolution.

The parametrisation used in PMC is an interpolation on pointwise Silicon calorimeter combined data, smooth and continuous and based on general 
classical assumptions suitable for sandwich calorimeters. The parametrisation including statistical, constant and linear terms is checked 
using GEANT3 simulation to be sufficient to describe the resolution of a realistic electromagnetic sampling calorimeter. This is also confirmed 
by test beam data, where a suitable description of the energy dependence is 
achieved using statistical and constant term only. It has to be noted that different read-out electronics has been used in 
the test beam, and it is assumed that constant and linear terms can not be directly compared. The measured total resolution at the Compton 
edge for converted photons is compatible with the resolution measured from test beam data. No further influence by the chosen parametrisation 
itself can be assumed from there.\\

{\noindent\bf Central spatial Description}\\
Spatial information for the resolution modelling is obtained from Silicon calorimeter combined table scan data spanning ranges up to 
$\pm 15$mm. In first order the resolution is found to be constant over large ranges in the horizontal as well as the vertical direction. Small 
structures are observed in data at high distances, which can be explained by the non-homogeneous average response of the calorimeter. 
By including this horizontal and vertical dependent average response in the parametrisations very similar structures in the spatial resolution 
can be generated. In consequence, also the spatial resolution as measured from data can be reproduced with the PMC. As the polarisation 
measurement takes place only in the very centre of the calorimeter in a range of $<3$mm, no influence of these far away regions can be 
expected.

In the very centre of the calorimeter, namely inside the gap decoupling the upper and the lower halves, a higher energy resolution is measured than 
off-centre. A modelling of this resolution excess is included by adapting $a$ and $b$ terms with a Gaussian excess in the very centre. 
Detailed GEANT3 simulations show that the height and width of the Gaussian excess is connected to the size of the gap. As Silicon calorimeter 
combined data are available only at some scarce points in $x$ and $y$, the width of the excess is derived from GEANT3 simulations, while the 
height is adapted to reproduce with PMC the measurements from Silicon calorimeter combined data in the very centre. The derived heights are in 
accordance with the values derived from the GEANT3 simulations itself. 
The influence of the Gaussian excess in the very centre has been studied in PMC, full chain, by scaling the height of the excess. 
For factors $\in [0.5,2.0]$, which is a large range to reflect the rather scarce knowledge of this phenomenon, the Analysing Power changes by 
$\delta AP/AP < 0.2\%$.\\

{\noindent\bf Difference between converted and non-converted Photons}\\
There is a substantial difference in energy resolution if a high energy photon hitting the calorimeter converts in the preradiator or in the 
calorimeter. GEANT3 simulations show that the main difference is given by the amount of leakage at the back plane of the calorimeter, as 
non-converting photons convert later and the shower maximum occurs later than for a shower starting already in the preradiator (the converted 
case). The contribution to the resolution as given by Eqn.~\ref{eqn:resolution} due to leakage is given by the constant term $b$. GEANT3 
simulations show that also the size of the gap between the calorimeter halves influences the constant term, adding a contribution which is 
the same for both photon classes, thus changing their total resolutions but not the difference between the two.

The size of the constant terms has been derived from GEANT3 simulations, where inner parameters\footnote{ILOSS=1,2, DRCUT values} of GEANT, 
influencing the shower development, have been varied. In addition the calorimeter geometry has been changed, by manually changing the 
scintillator and absorber densities. The chosen values represent the best adaption of GEANT simulation to the data and the value for converted 
photons is comparable to values measured in test beam data. The influence of the size of the constant terms itself is included in the
uncertainty estimation for the total resolution.

There is however, a difference in the total resolution between the two classes, which is measured from Silicon calorimeter combined data using the 
appearance or non-appearance of Silicon clusters. 
The measured difference between converted and non-converted photons is therefore diluted by a non-perfect purity of the samples derived 
through these cuts. However, the total resolution of both classes together, which is the class of all photons, implies no cuts on Silicon 
clusters at all and can therefore be used to cross check fine tuning of the two classes. The input values are tuned such, that by cuts on 
generator level of the simulation for the two classes and for the mixture the total resolution values at the Compton edges as measured in data 
are reproduced. As the simulation of Silicon clusters in GEANT is rather rudimentary, no statements on efficiencies concerning the appearance 
of clusters can be derived from there. Therefore, the purity of the two samples derived by cuts on cluster appearance has been estimated 
directly from data to be $\approx 89\%$, the measured difference should therefore be only little diluted. 

The total resolution of the class of all photons is higher than might be expected by the simple mixture of $54\%$ converted and $46\%$ 
non-converted photons, as the Compton edge position shifts due to the leakage at the back plane, thus adding an additional contribution to the 
edge resolution. After tuning of the resolution as well as the edge positions of the two classes the total resolution and the edge position of 
the mixture class are reproduced as well.
In conclusion, the additional influence due to the modelling of the difference in resolution of the two classes can be assumed to be very 
small with $\delta AP / AP < 0.1\%$, the limit of which being assigned as a conservative estimation for this source.\\

{\noindent\bf Resolution Correlations}\\
The above points consider the resolution of the reconstructed energies $E_{U\!D}=E_U+E_D$ and also $E_{LR}=E_L+E_R$. The second variable is 
given by the energy asymmetry $\eta_{\scriptscriptstyle U\!D}=(E_U-E_D)/E_{U\!D}$ and also $\eta_{\scriptscriptstyle LR} =(E_L-E_R)/E_{LR}$. 
The resolution of the energy asymmetry is 
not trivial as the asymmetry describes how the electromagnetic shower is shared between two calorimeter halves and the resolution therefore 
depends on the resolution of a part of a shower and on how it is correlated to the part of the shower in the other calorimeter half.
       
The resolution of the upper and the lower channels can be described by a covariance matrix:
\begin{equation} 
   V_{U\!D} = \begin{pmatrix} 
            \sigma_{\scriptscriptstyle U}^2 & \rho_{\scriptscriptstyle U\!D}\sigma_{\scriptscriptstyle U}\sigma_{\scriptscriptstyle D} \\ 
            \rho_{\scriptscriptstyle U\!D} \sigma_{\scriptscriptstyle U} \sigma_{\scriptscriptstyle D} & \sigma_{\scriptscriptstyle D}^2 
            \end{pmatrix}
\end{equation}
where the resolution widths for each half $\sigma_{\scriptscriptstyle U}$ and $\sigma_{\scriptscriptstyle D}$ are correlated by a correlation 
coefficient $\rho_{\scriptscriptstyle U\!D}$.
A correlation matrix also exists for the energy $E$ and the asymmetry $\eta$:
\begin{equation}
   V_{E\eta} = \begin{pmatrix} 
               \sigma_{\scriptscriptstyle E}^2 & \rho_{{\scriptscriptstyle E} \eta} \sigma_{\scriptscriptstyle E} \sigma_\eta \\ 
               \rho_{{\scriptscriptstyle E} \eta} \sigma_{\scriptscriptstyle E} \sigma_\eta & \sigma_\eta^2                     
               \end{pmatrix}
\end{equation}
connecting the total energy resolution $\sigma_E$ with a resolution of the energy asymmetry $\sigma_\eta$ and a correlation coefficient 
$\rho_{{\scriptscriptstyle E}\eta}$.
The variables $\rho_{\scriptscriptstyle U\!D}$, $\sigma_\eta^2$ and $\rho_{{\scriptscriptstyle E}\eta}$ can be expressed as functions of 
$\sigma_{\scriptscriptstyle E}$, $\sigma_{\scriptscriptstyle U}$ and $\sigma_{\scriptscriptstyle D}$ (needs 
also $E$ and $\eta$) which are then used to express the channel resolutions $\sigma_{\scriptscriptstyle U}$ and and $\sigma_{\scriptscriptstyle D}$ 
as a function of $\sigma_{\scriptscriptstyle E}$ and the 
two correlation coefficients, the latter three being easily accessible to parametrisations using functions of $E$ and $\eta$. For further 
details see the note on resolution correlation \cite{corrnote}. 

GEANT simulations are used to derive functions for $\rho_{\scriptscriptstyle U\!D}$ and $\rho_{{\scriptscriptstyle E}\eta}$ as a function of 
$E$ and $\eta$ over a wide range of $E$ and $\eta \in [-1,1]$, the total energy resolution $\sigma_{\scriptscriptstyle E}$ is parametrised 
as usual by a statistical and a constant term. Test beam data from 
CERN \cite{cerntb} at various energies using the vertical position $y$ derived from clusters in the vertical Silicon plane and the two 
calorimeter variables $E$ and $\eta$ confirm the principal behaviour of the GEANT simulated correlation coefficients and the width 
distributions $RMS(E_U)/RMS(E)$, $RMS(E_D)/RMS(E)$, $RMS(E)$ and $\sigma_\eta E$. Input to the PMC are parametrisations derived from GEANT, 
where the best adaption of the calorimeter model to general data is taken. The amount of available test beam data is not sufficient to derive 
parametrisation constants from there. But comparisons show that test beam data agree with the chosen parameters. To study the influence of the 
chosen correlation parameters the correlation coefficients $\rho_{{\scriptscriptstyle E}\eta}$ and $\rho_{\scriptscriptstyle U\!D}$ are varied 
by factors over a large range, both individually as well as together. 
Test beam data agree in general with the derived parametrisations within scaling factors $\in [0.5,1.5]$, and the influence on the Analysing 
Power in that range is estimated to be $\delta AP/AP \approx 0.5\%$ using PMC with simple histogram analysis. 

In addition, the GEANT calorimeter model has been changed using different scintillator, air, tungsten and lead contributions to change 
deliberately the shower development. The correlation parametrisation coefficients derived from there are also tested in PMC with simple 
histogram analysis, and surprisingly, although very different correlations can be generated when changing the sampling calorimeter 
substantially, the effect on the Analysing Power using these very different correlations is quite small, confirming that the estimation
$\delta AP/AP = 0.5\%$ is quite reasonable.\\

{\noindent\bf Digitisation}\\
The digitisation of the photomultiplier signals of the five calorimeter channels is implemented in the PMC using a detailed physical model. 
Four ADC samples are simulated with constant fractions of the total signal for each channel including an ADC pedestal and low and high 
frequency noise. Response is limited to the range of $[0,4095]$ ADC counts, introducing also well-known saturation effects in the Up and Down 
channels if the signal is very much off-centre and of high energy, meaning that one of the two channels needs to measure more than half of 
$30.4$GeV, the latter being the total range of $E_U+E_D$.  
The trigger is simulated using the channel with the analogue sum of Left and Right channels. The method with which the pulse shape is analysed 
is as close as possible to that used in data. All steering values like the pedestal positions and widths, the amount of signal amplitude in the 
pedestal, as well as the pulse shape fractions have been estimated from data directly.

No difference in $\eta(y)$ using the full $\eta(y)$ analysis chain has been found if digitisation is applied. It can be shown, that a 
contribution to the total resolution arises due to the digitisation which is linear in energy: $\sigma_E^2 = (cE)^2$, in accordance with the 
expectation from theory. Altogether the total resolution is tuned in the statistical terms to reproduce the measured values at the Compton 
edge. The change of Analysing Power due to the digitisation is expected to be very small, using PMC with own simple histogram analysis, 
comparisons of results for un-digitised and digitised signals show differences maximally $\delta AP/AP = 0.1\%$.\\

{\noindent\bf Cross Talk and Non-linearity}\\
There are no hints for additional effects occurring in the electronics during signal readout and digitisation, but there is also no evidence 
for their absence. The principal effects of cross talk in cables or readout or non-linearities occurring there have been studied in PMC, with 
own histogram analysis and with the full chain. 

As for cross talk in cables or readout, various cross talk models have been tried, including linear and quadratic energy dependencies as well 
as with $\sqrt(E)$ with $E_i = E_i + f E_j^k$, \mbox{$k=1,2,0.5$}, where either the other half of the calorimeter is taken to cross talk (\emph{ 
inside calorimeter} model) or all other four channels (\emph{in cables} model). The structures and behaviour cross talk, especially the most 
probable linear cross talk is generating to distort the Analysing Power is in some parts similar to that observed in data 
prior to the Silicon pitch calibration. A cross talk value of $0.007$ (\emph{inside calorimeter} model) is suitable to correct most of the 
observed energy dependence of the Analysing Power without the Silicon pitch calibration. 

However, the form of the energy asymmetry $\eta(y)$ would be changed too, not reaching the natural borders at $\pm 1$ at high $y$ values any 
more. The $\eta(y)$ measurement using Silicon calorimeter combined data excludes such types of cross talk down to a level of $<0.001$.  
The table scan used to measure the $\eta(y)$ transformation curve spans regions of $y$ up to $\pm 15$mm and the two shower lengths which are 
derived from this are well in accordance with the expectation from theory. No saturation of the curve below $\pm 1$ is observed and in case 
significant cross talk is present, a much longer halo shower length would have been measured. 
Due to this linear cross talk can be excluded down to a level of $<0.001$, leaving also no space for other cross talk configurations. The 
effect from linear cross talk at this level is negligible.  Therefore, no additional systematic uncertainty is added due to possible cross 
talk in cables or readout.

Non-linearities have been tried as a function of exponent $\alpha$ to the energies like 
\begin{equation}
E_{U,D}' = E_C \left( \frac{E_{U,D}}{E_C} \right)^\alpha
\end{equation}
where $E_C$ denotes the Compton edge energy, $\alpha$ being near but not equal to one. 
It can be shown, that the reconstruction is sensitive enough to exclude non-linearities on the percent level. 
The structure of the generated energy dependence and the general shifting of the overall Analysing Power level due to applied non-linearities 
show a behaviour which is not compatible to the behaviour observed in data. 
If it can be assumed that a large non-linearity is not hidden beneath another effect of similar or bigger impact, non-linearities can be 
excluded down to a level of $<0.005$, meaning that $\alpha \in [0.995,1.005]$, leaving a possible effect of $\delta AP/AP < 0.3\%$.\\

{\noindent \bf Horizontal Beam Position} \\
The horizontal position of the photon beam on the calorimeter face has influence in a direct and an indirect way. Directly the response of the
Up and Down channels changes when moving horizontally. Indirectly the response of the Left and Right channels change when moving horizontally,
influencing thus the calibration state of the Up and Down channels. 

In data the table position has been moved only once, most time of HERA II the table position has been at $-10$mm, changed in 2007 to $0$mm. 
According to combined Silicon calorimeter combined data the horizontal position of the beam has been relatively stable, the table position 
giving a good estimate for the beam position on the calorimeter surface. 
The beam position has also been confined by the horizontal aperture of $0.36$mrad. With a typical horizontal beam spread of $\approx 90\mu$rad, 
the beam can move maximally $2$ times its spread before cutting the beam within $2$ sigmas, which would be noticeable by luminosity and 
LR-calibration. This means, that the beam cannot move more than roughly about $\pm 1.5$cm without notice within the apertures.
 
The influence is studied in PMC, full chain, by varying the horizontal table position, calibrating at each position, including thus both 
possible effects. In addition the photon beam is moved horizontally with the table being fixed at different positions, so that the influence 
of apertures cutting into the beam are also taken into account.
First the pure table movement within $\pm1.5$cm is studied with centred beam and then the table is moved within $\pm1$cm and the beam is 
moved on its surface within $\pm1.5$cm. The change of Analysing Power observed is $\delta AP/AP < 0.2\%$, the effect on single energy bins 
being a bit higher, but still $<0.5\%$. The reason for this is that the largest influence is seen in the very low and very high energies 
(i.e.~the low and high energy bin), where the Analysing Power is reduced most, while the inner energy bins keep being stable. The 
reconstruction is influenced by this only little, so that the overall Analysing Power therefore stays rather constant.

As information on the exact beam position on the calorimeter surface is available only from occasional Silicon data, the limit of 
$\delta AP/AP = 0.2\%$ is applied as an uncertainty due to this systematic source.

\subsubsection{Data Calibration}
{\noindent\bf Gain and Gain Difference}\\
The high voltage for the photomultiplier channels of the calorimeter is set so that the Compton edge as determined online from the energy 
spectrum measured by the sum of up and down channels lies at the expected value given by the HERA beam energy.
The gain factors applied by the data analysis are thus $\approx 1$, deviations of $O(0.01)$ at maximum are corrected for by applying a 
resampling technique to the collected data energy $E$ vs $\eta$ histograms per minute.
The lepton beam energy of HERA varied with time, being on average about $27.6$GeV with variations around this value well below $100$MeV.
During the low and middle proton energy runs in 2007 the beam energy was a bit lower with on average $27.5$GeV. 

The Analysing Power depends on the energy, as can be seen from Fig.~\ref{bins}. A mismatch of the position of the Compton edge with respect to 
that in the simulated templates introduces a strong effect on the Analysing Power in each bin, leading to a relative energy dependence of the 
template Analysing Power. Strictly speaking, the polarisation derived from each bin using the mismatched templates would be different, showing 
a rising or falling behaviour with energy, depending if the Compton edge of the analysed sample is moved down or upwards with respect to that 
of the templates. 

The energy calibration of the template maps is tuned to give with the online differentiation method the expected Compton edge corresponding to 
a HERA beam energy of $27.6$GeV. 
For a final energy calibration, to adjust the Compton edge of the data samples to the templates in the maps, the data samples, collected in 
time periods comprising typically $1-3$ months, are calibrated to a fixed energy slightly off from $27.6$GeV to give for that period the 
minimal energy dependence of the Analysing Power.
The final gain factors have per period an average deviating from $1$ by $< \pm 0.005$, the average of all periods being at $1.001$. The widths
of all periods are $<0.008$, being on average $\approx 0.005$.

The effect of gain factors $\neq 1$ and subsequent resampling of the data histograms has been studied in PMC, full chain, as a function of the 
applied gain difference $\delta g$ and gain $g$. These two variables are represented by the calibration factors $(f_U-f_D)/2$ and 
$(f_U+f_D)/2$ in data. With the final energy calibration the reconstructed gain difference in data is in every period fully within 
$\pm 0.005$ and the reconstructed gain mostly within $\pm 0.01$. Only for the year 2003 a higher gain spread is observed, here data lie 
within $\pm 0.015$ around the average $1.0$.

The influence of gain and gain difference is studied by varying from a fully calibrated state at $g=1$ and $\delta g=0$, varying both 
variables together with $g \in [0.97,1.03]$ and $\delta g \in [-0.02,0.02]$. The influence on the Analysing Power is found for each variation 
direction to be $\delta AP/AP < 0.3\%$, no specific correlation is observed when applying both variations. It is thus concluded that the total 
effect of both can be added in quadrature, to be $\delta AP/AP < 0.4\%$ and each contribution alone to be $<0.3\%$.

The evaluation of the calibration constants could be influenced by the (real) IP distance and the beam spot size, feeding back into the 
reconstructed parameter values, thus distorting the reconstruction. This has been studied at different phase space points in PMC, full chain, 
and no significant influence has been found.\\

{\noindent\bf Vertical Table Centring}\\
The table is centred during polarisation measurement by an autopilot by measuring the off-centring from the average in energy asymmetry 
spectra. The gain difference also influences this average $\eta$ value, but this variable is measured from parabola fits to the ratio 
$E_U/(E_L+E_R)$ vs $\eta$ profiles, making thus the gain difference nearly uncorrelated to the centring.    
The table is kept by the autopilot within a certain range, the measured spread in data being mostly $\approx 20\mu$m (applying a conversion 
function to calculate average $\eta$ values into vertical offsets). A decentred table dilutes the vertical asymmetry, but because of the 
uncorrelation to the gain difference this effect does not induce resampling of the data histograms and is therefore largely unconnected to 
resampling effects. 
The effect of a decentred table has been studied in PMC, full chain. The centring estimate from data is found to show a $71\%$ slope 
compared to the real applied decentring. A measured spread of $20\mu$m thus corresponds to a decentring of $\approx 28\mu$m. For the year 
2003 a higher spread of $\approx 33\mu$m has been measured, translating to estimated $\approx 47\mu$m. The influence on the Analysing Power 
has been studied by applying a decentring to a pure calibrated state of the simulation, keeping the calibration fixed at their calibrated 
values.
It is found, that the Analysing Power changes almost nothing for decentrings of up to $25\mu$m, and $\delta AP/AP < 0.1\%$ for $50\mu$m 
decentring, rising quadratically with the decentring value.
It can therefore be concluded that for the range of spot sizes as measured in data, the influence on the Analysing Power is at most 
$\delta AP/AP=0.1\%$.\\

{\noindent\bf Background Subtraction}\\
Background is subtracted in data analysis on a statistical basis by subtracting data taken with the laser being Off from data taken with the 
laser being On with proper scaling factors. Nevertheless, there could be a possible influence by changing or adding fluctuations to the RMS 
values leading to different reconstructed setup values and consequently biases in the subsequent derived Analysing Power values. 
This possible effect has been studied in PMC, full chain, for laser ON rates varying from $1$kHz to $90$kHz with different fractions of laser 
Off rate to laser On rate $R_{Off}/R_{On} \in [0.02,0.24]$, thus spanning the total range of laser On rates and On/Off fractions that might 
occur in data. 
Two different background modellings with or without off-centre Bremsstrahlung satellites have been studied to explore the dependence on the 
background modelling. No biases have been found above laser On rates of $10$kHz, the change in Analysing Power is therefore estimated 
to be at most $\delta AP/AP = 0.1\%$.

\subsubsection{Fitting Procedure}
The analysis method relies centrally on mapping functions for the first and second moments (mean and RMS values) of the energy asymmetry 
distributions in bins of energy, which are derived from template distributions generated with the parametrised simulation on a regular grid, 
smoothed and then interpolated. These mapping functions are used in a fit to find setup parameters consisting of IP distance, beam spot size 
and pedestal shift giving the best description of the widths of the energy asymmetry distributions (RMS) in the different energy bins. On 
convergence of the fit mapping functions for the first moment, i.e. the mean of the energy asymmetry distribution, for both helicities are 
consulted to calculate the Analysing Power for the given set of parameters.

In the final maps a grid with $20 \times 22 \times 4 = 1760$ phase space points have been generated covering relative IP distances from 
$-200$cm (farther away) to $275$cm (nearer) around the nominal IP distance of $66$m, beam spot sizes from $250\mu$m to $1300\mu$m and pedestal 
shifts from $-300$MeV to $600$MeV.
For each grid point energy asymmetry distributions for the four helicity states with Stokes components $S_3=\pm 1$ and $S_1=\pm1$ are 
accumulated with a total of $100$M photon events per helicity state.

Although great effort has been taken to produce as much statistics as possible for the underlying energy asymmetry distributions, there is 
still finite statistics in the different energy bins, leading to fluctuations in the derived first and second moments. 
A combination of Savitzky-Golay filters and cubic splines algorithms is used to smooth the gridded Monte Carlo templates for both the first 
and second moments, which are then interpolated using basic splines and linear regression algorithms to create 3-dimensional continuous, 
smooth and differentiable mapping functions. 

As is always the case with smoothing, there is a danger to introduce biases by flattening significant structures of the maps. The strength of 
the smoothing has to be chosen such as to give the best compromise in removing short-scale statistical fluctuations while preserving structures 
changing on a longer scale. The quality of the maps is ensured by choosing a grid as fine as possible with as much statistics as possible and 
smoothing algorithms which are known to introduce as little biases as possible on relatively short scales.

A detailed study of the fitting method has been performed using independent simulated data to ensure that the method is self consistent and 
that no biases exist. Covering the complete available phase space of IP distance, beam spot size and pedestal shift, no significant systematic 
biases or problems have been observed, within the expected statistical errors, which are exceeding a level of $\delta AP/ AP = \pm 0.5\%$, 
which is assigned as an estimate of the intrinsic error of the method with the given templates and the chosen smoothing parameters. 
It has been studied that a variation of the smoothing introduces changes not larger than $\pm 0.2\%$, that the number of degrees of freedom 
and the number of energy bins in the fit does not change or bias the results. 
However, the fits, done with the MINUIT package \cite{Minuit:1994}, are based on MIGRAD, which is a local minimiser. It cannot be excluded that
depending on the starting values convergence is reached in different local minima. 
The analysis has been repeated with simulated as well as real data, repeating the fit after changing the starting values over a sensible 
range. The results are observed to change by less than $\pm 0.2\%$, which is assigned as an error from this source. 

The simulated data templates for the mapping functions assume a pedestal shift distribution with equal sharing of the pedestal shift value 
between the Up and the Down channel. 
The influence of this distribution has been studied in PMC, full chain, varying the fraction of the total pedestal shift in the Up channel 
from $0$ to $100\%$ for different total pedestal shift values $E_{p}\in [-100,200]$MeV. No influence on the reconstructed parameters IP 
distance, beam spot size or pedestal shift or the derived Analysing Power values could be found. 

\subsubsection{Laser Light Properties}
The linear polarisation component of the laser light has been measured and minimised in-between the fills by measuring the light intensity 
behind a rotating Glan prism downstream of the interaction point at different high voltage settings of the Pockels cell upstream. 
Using the phase of the Glan prism $\phi_{optic}$, i.e.~the initial angle of its optical plane in the tunnel, and its rotation direction, the 
Stokes components with respect to the horizontal and vertical direction of the HERA plane can be calculated, assuming that the exit window 
between the interaction point inside the HERA vacuum and the analyser box with its optical path in air does not change the linear polarisation,
e.g. through induced birefringence. The Stokes components $S_1$ and $S_2$ and with this the total circularity $S_3$ are needed to create the 
proper mean and RMS maps out of the maps with total polarisations by reweighting methods. 
The initial parameters have been determined to be $\phi_{optic} = -52 \pm 3^\circ$, rotating counter clockwise when looking towards the laser 
beam. The phase offset measured in the light polarisation data is consistent with this observation. 

The uncertainties arising from this determination are studied directly with data, assuming $\phi_{optic} = 45^\circ$ and $67.5^\circ$, 
introducing thus maximal and minimal influence of the measured linear polarisation onto data. 
The change in Analysing Power is then deduced by comparing the resulting polarisation measurements in every energy bin for all data periods of 
HERA II. The maximum change observed is $\delta P / P = 0.5\%$ in the low energy bin and only $\delta P / P = 0.3\%$ in the large central 
energy bin. Within the uncertainties of $\phi_{optic}$ a change of Analysing Power of $\delta P/P = 0.2\%$ is thus estimated.

During the HERA polarisation measurement the Pockels cell is switched at $80$Hz, while during the light polarisation measurement each helicity 
is measured separately at different high voltage settings without such a high rate switching. In the transition time between the two helicities 
the linear light component is mainly undefined and a veto is applied to inhibit the taking of data during this time. However, it the timing of 
the veto relative to the switching is not stable or if the length of the veto is not sufficient to cover the complete transition period, the 
effective linear light polarisation during the polarisation measurement would be larger than measured in-between the fills. A small number of 
fills have been identified where the high voltage supply was not functioning properly, resulting in an incomplete switching of the Pockels cell.
These data are flagged as bad.

The ratio of event rates in a central part of the $\eta$ spectrum over that of a broader range in $\eta$ in the large central energy bin has 
been studied in data for each helicity separately. This ratio should have some sensibility to the linear light contamination of the Compton 
cross section. 
The behaviour of this ratio is compared to that of the event rate asymmetry in a central $\eta$ spectrum, which should have sensibility to the
difference of the linear light components of the two helicities $\delta S_1$. 
In data runs where it is known that the high voltage of at least one helicity was broken, implying $S_\text{lin}=100\%$, a significant change of the
rate asymmetry and the single helicity event rate ratios is observed. Otherwise no such changes or correlations could be observed, concluding 
that no hint can be found on different linear components $S_1$ or $S_\text{lin}$ during polarisation measurement otherwise than measured in-between 
the fills. Also, there is no indication that other problems as described above have happened. 

\subsubsection{Trigger Threshold}
Depending on the threshold settings the trigger thresholds in data are moving considerably from time to time, overall changing in a range
$\approx [2.5,3.8]$GeV, with a finer movement and fluctuation for a given setting. No distortion of the energy spectrum for different trigger 
edge configurations could be found in data spectra and no dependence on the threshold setting, timing, etc.~has been found.

For very high trigger thresholds, near the lower edge of the low energy bin, the trigger threshold might cut into the low energy bin spanning 
the energy range $\approx [4.275,6.175]$GeV, resulting in a distortion of the shape of the $\eta$ distribution in that bin. The consequence 
would be a change of the reconstructed interaction point distance and pedestal shift and therefore the derived Analysing Power. 

The influence of a trigger threshold cutting into the low energy bin is studied in PMC, full chain, by varying the trigger edge in a range
$[2.8,4.4]$GeV and trigger threshold widths $[0.1,0.3]$GeV. The position of the trigger edge can be well reconstructed by calculating the 
maximum in a differentiation method. Interaction point distance and pedestal shift are influenced significantly only for trigger thresholds 
above $4$GeV, the influence on the asymmetry in the low energy bin starting already earlier at $\approx 3.4$GeV, independent of the trigger 
edge width. The next higher energy bin is influenced for thresholds $>3.6$GeV. The Analysing Power in the large central energy bin remains 
unchanged in the complete range of thresholds within $\delta AP/AP = 0.2\%$, and no correction of the Analysing Power as function of the 
threshold position is needed.

\subsubsection{Data related Effects}
All studies presented so far assume that the overall system is stable over time. A number of effects however have been observed in data and are
expected which might vary with time and which cannot be reproduced or understood from simulations.

The reconstructed IP distance shows some small jumps in different periods, of order of $50$cm. These jumps are not correlated to any known 
observables besides the auxiliary pedestal shift variable. A priori it is not excluded that the IP distance did jump in reality, but the 
absence of sufficient monitoring precludes the experimental validation of this assumption. To estimate the impact of the unstable IP distance 
an error has been added which accounts for the typical jump size. This results in an uncertainty of $\delta P/P = 0.5\%$. 

Throughout a time period with stable optics setup, where no explicit machine studies have been performed, the emittance of the lepton beam 
is fairly constant at $2-3$nm, though varying a bit from fill to fill and changing on small scales throughout a fill.
At the end of HERA throughout the low and middle energy proton runs, the emittance is expected to be around $6-7$nm.
Indeed studying the correlation between IP distance and beam spot size, data points move mostly on curves of constant emittance. However some
fluctuations and outliers with too small emittance are observed, which again can not be explained, nor can they be accounted for by problems 
in the reconstruction. To estimate its impact the typical shift in emittance has been found to be around $1$nm, typically connected to very 
large IP distances and thus corresponding to a change in the IP distance of around $70$cm. This translates into an error of the Analysing 
Power of $\delta AP/AP = 0.9\%$. 
Many of the effects discussed above have an impact on the final determination of IP distance and beam spot size, and effectively could move
the fit results in the plane. Thus the observed variation of the emittance is already covered to some extent by other errors. Nevertheless, as 
a conservative estimation, the observed variation is assigned as an additional systematic uncertainty.

The variable pedestal shift has been introduced to describe the possible influence of additional low energy contributions as they might 
arise from physical sources like on-time synchrotron radiation, but also from more complicated electronic effects which are not covered by the 
online pedestal shift subtraction using late digitisation samples which are off-time to the beam pass. Although the reconstructed pedestal 
shift for most of the HERA II data fluctuates around $0$MeV, giving confidence in both the description of the polarimeter setup in the 
parametrised simulation as well as in the absence of such more complicated error sources, there are some periods which are reconstructed with 
pedestal shifts significantly shifted from zero. 
The observed pedestal shifts at those times would correspond to synchrotron energies of less than $3$MeV. As there is no data with very low 
energies available, i.e. no untriggered data is available, the existence of synchrotron radiation can neither be proven nor excluded. In 
addition, there is no independent hint pointing to electronic problems connected to the pedestals of the signals. In that sense, the pedestal 
shift remains an auxiliary parameter of the new analysis. Its total influence is estimated by taking this parameter out of the fit, fixing it 
to zero and running the analysis over the complete HERA II data set with only IP distance and beam spot size as free parameters to the fit. 
The global influence that can be observed in the measured polarisations $P$ with and without pedestal shift or just as well in the ratio to 
the LPOL polarisation for the two configurations is within $\delta P/P = 0.5\%$, which is taken to be the global systematic uncertainty 
connected to this auxiliary.

\subsection{Polarisation Scale from Rise Time Measurements}
\begin{figure}
	\centering
		\includegraphics[width=0.5\textwidth]{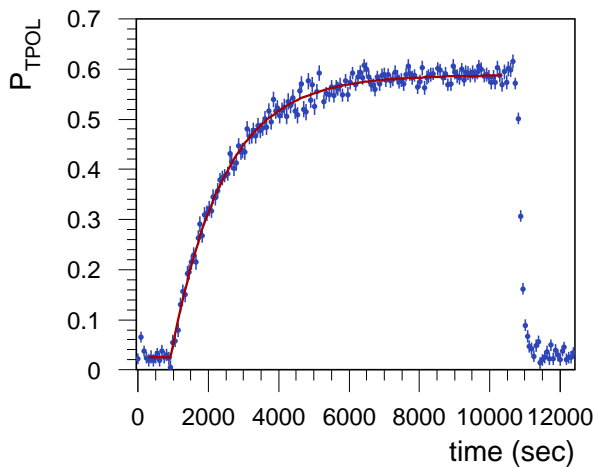}
        \caption{\sl \label{tpol_risetime} Example of a rise time curve, data taken on 26 June 2007, together with a fit to the curve using 
Eqn.~\ref{risetimefit}. The beam is depolarised prior to time $t_0$ with a residual polarisation level $P_0$. After the depolarising kicker 
magnet has been turned off, the polarisation rises with rise time $\tau$, reaching the asymptotic level $P_\mathrm{max}$. Towards the end the 
kicker is switched on again, destroying the polarisation.}
\end{figure}

The maximal polarisation and its rise time according to the Sokolov-Ternov effect in a storage ring are reduced in the presence of
depolarising effects.
However, for most depolarising effects the ratio of asymptotic polarisation limit and rise time $P_\mathrm{max}/\tau$ is still constant
\begin{equation}
   \frac{P_\mathrm{max}}{\tau} = \frac{P_\mathrm{st}}{\tau_\mathrm{st}}
\end{equation}
with $P_\mathrm{st}$ and $\tau_\mathrm{st}$ being the asymptotic polarisation level and rise time given by a pure Sokolov-Ternov effect in 
the absence of any depolarising effects.
The rise time in the presence of depolarising effects can be written as \cite{barberbook}
\begin{equation}
   \frac{1}{\tau} = \frac{1}{\tau_\mathrm{st}} + \frac{1}{\tau_\mathrm{depol}}
\end{equation}
This intrinsic correlation between the maximal polarisation value and the rise time allows \emph{the absolute scale} of the polarisation 
measurement to be checked. Measuring the complete build-up of polarisation from a baseline till saturation and applying a fit of a function 
\begin{equation}
   P(t) = \begin{cases} P_0 + (P_\mathrm{max} - P_0) \cdot (1- e^{-(t-t_0)/\tau}) & \mbox{if $t\geq t_0$}\\ P_0 & \mbox{otherwise} \end{cases} 
   \label{risetimefit}
\end{equation}
allows the determination of the ratio $P_\mathrm{max}/\tau$ and the checking of or even the calibration of the Analysing Power by comparison 
with theoretical expectations. An example for such a rise time measurement is given in Fig.~\ref{tpol_risetime}. 

However, the above ratio might change in the presence of depolarising effects and calibration using rise time measurements will be 
biased unless the strength of the effects can be estimated, e.g.~from machine simulations \cite{barberbook}, or even be measured. Such effects 
become especially important in a non-flat machine as is the case in the presence of spin rotators.

In June 2007 a series of rise time measurements with all three spin rotator pairs has been taken. In the case of three rotator pairs the
theoretical uncertainty in estimation attempts of changes to the above ratio are surmised to be of the order of $3\%$ 
\cite{privcomm:Barber}. 
Dedicated machine simulations, see \cite{barberbook}, based on a detailed model of the spin properties of the HERA collider estimated an 
average value of the ratio of 
\cite{risetimecalcs}
\begin{equation}
   \left< \frac{P_\mathrm{max}}{\tau} \right> = (4.08\pm0.03)\cdot 10^{-4} \mathrm{s}^{-1}.
   \label{risetimetheory}
\end{equation}
\begin{figure}
        \centering
		\includegraphics[width=0.5\textwidth]{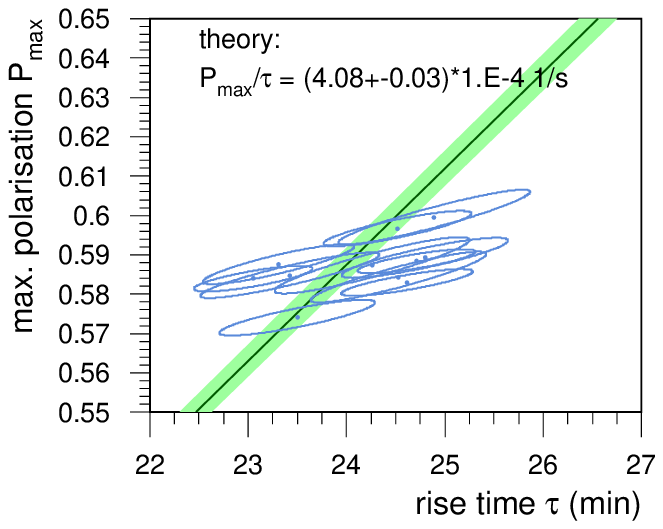}\hfill
		\includegraphics[width=0.48\textwidth]{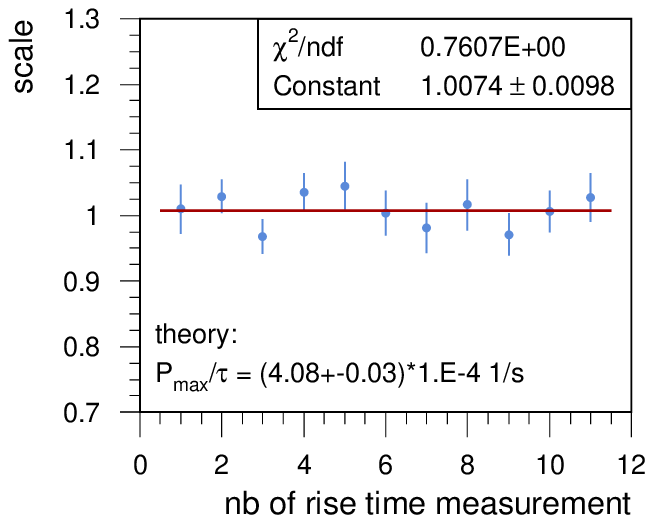}
	\caption{\sl \label{tpol_risetime_results}Fit results for the rise time curves from June 2007. The rise time curves were fitted using
Eqn.~\ref{risetimefit}. The plateau polarisation obtained with the new analysis is plotted against the corresponding rise time 
(left) together with the theory prediction from Eqn.~\ref{risetimetheory}. On the right the absolute scale factor using this theory 
prediction is shown. The average scale factor for the polarisation from the new analysis is $1.007 \pm 0.010$.}

\end{figure}
In Fig.~\ref{tpol_risetime_results} (left) the plateau polarisation and the rise time determined from the analysis of all available rise time 
curves are compared with the theoretical expectation. The right plot shows a summary of the scale factors derived from this analysis including 
the statistical errors from the fit as well as the error from theory.
The average scale is found to be $1.007 \pm 0.010$.
This gives an independent confirmation that the scale determined in the analysis of the TPOL is correct and valid for 
beam conditions, energy and polarimeter conditions as were given during the rise time measurements. 

It should be pointed out that the results from the rise time measurement are not used any further, and in particular are not used to calibrate 
the scale of the TPOL. Beam and polarimeter conditions are different during normal HERA II running and the systematic uncertainty due to these
changes is unknown. However, the results from the rise time measurements give confidence, that the systematic uncertainties as assigned to
the new analysis are complete and sufficient.

In an entirely independent analysis based on the cavity LPOL \cite{cavity} the same data have been analysed using the cavity measurements, and 
a similar agreement has been found. Unfortunately a slightly different theoretical model for the polarisation rise time was used in this 
analysis, which makes a final quantitative comparison difficult. However within the relative error between the two theories of $4\%$ both 
results are compatible. 

\section{Comparison of LPOL and TPOL Measurements}\label{sec:compare}
\begin{figure}
	\centering
		\includegraphics[width=0.97\textwidth]{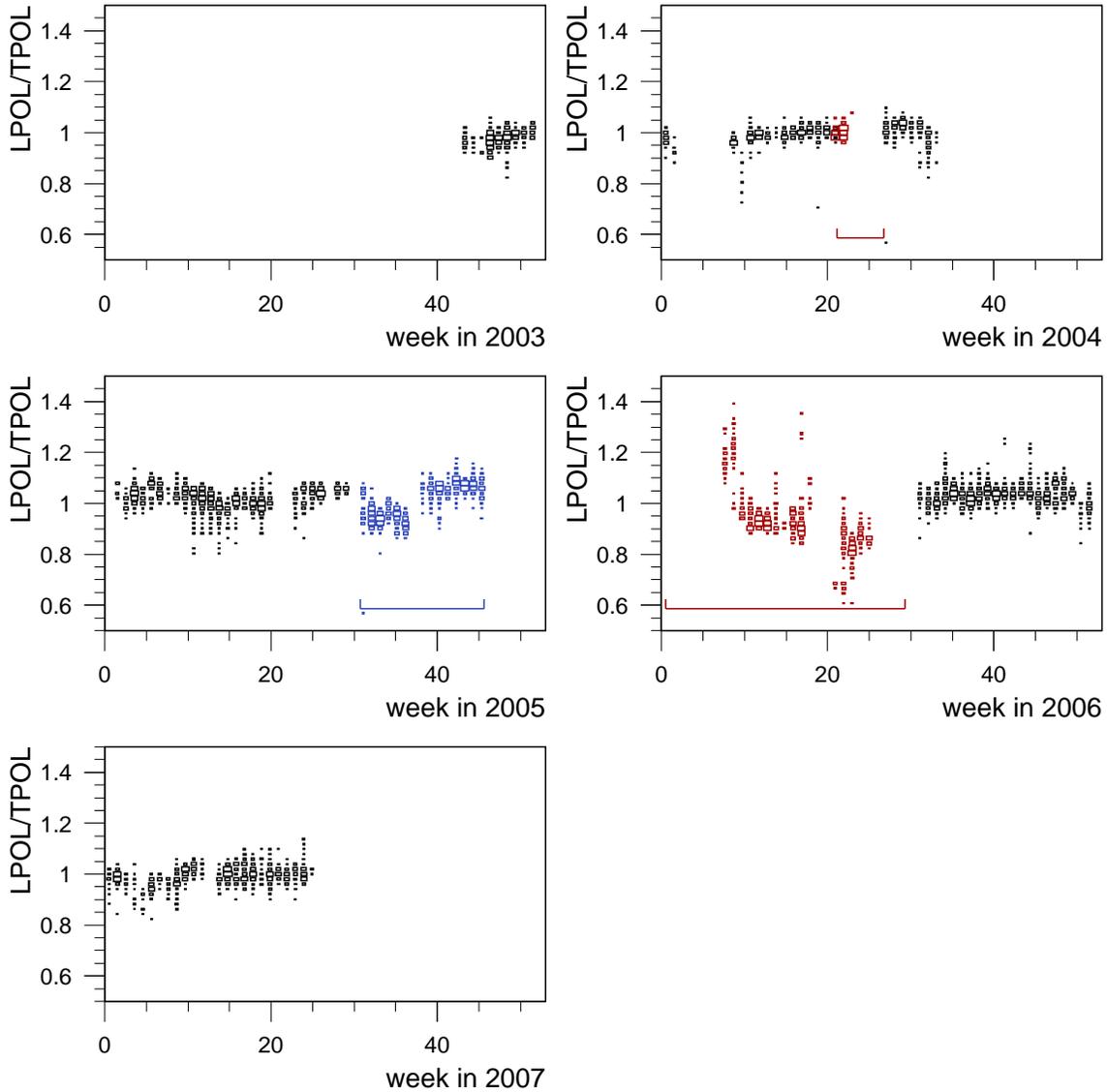}
	
	\caption{\label{LTratiovstime}Top: Ratio LPOL/TPOL for the years 2003-2007 for colliding bunches. Periods where LPOL is known to have 
had problems and should be discarded are indicated by red marks and the brackets at the bottom of the plots.
The period marked in blue and the bracket at the bottom shows polarimeter problems too, the reasons for which are unknown. It is recommended 
to increase the systematic uncertainty for both polarimeters during this period. 
See also Tab.~\ref{badperiods} for details.}
\end{figure}

\begin{figure}
	\centering
		\includegraphics[width=0.9\textwidth]{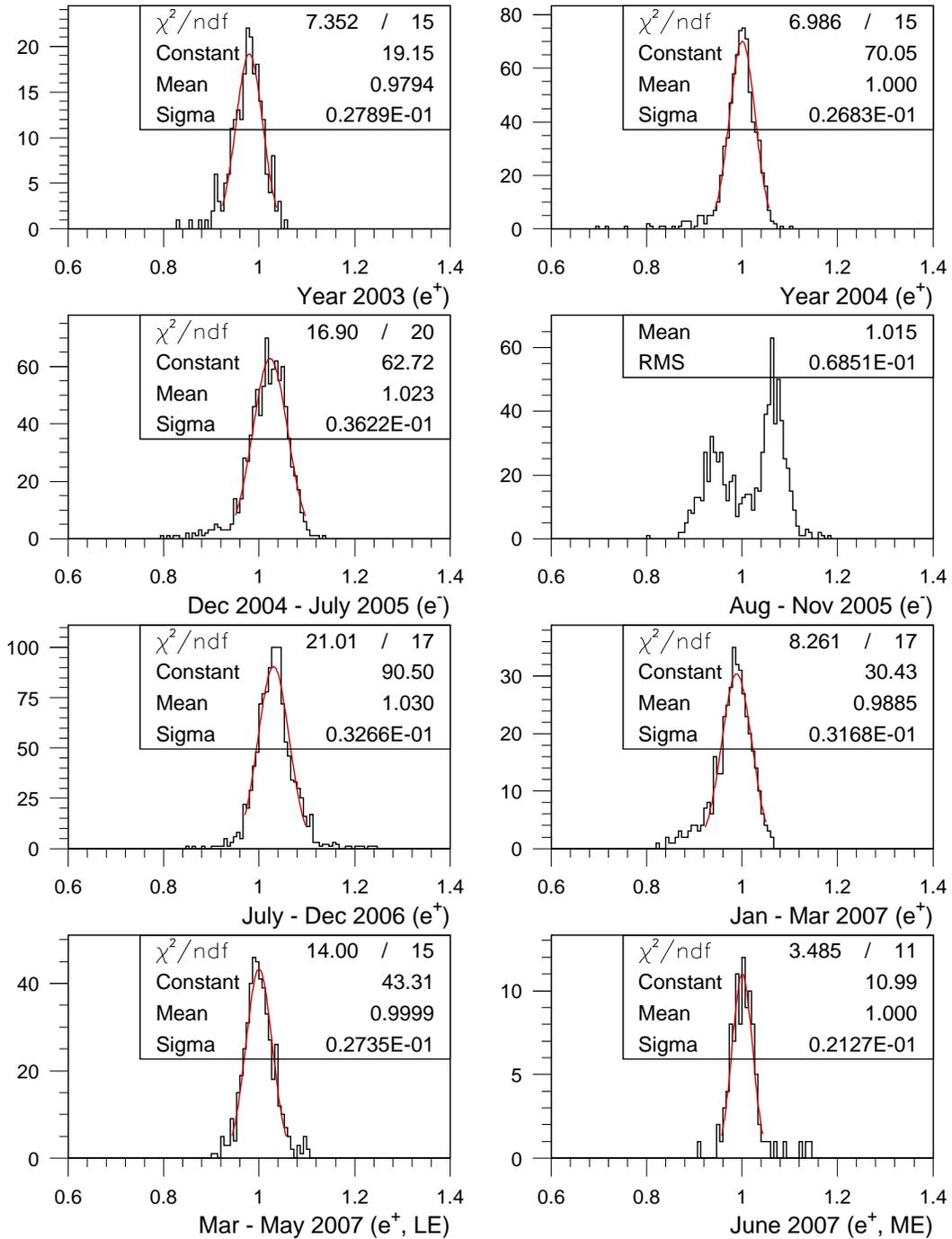}
	
	\caption{\label{LTratio}Ratio LPOL/TPOL for the complete HERA II data set using the new TPOL analysis. Data used are 1 hour average 
data for colliding bunches only. Superimposed is a fit to a single Gaussian curve, with the fit range restricted to be within $\pm 2\sigma$.
Time periods with known LPOL problems are not shown, see Tab.~\ref{badperiods}. The period Aug.-Nov.~2005 shows polarimeter problems too and 
it is recommended to increase the systematic uncertainty for both polarimeters here.} 
\end{figure}

To compare the results of LPOL and TPOL the ratio of the LPOL polarisation to the TPOL polarisation is shown in Fig.~\ref{LTratiovstime} 
and Fig.~\ref{LTratio}. 
Normal data taking periods are presented in black points, while red and blue marks indicate periods where serious problems with at least one 
of the two polarimeters must have existed. For these plots only data taken during luminosity operation of the experiments are included, and 
only with a minimal polarisation of $20\%$ to ensure exclusion of the initial polarisation build-up period at the start of the fill. 

The known periods of poor data quality and the corresponding recommendations are reported in Tab.~\ref{badperiods}.

\begin{table} \begin{center}
\begin{tabular}[h]{llll}
\hline
Year & Start & End & POL2000 Recommendation \\
\hline
2004 & 1085373000 (May 24)  & 1088786591 (July 2)  & Discard LPOL \\
2005 & 1122990001 (Aug.~2)  & 1136073599 (Dec.~31) & \begin{minipage}[t][5.5ex]{4.5cm}{Increase systematic error LPOL and TPOL}\end{minipage}\\
2006 & 1136073600 (Jan.~1)  & 1153506512 (July 21) & Discard LPOL \\
\hline
\end{tabular}
      \caption{Long periods with polarimeter problems. The start/stop time is given as a UNIX timestamp (seconds since 1/1/1970).}
      \label{badperiods}
\end{center} \end{table}
Towards the end of the first half of 2004 it was observed that significant damage has been done to the LPOL crystal calorimeter, which made 
the complete replacement of the device necessary. It is not clear when the quality of the data started to deteriorate. During this period the 
TPOL was functioning properly, and only measurements from the TPOL should be used. As described in the section on the LPOL the replacement 
of the LPOL calorimeter necessitated the introduction of an additional scale error of the LPOL of $1.4\%$. 

Starting in the summer of 2005 the LPOL/TPOL ratio displayed a number of spurious jumps which have not been explained in detail. During this 
time, from weeks 31 to 36 a drop is visible in the LPOL/TPOL ratio, followed by an increase in weeks 38 to 45. 
This behaviour continued until end of 2005. Data in this period need to be treated specially.

During the first half of 2006 the LPOL encountered a number of serious problems. After the winter shutdown a mistake in the laser transport 
system resulted in a wrong laser spot size, which introduced spurious and unreliable measurements. These data have been flagged as unreliable 
for the LPOL and should be excluded from analyses, see Tab.~\ref{badperiods}. Only the TPOL measurements should be used throughout this period.

A summary of all good periods is given in Tab.~\ref{yearcomp}. The means vary from 1, but are compatible with the quoted error of the 
individual measurements of $2\%$. From the combination of the two devices we expect a width of the distribution of around $3\%$. This is 
compatible with the observation given in the table. Compared to the previous analysis the width has been reduced significantly. 

\begin{table} \begin{center}
\begin{tabular}[h]{c|cc|cc}
\hline    & \multicolumn{2}{|c|}{Old TPOL} & \multicolumn{2}{|c}{New TPOL} \\
Year & Mean & Width & Mean & Width   \\
\hline
2003 & 0.979 & 0.033 & 0.979 & 0.028 \\ 
2004 & 0.997 & 0.024 & 1.000 & 0.027 \\
2005 & 1.004 & 0.039 & 1.023 & 0.036 \\
2006 & 1.017 & 0.036 & 1.030 & 0.033 \\
2007 & 0.989 & 0.040 & 0.996 & 0.029 \\
\hline
\end{tabular}
\caption{\sl \label{yearcomp} Yearwise comparison of mean and width of the LPOL/TPOL ratio distributions for the old and the new TPOL analysis.
For the old analysis values are taken from \cite{pol-summer2007}.}
\end{center} \end{table}

\section{Recommendations for Polarisation Values Treatment}
In this section the recommended way to treat the polarisation measurements from the TPOL and the LPOL is discussed. 

\subsection{Systematic Error Classes}\label{sec:class}
The systematic errors for both the TPOL and the LPOL enter into any discussion of combined measurements. Two different types of correlations 
between the errors are considered: self correlations within one polarimeter, and correlations between the two polarimeters. When combining 
polarisation measurements within one and with the other polarimeter both types of correlations need to be taken into account. 
The results from both the TPOL and the LPOL can be written as: 

\begin{equation}\begin{gathered}
    P^\text{LPOL} \pm \delta P^\text{LPOL}_\text{stat} \pm \delta P^\text{LPOL}_{L u,T u} \pm \delta P^\text{LPOL}_{L c,T u} 
                  \pm \delta P^\text{LPOL}_{L c,T c}   \pm \delta P^\text{LPOL}_{L u,T c}\\
    P^\text{TPOL} \pm \delta P^\text{TPOL}_\text{stat} \pm \underbrace{\delta P^{TPOL}_{T u,L u}}_\text{class I} 
                  \pm \underbrace{\delta P^\text{TPOL}_{T c,L u}}_\text{class II} 
                  \pm \underbrace{\delta P^\text{TPOL}_{T c,L c}}_\text{class III} 
                  \pm \underbrace{\delta P^\text{TPOL}_{T u,L c}}_\text{class IV}
\end{gathered}\end{equation}
where a measurement value is accompanied by a statistical error and up to four different types of systematic errors:
\begin{itemize}
\item {\bf Class I} errors are totally uncorrelated systematic errors. They apply only to the polarimeter in question, and are neither 
correlated to a measurement of the same polarimeter at a different time (on a sufficiently large time scale), nor to a measurement of the 
second polarimeter.
\item {\bf Class II} errors correlate measurements of the same polarimeter (at a different time, on a sufficiently large time scale), 
introducing a so-called self correlation. The errors of this class divide further into \emph{period dependent} errors, i.e.~where correlations 
over large time ranges are assumed to vanish and \emph{period independent} errors, where correlations over large time ranges are assumed to 
persist. 
Examples are given e.g.~by global scale factors of each polarimeter (period independent), which correlate the measurements of a 
polarimeter possibly over a large time range, but are independent of the other polarimeter, or errors relevant only for one polarimeter, 
e.g.~laser beam alignment, etc.~(period dependent).
\item {\bf Class III} errors correlate the measurement of one polarimeter to that of the other one. By definition such errors also imply self 
correlation within one polarimeter. Examples are given e.g.~by machine dependent errors, e.g.~the HERA optics, which might affect both 
polarimeters at once. Errors of this class are period dependent.
\item {\bf Class IV} errors are those which describe internally uncorrelated errors, but result in a correlation to the other experiment. This 
would imply that a measurement of one polarimeter fluctuates quasi statistical, but in a correlated manner with the other polarimeter. 
As both polarimeters measure the polarisation independently there is no reason to believe that such errors exist.
\end{itemize}

The internal (self) correlation affects measurements from one polarimeter at different times in a similar way. Thus when making averages over 
long times, as they are needed by the experiments, this correlation needs to be taken into account. It will affect the error, but potentially 
also the value of the average. Sources of systematic error implying such long correlations are called \emph{period independent}, examples are 
given by e.g.~the global scale errors of both polarimeters.  
If there are reasons to believe that correlations are changing over sufficiently short times, so that two very long periods can be regarded as 
uncorrelated, the sources of systematic uncertainty are called \emph{period dependent}.  Class II errors can display both types of period 
dependencies. 

Being of quasi-statistical nature, the period dependence classification is not applicable to class I errors.

In the absence of class IV errors, correlation with the other polarimeter also implies self correlation (class III errors). This type of 
errors is induced by external sources which are common to both polarimeters, thus not only correlating two associated measurements of both 
polarimeters, but equally the measurements of each polarimeter with themselves. Being induced by external sources which are subject to change 
over time, class III errors are taken to be period dependent.

Possible period dependence needs to be distinguished from the possible time dependence of the assigned value of the uncertainty.


\begin{table}\begin{center}
\begin{tabular}[t]{c|cc}
\hline 
     & \multicolumn{2}{|c}{$\delta P / P \,(\%)$} \\   
Class& LPOL & TPOL\\
\hline
I    & 0.62 & 0.89   \\ 
IIu  & 1.70 & 0.99   \\
IId  & 0.40 & 1.20   \\
IIId & 0.80 & 0.53   \\
\hline
Sum  & 2.01 & 1.87   \\
\hline

\end{tabular}
\caption{\sl \label{tab:classsyst} Summary of systematic uncertainties for the LPOL and the TPOL, in percent. The systematic uncertainties of 
both polarimeters are given for three classes with definitions as described in Sect.~\ref{sec:class} (uncorrelated, internally correlated, 
correlated between both polarimeters) and period dependence (u: independent, d: dependent).}
\end{center} \end{table}
Tab.~\ref{tab:lpol} lists the known sources of systematic uncertainties for the LPOL and Tab.~\ref{tab:TPOLsys1} and Tab.~\ref{tab:TPOLsys2} 
those of the TPOL, with the class of the systematic errors and possible period dependence (u: independent, d: dependent) indicated.
Based on this table the final list of systematic errors of the two polarimeters split into the different categories are given in 
Tab.~\ref{tab:classsyst}.
 
\subsection{Error Scale Factors}\label{subsec:scales}
In Sect.~\ref{sec:compare} the ratio of LPOL/TPOL has been discussed. For a given period this distribution should be centred around 1, and 
should have a width compatible with the total systematic error within this period. Following the procedure for averaging results given in the 
PDG \cite{pdg} a $\chi^2$ is formed for each of the different periods: 
\begin{equation}
   \chi^2 = \sum_i { \frac{(\langle R\rangle - R_i)^2}{dR_i^2}},
\end{equation}
where the sum runs over $N$ measurements of the LPOL/TPOL ratio $R$ with systematic errors $dR$ in this period, which is formed from the 
separate systematic errors of the two polarimeters assuming full correlation between the class III errors. The expectation value of the 
ratio is $\langle R \rangle = 1$.
If  $\chi^2/{N-1}$ is less or equal to 1, there are no problems with the errors. Where $\chi^2/{N-1}$ is somewhat larger than 1, 
a scale factor is calculated according to 
\begin{equation}
   S = \sqrt{ \frac{\chi^2}{N-1} },
\end{equation}
by which the error of the ratio is scaled. By means of error propagation the same scale factor needs to be applied to the systematic 
contributions of each polarimeter too. This scale factor is calculated for all periods individually which are considered for the polarisation 
measurement. 
It is based on measurements taken within a given period, 
excluding those which are outside $\pm 2 \sigma$ around the mean of the distribution, to stabilise the results 
and remove outliers. As class I errors are considered as being quasi-statistical errors, they do not contribute in moving the ratio from 1, 
the scale factors are therefore recalculated to be applied only to the class II and III errors:
\begin{equation}
 S' = \sqrt{\frac{ S^2 \left(\frac{dR^2}{R^2}\right)_\text{tot} - \left(\frac{dR^2}{R^2}\right)_{I} }
                 {\left(\frac{dR^2}{R^2}\right)_{IIu} + \left(\frac{dR^2}{R^2}\right)_{IId} + \left(\frac{dR^2}{R^2}\right)_{IIId}}}
\end{equation}
The results of this procedure are summarised in Tab.~\ref{tab:perioderrors}.
\begin{sidewaystable}
\begin{tabular}[t]{clll|c|c|cccc|cccc|l}
\hline
\multicolumn{4}{c|}{Time Period}    & & Scale $S'$ from
                          & \multicolumn{4}{c|}{$(\delta P / P)_\text{LPOL}\,(\%)$} 
                          & \multicolumn{4}{c|}{$(\delta P / P)_\text{TPOL}\,(\%)$} & \\ 
Nb.&Year & Start & End & Est.~$\cal L$& $\chi^2/(N-1)$ & I & IIu & IId & IIId & I & IIu & IId & IIId & Comment\\
\hline
1&2003         & -                                                                & \begin{minipage}[t]{2cm}{1072915199 (Dec.~31)}\end{minipage} &$\sim 4 \%$ &1.186&0.62&2.01&0.47&0.95&0.89&1.17&1.42&0.63& \\
2&2004         & \begin{minipage}[t]{2cm}{1072915200 (Jan.~1)}\end{minipage}      & \begin{minipage}[t]{2cm}{1101859199 (Nov.~30)}\end{minipage} &$\sim 14\%$ &1    &0.62&1.70&0.40&0.80&0.89&0.99&1.20&0.53& \begin{minipage}[t]{2.5cm}{Discard LPOL in-between}\end{minipage}\\
3&2004/05      & \begin{minipage}[t]{2cm}{1101859200 (Dec.~1)}\end{minipage}      & \begin{minipage}[t]{2cm}{1122990000 (Aug.~2)}\end{minipage}  &$\sim 17\%$ &1.327&0.62&2.25&0.53&1.06&0.89&1.31&1.59&0.70& \\
4&2005         & \begin{minipage}[t]{2cm}{1122990001 (Aug.~2)}\end{minipage}      & \begin{minipage}[t]{2cm}{1126137599 (Sept.~7)}\end{minipage} &$\sim 6 \%$ &2.551&0.62&4.33&1.02&2.04&0.89&2.52&3.05&1.35& \\
5&2005             & \begin{minipage}[t]{2cm}{1126137600 (Sept.~8)}\end{minipage}     & \begin{minipage}[t]{2cm}{1136073599 (Dec.~31)}\end{minipage} &$\sim 8 \%$ &2.402&0.62&4.08&0.96&1.92&0.89&2.38&2.87&1.27& \\
6&2006         & \begin{minipage}[t]{2cm}{1136073600 (Jan.~1)}\end{minipage}      & \begin{minipage}[t]{2cm}{1151711999 (June 30)}\end{minipage} &$\sim 11\%$ &1.397&-   &-   &-   &-   &0.89&1.38&1.67&0.74& Discard LPOL\\
7&2006             & \begin{minipage}[t]{2cm}{1151712000 (July 1)}\end{minipage}      & \begin{minipage}[t]{2cm}{1167609599 (Dec.~31)}\end{minipage} &$\sim 20\%$ &1.397&0.62&2.37&0.56&1.12&0.89&1.38&1.67&0.74& \begin{minipage}[t]{2.5cm}{Discard LPOL at beginning}\end{minipage}\\
8&2007         & \begin{minipage}[t]{2cm}{1167609600 (Jan.~1)}\end{minipage}      & \begin{minipage}[t]{2cm}{1174521599 (Mar.~21)}\end{minipage} &$\sim 7 \%$ &1.094&0.62&1.86&0.44&0.87&0.89&1.08&1.31&0.58& \\
9&2007             & \begin{minipage}[t]{2cm}{1174521600 (Mar.~22)}\end{minipage}     & \begin{minipage}[t]{2cm}{1180693800 (June 1)}\end{minipage}  &$\sim 9 \%$ &1    &0.62&1.70&0.40&0.80&0.89&0.99&1.20&0.53& \\
10&2007             & \begin{minipage}[t][5.5ex]{2cm}{1180693800 (June 1)}\end{minipage}      & - (July 1)                                           &$\sim 2 \%$ &1    &0.62&1.70&0.40&0.80&0.89&0.99&1.20&0.53& \\
\hline
\end{tabular}
\caption{\sl \label{tab:perioderrors}Table summarising the results from TPOL and LPOL with systematic errors, for the different run periods 
of HERA II. Indicated is also the integrated luminosity collected for each of the periods. Note that this number is only indicative and 
included for information only, and should not be used in any analysis.
The scale factor $S'$ applies to the error classes II and III of both polarimeters, resulting in the effective systematic values for LPOL
and TPOL as given in this table.
The start/stop times are given as a UNIX timestamp (seconds since 1/1/1970).}
\end{sidewaystable}
The helicity changes of the longitudinal polarisation of the lepton beam as well as the particle type in the different running periods of 
HERA II are summarised in Tab.~\ref{tab:helicities}.
\begin{table}\begin{centering}
\begin{tabular}[t]{llllll|c|c}
\hline
\multicolumn{6}{c|}{Time Period}& & LPOL\\
Nb. & Year &  \multicolumn{2}{c}{Start} & \multicolumn{2}{c|}{End} & Particle & Sign\\
\hline
1 & 2003   &-		&               &1055000000	&(June 7)	&$e^+$&	+1\\
1 & 2003   &1055000001	&(June 7)	&1072915199	&(Dec.~31)	&$e^+$&	-1\\
2 & 2004   &1072915200	&(Jan.~1)	&1080850000	&(Apr.~1)	&$e^+$&	-1\\
2 & 2004   &1080850001	&(Apr.~1)	&1088040000	&(June 24)	&$e^+$&	+1\\
2 & 2004   &1088040001	&(June 24)	&1101859199	&(Nov.~30)	&$e^+$&	-1\\
3 & 2004/05&1101859200	&(Dec.~1)	&1107200000	&(Jan.~31)	&$e^-$&	-1\\
3 & 2005   &1107200001	&(Jan.~31)	&1116780000	&(May 22)	&$e^-$&	+1\\
3 & 2005   &1116780001	&(May 22)	&1122990000	&(Aug.~2)	&$e^-$&	-1\\
4 & 2005   &1122990001	&(Aug.~2)	&1126137599	&(Sept.~7)	&$e^-$&	-1\\
5 & 2005   &1126137600	&(Sept.~8)	&1136073599	&(Dec.~31)	&$e^-$&	+1\\
6 & 2006   &1136073600	&(Jan.~1)	&1146620000	&(May 3)        &$e^-$&	+1\\
6 & 2006   &1146620001	&(May 3)	&1151711999	&(June 30)	&$e^-$&	-1\\
7 & 2006   &1151712000	&(July 1)	&1165480000	&(Dec.~7)	&$e^+$&	-1\\
7 & 2006   &1165480001	&(Dec.~7)	&1167609599	&(Dec.~31)	&$e^+$&	+1\\
8 & 2007   &1167609600	&(Jan.~1)	&1174521599	&(Mar.~21)	&$e^+$&	+1\\
9 & 2007   &1174521600	&(Mar.~22)	&1180693800	&(June 1)	&$e^+$&	-1\\
10& 2007   &1180693800	&(June 1)	&-              &(July 1)	&$e^+$&	-1\\
\hline
\end{tabular}
\caption{\sl \label{tab:helicities} Table summarising the longitudinal polarisation helicity and 
the particle type of the lepton beam for the different run periods of HERA II as defined in 
Tab.~\ref{tab:perioderrors}. 
The helicity is given by the LPOL polarisation sign, which is defined by the HERMES spin rotator setting.
The spin rotators at H1 and ZEUS were operational from Oct.~2003 onwards and were usually 
flipped together with the HERMES rotator except for the last helicity flip in March 2007. In addition an 
overall sign change might apply and
possible additional helicity changes in the H1 or ZEUS spin rotators are not comprised.
The start/stop time is given as a UNIX timestamp (seconds since 1/1/1970).}
\end{centering}\end{table}

\subsection{Averaging of Polarimeter Measurements}
\subsubsection{Error Classes in Averaging Procedures}
The polarimeter uncertainties as defined in Sect.~\ref{sec:class} can be categorised according to the type of correlation between pairs of
distinct polarisation measurements.
In the following, the term \emph{fully correlated} shall be used for two polarisation values $P_i$ and $P_j$, if the following condition for the 
correlation coefficient between the two is fulfilled: $\rho(P_i,P_j)=\pm 1$. 
It is important to consider the sign of the correlation coefficient, as the LPOL measurements carry a sign depending on the longitudinal 
polarisation helicity. A pair of polarimeter measurements of opposite helicity $(P_L,P_R)$ which are considered \emph{fully correlated} are in 
fact negatively correlated, i.e.~$\rho(P_L,P_R)=-1$.
On the other hand the term \emph{uncorrelated} shall correspond to the case where $\rho(P_i,P_j)=0$. 

The relationship between the polarimeter uncertainties and the type of correlation is outlined in more detail:
\begin{itemize}
\item {\bf stat:} Statistical errors are uncorrelated between any pair of polarimeter measurements. When forming averages over longer 
periods these errors decrease and can be eventually neglected in long periods.
\item {\bf local:} \emph{Local} systematic errors correspond to errors with short-range correlations.
For two measurements taken in the same run period and with the same polarimeter the corresponding \emph{local} uncertainties are considered fully 
correlated. If the measurements originate from different run periods or from different polarimeters, the corresponding \emph{local} uncertainties 
are considered uncorrelated. According to this \emph{local} errors are formed by the error classes I and IId.
\item {\bf hera:} \emph{Hera} systematic errors originate from machine dependent effects. The \emph{hera} uncertainties are fully correlated for 
any measurement of either polarimeter within one run period. Only if the measurements originate from different run periods their \emph{hera} 
uncertainties are considered uncorrelated. \emph{Hera} errors are represented by the error class IIId.
\item {\bf scale:} \emph{Scale} uncertainties are systematic errors affecting the overall scale of all measurements of a given polarimeter.
The \emph{scale} uncertainties of any pair of measurements taken with the same polarimeter are considered fully correlated. When dealing with 
measurements of different polarimeters, the corresponding \emph{scale} uncertainties are considered uncorrelated.
\emph{Scale} errors are represented by the error class IIu.
\end{itemize}
Using the notation $\delta := \delta P/P$ for the relative systematic errors, the above description can be summarised adding class I and IId errors
in quadrature:
\begin{equation}\begin{split}\label{eqn:class}
    \delta_\text{local} &= \sqrt{\delta_I^2 + \delta_{IId}^2}\\
    \delta_\text{hera}  &= \delta_{IIId}\\
    \delta_\text{scale} &= \delta_{IIu}
\end{split}\end{equation}

\subsubsection{Averaging Procedure}\label{sect:avgproc}
For use in the experiments luminosity weighted time averages of the polarimeter measurements are required. A possible procedure might look like
\begin{enumerate}
\item Analyse each run period with constant helicity separately. The information about the systematic errors of the run periods is given in 
Tab.~\ref{tab:perioderrors} and the information about helicity changes is given in Tab.~\ref{tab:helicities}.

\item Group the experiment data for a given run period into three samples, depending on the type of polarisation measurement available in a 
      certain time window (e.g.~5 minutes). Calculate for each data sample the contributing integrated luminosity ${\cal L}_i$:
  \begin{itemize}
    \item ${\cal L}_{L}$: integrated luminosity of the data sample data with LPOL measurements only
    \item ${\cal L}_{T}$: integrated luminosity of the data sample data with TPOL measurements only
    \item ${\cal L}_{LT}$: integrated luminosity of the data sample data with measurements available from both polarimeters
    \end{itemize}

\item Calculate for the three data samples four luminosity weighted average polarisation values:
  \begin{itemize}
    \item $P_{L\text{,\,avg}}^\text{LPOL}$: luminosity weighted average of LPOL polarisation corresponding to the integrated 
          luminosity ${\cal L}_{L}$
    \item $P_{T\text{,\,avg}}^\text{TPOL}$: luminosity weighted average of TPOL polarisation corresponding to the integrated 
          luminosity ${\cal L}_{T}$
    \item $P_{LT\text{,\,avg}}^\text{LPOL}$: luminosity weighted average LPOL polarisation corresponding to the integrated 
          luminosity ${\cal L}_{LT}$
    \item $P_{LT\text{,\,avg}}^\text{TPOL}$: luminosity weighted average TPOL polarisation corresponding to the integrated 
          luminosity ${\cal L}_{LT}$
  \end{itemize}
  At this step, the statistical uncertainties of the polarisation averages over a full run period are small and can be neglected safely.
  If desired, they can still be calculated, taking into account that distinct polarimeter measurements are statistically independent.

\item Calculate the luminosity weighted average of the four polarimeter averages for the given run period as
  \begin{equation}
    P_\text{avg} = \frac{{\cal L}_{L}\vert P_{L\text{,\,avg}}^\text{LPOL} \vert
                       + {\cal L}_{T}      P_{T\text{,\,avg}}^\text{TPOL}
                       + {\cal L}_{LT} \left( f_L \vert P_{LT\text{,\,avg}}^\text{LPOL} \vert 
                            + f_T P_{LT\text{,\,avg}}^\text{TPOL} \right)}
                        {{\cal L}_{L} + {\cal L}_{T} + {\cal L}_{LT}} \, .
  \end{equation}
  Here, $f_L$ is the relative weight of the LPOL average when averaging with the TPOL and $f_T=1-f_L$. We recommend to simply use $f_L=1/2$,
  because the systematic errors of LPOL and TPOL are of similar size. However, if desired, the parameter $f_L$ can be varied such that the 
  overall systematic uncertainty is minimised. 
  Note, that the signed LPOL averages need to enter with their absolute values, while the TPOL values are helicity free. The average polarisation 
  $P_\text{avg}$ calculated in this way is helicity free as well.
  Alternatively, the TPOL averages can be multiplied by a helicity sign $h=\pm 1$ corresponding to the LPOL average signs, resulting then in a 
  signed average polarisation value $P_\text{avg}$.

  The average polarisation $P_\text{avg}$ can be rewritten as
  \begin{equation}
     P_\text{avg} = w_{L} \vert P_\text{avg}^\text{LPOL} \vert + w_{T} P_\text{avg}^\text{TPOL}
  \end{equation}
  using weights
  \begin{equation}\label{eqn:periodweightlpoltpol}
    w_{L} =  \frac{{\cal L}_{L} + f_{L} {\cal L}_{LT}}{{\cal L}_{L}+{\cal L}_{T}+{\cal L}_{LT}} 
    \qquad\mbox{and}\qquad    
    w_{T} = 1 - w_{L}
  \end{equation}
  and average LPOL and TPOL polarisations
  \begin{equation}\begin{split}\label{eqn:periodlpoltpol}
    \vert P_\text{avg}^\text{LPOL} \vert &= \frac{{\cal L}_{L} \, \vert P_{L\text{,\,avg}}^\text{LPOL} \vert
                                                + f_{L} \, {\cal L}_{LT} \, \vert P_{LT\text{,\,avg}}^\text{LPOL} \vert}
                                                 {{\cal L}_{L} + \, f_{L} \, {\cal L}_{LT}}\\
    P_\text{avg}^\text{TPOL} & = \frac{{\cal L}_{T} \, P_{T\text{,\,avg}}^\text{TPOL}
                                       +f_T \, {\cal L}_{LT} \, P_{LT\text{,\,avg}}^{TPOL}}
                                       {{\cal L}_{T}+ \, f_T \, {\cal L}_{LT}} \, .
  \end{split}\end{equation}
    
\item Calculate the corresponding systematic uncertainties as
  \begin{equation}\begin{split}
    \Delta P_\text{scale,\,avg}^\text{LPOL} &= w_L \, \vert P_\text{avg}^\text{LPOL}\vert \, \delta_\text{scale}^\text{LPOL} \\
    \Delta P_\text{local,\,avg}^\text{LPOL} &= w_L \, \vert P_\text{avg}^\text{LPOL}\vert \, \delta_\text{local}^\text{LPOL} \\
    \Delta P_\text{scale,\,avg}^\text{TPOL} &= w_T \, P_\text{avg}^\text{TPOL} \, \delta_\text{scale}^\text{TPOL} \\
    \Delta P_\text{local,\,avg}^\text{TPOL} &= w_T \, P_\text{avg}^\text{TPOL} \, \delta_\text{local}^\text{TPOL} \\
    \Delta P_\text{hera,\,avg}              &= w_L \, \vert P_\text{avg}^\text{LPOL} \vert \, \delta_\text{hera}^\text{LPOL}
                                                        \,+\, w_T \, P_\text{avg}^\text{TPOL} \, \delta_\text{hera}^\text{TPOL} 
  \end{split}\end{equation}
  with the relative uncertainties $\delta^\text{LPOL,TPOL}_\text{scale,local,hera}$ given by Eqn.~\ref{eqn:class} and the values as summarised
  in Tab.~\ref{tab:perioderrors} for each run period and adding the \emph{hera} errors linearly to take their correlation into account.

\item Add the \emph{local} and the \emph{hera} errors quadratically as those two classes are uncorrelated to each other and across run periods:
\begin{equation}
  \Delta P_\text{uncorr,avg}=\sqrt{(\Delta P_\text{local,\,avg}^\text{LPOL})^2 
                                 + (\Delta P_\text{local,\,avg}^\text{TPOL})^2 
                                 + (\Delta P_\text{hera,\,avg})^2}\, .
\end{equation}
The result are an absolute value for the average polarisation and three sources of systematic uncertainty given for each run period:
\begin{equation}
   P_\text{avg} \, \pm \, \Delta P_\text{scale,\,avg}^\text{LPOL} 
                \, \pm \, \Delta P_\text{scale,\,avg}^\text{TPOL} 
                \, \pm \, \Delta P_\text{uncorr,avg} \,.
\end{equation}

\item If desired, average run periods with same particle type ($e^+$ or $e^-$) and same longitudinal polarisation helicity.
The average polarisation of $i=1,N$ run periods with integrated luminosities ${\cal L}_i$ and average polarisations $P_{i}$ 
derived using the preceding steps, calculates as
\begin{equation}
   P_\text{avg} = \sum_{i=1}^N W_i P_i  \qquad \mbox{with weights} \qquad  W_i = \frac{{\cal L}_i}{\sum_j^N {\cal L}_j}
\end{equation}
which can be expressed again in terms of average LPOL and TPOL polarisations as
\begin{equation}
   P_\text{avg} = W_L \, \vert P_\text{avg}^\text{LPOL} \vert + W_T \, P_\text{avg}^\text{TPOL}
\end{equation}
with LPOL and TPOL polarisations averaged over the run periods and corresponding weights
\begin{equation}\begin{split}
   \vert P_\text{avg}^\text{LPOL} \vert &= \frac{\sum_i W_i \, w_{L,i} \, \vert P_i^\text{LPOL} \vert}{\sum_i W_i \, w_{L,i}}\\
   P_\text{avg}^\text{TPOL} &= \frac{\sum_i W_i \, w_{T,i} \, P_i^\text{TPOL}}{\sum_i W_i \, w_{T,i}}\\
   W_L = \sum_{i=1}^N W_i \, &w_{L,i} \qquad \mbox{and} \qquad W_T = \sum_{i=1}^N W_i \, w_{T,i}
\end{split}\end{equation}
where the weights $w_{L,i}$ and $w_{T,i}$ denote the relative weights between LPOL and TPOL averages for each run period $i$ following 
Eqn.~\ref{eqn:periodweightlpoltpol} and Eqn.~\ref{eqn:periodlpoltpol}. Of course, the overall weights obey $W_L + W_T = 1$. 

The corresponding systematic uncertainties of the average polarisation $P_\text{avg}$ are given by
  \begin{equation}\begin{split}
    \Delta P_\text{scale,\,avg}^\text{LPOL} &= \sum_i W_i \, \Delta P_{\text{scale},i}^\text{LPOL}\\
    \Delta P_\text{scale,\,avg}^\text{TPOL} &= \sum_i W_i \, \Delta P_{\text{scale},i}^\text{TPOL}\\
    \Delta P_\text{uncorr,\,avg}            &= \sqrt{\sum_i \left( W_i \, \Delta P_{\text{uncorr},i} \right)^2}\\
  \end{split}\end{equation}
where the \emph{scale} errors of either polarimeter are considered correlated across periods.
\end{enumerate}

{\noindent \bf Example: Calculate the total error on the quantity $P_L \pm P_R$ with polarisation averages from two different run periods 
with opposite helicity.}\\
For data analysis one has to consider the polarisation values from each run period together with the three types of error as discussed above. 
The \emph{scale} errors of each polarimeter are fully correlated across run periods to themselves but uncorrelated to the \emph{scale} errors of the
other polarimeter, whereas the \emph{uncorr} errors are uncorrelated.

Consider two run periods with opposite helicity (denoted by $L$ and $R$):
\begin{eqnarray*}
   P_{L} & \pm & \Delta P_{L}^\text{LPOL,\,scale} \pm \Delta P_{L}^\text{TPOL,\,scale} \pm \Delta P_{L}^\text{uncorr} \\
   P_{R} & \pm & \Delta P_{R}^\text{LPOL,\,scale} \pm \Delta P_{R}^\text{TPOL,\,scale} \pm \Delta P_{R}^\text{uncorr}
\end{eqnarray*}
Note, that by convention all errors are positive numbers here and that in the case of different helicities, where $P_L>0$ and $P_R<0$, 
the \emph{scale} errors are negatively correlated. The total error on the quantity $P_L \pm P_R$ thus is given by

\begin{equation}\begin{split}
   \Delta(P_L \pm P_R) = \Bigl( &\bigl(\Delta P_{L}^\text{uncorr})^2 + (\Delta P_{R}^\text{uncorr}\bigr)^2 \Bigr.\\
                       &+ \bigl(\Delta P_{L}^\text{LPOL,\,scale} \mp \Delta P_{R}^\text{LPOL,\,scale} \bigr)^2 \\
                       &+\left.\!\bigl(\Delta P_{L}^\text{TPOL,\,scale} \mp \Delta P_{R}^\text{TPOL,\,scale}\bigr)^2 \right)^\frac{1}{2}
\end{split}\end{equation}


\section{Summary}
In this note the final analyses of the polarisation in the TPOL and the final results of the LPOL polarimeter at HERA II are presented. 
The analysis for the transverse polarimeter TPOL is based on a complete reanalysis of the data recorded during the HERA II running. Effects 
from the position and property of the interaction region between the laser beam and the lepton beam are taken into account. For the first 
time the effect from the non-vanishing linear light polarisation are considered as well. An internal systematic error of the TPOL of $1.9\%$ 
has been found. 
The analysis of the LPOL has been reevaluated and the systematic errors have been essentially confirmed to be $2.0\%$.

The polarisation scale for the TPOL has been confirmed in a totally independent way using rise time measurements, which were done at the end 
of the HERA II running in 2007. Good agreement with the polarisation scale intrinsic to the analysis is found, with no indications of a 
systematic shift or bias.

A procedure to average the results from the two polarimeters TPOL and LPOL is described. It takes into account possible correlations between 
the two polarimeters, and between periods of different running conditions. A prescription is presented how the experiments should use the 
polarisation values in their analyses. 

\begin{table} \begin{center}
\begin{tabular}[h]{lcccc|c}
\hline
Period & $P_\text{avg}\,(\%)$ & $\delta P_\text{avg}^\text{uncorr}\,(\%)$ & $\delta P_\text{avg}^\text{LPOL,scale}\,(\%)$ 
                              & $\delta P_\text{avg}^\text{TPOL,scale}\,(\%)$ &  $\delta P_\text{avg}^\text{total}\,(\%)$\\
\hline
 $e^+_L$ &-36.3& 0.73& 0.70& 0.67& 1.21\\ 
 $e^+_R$ & 30.4& 0.75& 0.98& 0.64& 1.39\\
 $e^-_L$ &-26.0& 1.10& 0.99& 1.10& 1.84\\
 $e^-_R$ & 30.3& 1.03& 0.91& 1.15& 1.79\\
\hline
\end{tabular}
	\caption{\sl Table summarising estimated average results for the four different periods with either positron or electron running and
left or right handed polarisation state. Here, left/right helicity refers to a negative/positive LPOL polarisation sign, following the convention at 
H1. For this estimation the suggested averaging procedure has been applied to all of HERA II polarimeter data
using equal integrated luminosity weights for all 5 minute buckets as described in Sect.~\ref{sect:avgproc}, enumeration point 2. The total 
systematic uncertainty is calculated by adding all three distinct sources in quadrature. The values are for illustration only and will vary in the 
analyses when the experiments' integrated luminosities and data selection cuts are properly taken into account.
\label{tab:avgresults}}
\end{center} \end{table}
When applying the suggested averaging procedure to the complete set of HERA II polarimeter data using equal contributing integrated luminosity 
values ${\cal L}_i$ for every 5 minute bucket as described in Sect.~\ref{sect:avgproc}, enumeration point 2, the contributions of systematic
uncertainty for the four running periods with either electrons or positrons and left or right handed polarisation can be estimated. The results 
are shown in Tab.~\ref{tab:avgresults}.
Following this estimation the total systematic uncertainty, calculated by adding the three systematic contributions in quadrature, is for all four 
periods $< 1.9\%$ and for the positron run periods even $< 1.4\%$. These estimated results are for illustration only. It is expected that the precise
results will vary somewhat for the experiments as their integrated luminosities and data selection cuts need to be taken into account properly.

\section*{Acknowledgements}
We are grateful to  P.~Sch\"uler and V.~Gharibyan as well as all members of the POL2000 group for their support and discussions.
The support of the H1, HERMES and ZEUS experiments is warmly acknowledged. 
We thank the HERA machine group whose outstanding efforts have made this experiment possible and the DESY directorate for its support.
We thank also M.~Voigt and D.~P.~Barber for discussions and J.~List for technical help.



\begin{thebibliography}{99}
     \bibitem{privcomm:Olsson}{J.~Olsson, H1 Collaboration, \
              {private communication, 2011}}
     \bibitem{cavity}{M.~Jacquet, \
             {Precise measurement of the longitudinal polarisation at
              HERA with a Fabry-Perot cavity polarimeter},
              \mbox{J.Phys.Conf.Ser. 298 (2011) 012023}}
     \bibitem{LPOL_paper}{M.~Beckmann et al., \ 
              {The Longitudinal Polarimeter at HERA}, 
              \mbox{NIM {\bf A 479} (2002) 334-348}}
     \bibitem{REPO_DESCRIPTION}{R.~Fabbri, \ 
              {Description of the Reorganization of the LPOL Data},
              \mbox{\url{http://www.desy.de/~pol2000/analysisarchive}}}
     \bibitem{PRC_2009}{The Pol2000 Group, \ 
              {The HERA Polarimeters, Status of Analyses}\\
              Note submitted to the $68^\text{th}$ DESY Physics Research Committee,
              DESY, Hamburg, Germany, November 2009, 
              POL2000-2009-001}
     \bibitem{menden}{F.~M.~Menden, \
              {Determination of the Gluon Polarization in the Nucleon}, 
              PhD Thesis 2001, DESY-THESIS-2001-060}
     \bibitem{STATISTICS}{T.~H.~Wonnacott and R.~J.~Wonnacott, \
              {Introductory Statistics},
              John Wiley \& Sons Inc.}
     \bibitem{lpoloverview}{A.~Airapetian et al., \
              {Overview on Systematic Studies of the HERMES Longitudinal Polarimeter}, 
              HERMES Internal Report 04-014,
              \mbox{\url{http://www.desy.de/~pol2000/documents}}}
     \bibitem{pol-summer2007}{A.~Airapetian et al., \
              {Using the HERA Polarisation Measurements, Recommendations for the Summer 2007 Conferences}, 
              HERMES-07-016, POL2000-2007-001}
     \bibitem{focusnote}{F.~Corriveau et al., \
             {A Calibration of the HERA Transverse Polarimeter for the 2003/2004 Data}, internal note, unpublished,
              2004, \mbox{\url{http://www.desy.de/~pol2000/documents}}}
     \bibitem{geant3}{R.~Brun, F.~Carminati and S.~Giani,\
             {GEANT Detector Description and Simulation Tool},
              Reference Manual, {CERN} Program Library Entry {\bf W5013}, 1994,
              \mbox{\url{http://wwwasd.cern.ch/wwwasd/geant}}}
     \bibitem{privcomm:Voigt}{M.~Voigt, DESY, \
              {private communication, 2007}}
     \bibitem{privcomm:Schueler}{P.~Sch\"uler, DESY, \ 
              {Measurements of the approximate laser size in Nov.~1999},
              private communication, 2007}
     \bibitem{privcomm:Vahag}{V.~Gharibyan, DESY, \
              {private communication, 2007}}
     \bibitem{Wigmans}{R.~Wigmans, \ 
              {Calorimetry: Energy Measurement in Particle Physics},
              Clarendon Press, Oxford, UK, 2000}
     \bibitem{corrnote}{S.~Schmitt, \
             {TPOL $\eta$ resolution studies}, internal note, unpublished,
              2010, \mbox{\url{http://www.desy.de/~pol2000/ analysisarchive}}}
     \bibitem{cerntb}{T.~Behnke et al, \
             {The Transverse Polarimeter (TPOL) Test Beam at CERN in July-August 2001},
              2002, ZEUS-02-019,
              \mbox{\url{http://www.desy.de/~pol2000/documents}}}
     \bibitem{Minuit:1994}{F.~James, \
              {{MINUIT} - Function Minimization and Error Analysis},
              Reference Manual, {CERN} Program Library Entry {\bf D506}, 1994}
     \bibitem{barberbook}{D.P.~Barber and G.~Ripken, \
             {Handbook of Accelerator Physics and Engineering}, 
              Eds.~A.W.~Chao and M.~Tigner, 1st edition, 3rd printing, 
              World Scientific, 2006}
     \bibitem{privcomm:Barber}{D.~P.~Barber, DESY, \
              {private communication, 2007}}
     \bibitem{risetimecalcs}{S.~Schmitt, \
              {Risetime analysis: machine calculations}, 
              talk presented at the POL2000 meeting, Nov.~6th 2007,
              \mbox{\url{http://www.desy.de/~pol2000/talks}}}
     \bibitem{pdg}{K.~Nakamura et al. (Particle Data Group), \
              \mbox{J.~Phys.~G 37, 075021 (2010)}
              and 2011 partial update for the 2012 edition}


\end{thebibliography}
\end{document}